\documentclass[useAMS,usenatbib]{mn2e}

\usepackage{graphicx}
\usepackage{gensymb}
\usepackage{epstopdf}
\usepackage{rotating}
\usepackage{lscape}
\usepackage{aas_macros}
\usepackage{hyperref}

\newcommand{\unit}[1]{\ensuremath{\, \mathrm{#1}}}
\newcommand{\change}[1]{{#1}}
\newcommand{\changeb}[1]{{\bf #1}}

\title[Transient and variable radio sources with MOST]{A 22-year Southern Sky Survey for Transient and Variable Radio Sources using the Molonglo Observatory Synthesis Telescope}
\author[K. Bannister et al.]{K. Bannister$^{1}$\thanks{E-mail:
k.bannister@physics.usyd.edu.au}, T. Murphy$^{1, 2}$,  B. M. Gaensler$^{1, 3}$, R. Hunstead$^{1}$, S. Chatterjee$^{4}$ \\
$^{1}$Sydney Institute for Astronomy, School of Physics A29, The University of Sydney, NSW 2006, Australia\\
$^{2}$School of Information Technologies, The University of Sydney, NSW 2006, Australia\\
$^{3}$Australian Research Council Federation Fellow\\
$^{4}$Department of Astronomy, Cornell University, Ithaca, NY 14853, USA}

\begin{document}
\date{Accepted 2010 XXX XXX. Received XXX December XX; in original form XXX XXX XXX}
\pagerange{\pageref{firstpage}--\pageref{lastpage}} \pubyear{2010}

\maketitle

\label{firstpage}

\begin{abstract}
We describe a 22-year survey for variable and transient radio sources, performed with archival images taken with the Molonglo Observatory Synthesis Telescope (MOST). This survey covers $2775 \unit{deg^2}$ of the sky south of $\delta < -30\degree$ at an observing frequency of 843~MHz, an angular resolution of $45  \times 45  \csc | \delta| \unit{arcsec^2}$ and a sensitivity of $5 \sigma \geq 14 \unit{mJy~beam^{-1}}$.  We describe a technique to compensate for image gain error, along with statistical techniques to check and classify variability in a population of light curves, with applicability to any image-based radio variability survey. Among radio light curves for almost 30000 sources, we present 53 highly variable sources and 15 transient sources. \change{Only 3 of the transient sources, and none of the variable sources have been previously identified as transient or variable.} Many of our variable sources are suspected scintillating Active Galactic Nuclei. \change{We have identified three variable sources and one transient source that are likely to be associated with star forming galaxies at $z \simeq 0.05$, but whose implied luminosity is higher than the most luminous known radio supernova (SN1979C) by an order of magnitude.} We also find a class of variable and transient source with no optical counterparts.
\end{abstract}


\section{Introduction}

Many of the most extreme events in the universe manifest themselves through variability. For example, explosive events such as Gamma Ray Bursts (GRBs) are thought to originate from stellar collapses which can momentarily out-shine the entire universe \citep{MacFadyen99}; the magnetosphere of the Crab pulsar can produce nanosecond-duration pulses with a brightness temperature of $10^{38}$~K \citep{Hankins03}; thermonuclear explosions up to a month long have been observed in X-ray binaries (\citeauthor{Sokoloski06} \citeyear{Sokoloski06},  \citeauthor{Bode06} \citeyear{Bode06}); and large variations in accretion rate have been observed in accreting black holes, ranging in size from Galactic microquasars \citep{Fender99}, to supermassive black holes (e.g. \citeauthor{Gezari06}  \citeyear{Gezari06}, \citeauthor{Blaz05} \citeyear{Blaz05}).

In addition to transient emission indicating extreme events, many aspects of the interstellar and intergalactic media, and of the universe as a whole, can be revealed through measurements in the time domain. For example, the presence of massive compact halo objects \citep{Alcock97} and planets \citep{Gould05} can be inferred from gravitational microlensing of starlight around massive objects; the expansion history of the universe can be measured using light curves of Type 1a supernovae (\citeauthor{Riess98} \citeyear{Riess98}, \citeauthor{Perlmutter99} \citeyear{Perlmutter99}); and interstellar scintillation (ISS) of compact radio sources can be used to infer the distance, velocity and turbulence of otherwise invisible ionised clouds in the Milky Way \citep{Bignall2003IDV}.

Low-energy radio photons can be easily produced by both thermal and non-thermal processes, and in contrast to optical and X-ray wavelengths, the universe is essentially transparent to these photons. As a vast range of objects produce radio emission, and these signals are detectable over much of the observable universe, radio variability surveys are an effective method of discovering and characterising such sources and propagation phenomena.  Furthermore, important propagation effects such as interstellar scintillation are only observed in the radio.

While radio observations are often used for follow-up of detections at other wavelengths, blind radio surveys have also been a fruitful method for discovering new objects and phenomena in their own right. \citet{gregory1986radiopatrol} discovered a number of flaring radio stars in a survey of the Galactic plane, while \citet{bower2007sta} with 944 images of a single field, discovered ten faint transient sources, eight of which had no optical counterparts. \citet{carilli2003vsrs} also discovered a number of highly variable sources with observations of a single field. \citet{levinson2002} covered a large fraction of the sky with two epochs by comparing the NVSS \citep{NVSS} and FIRST \citep{FIRST} images, and was able to set constraints on the GRB beaming fraction. \citet{Croft10} set an upper limit on the rate of transients with flux densities greater than 40~mJy of $0.004 \unit{deg^{-2}}$ by comparing a 690 square degree image obtained with the Allan Telescope Array (ATA) at 1.4~GHz with NVSS.

A number of transient source detections remain unexplained. \citet{Matsumura09} have discovered a number of $>1\unit{Jy}$  transient sources with $\sim$ 1 day time-scales at a range of Galactic latitudes. \citet{lorimer2007bmr} discovered a 30~Jy transient source at 1.4~GHz with a 5~ms time-scale, but there is evidence this may be atmospheric \citep{BurkeSpolaor10lg}. \citet{lec2008deep} found a $\sim 100 \unit{mJy}$ transient at 330~MHz during deep Very Long Baseline Interferometer (VLBI)  observations, and \citet{Becker10} found 39 variable sources in the Galactic Plane with 3 epochs of 5~GHz Very Large Array (VLA) observations, most of which had no known counterparts at other wavelengths.

In this paper we present a survey for transient and variable sources  at 843~MHz with characteristic time-scales from days to years. We aim to characterise the variable and transient radio sky at this frequency, enumerate the most extreme variable sources, find transient sources and develop techniques suitable for upcoming radio surveys.

For our analysis we are using the Molonglo Observatory Synthesis Telescope (MOST). The MOST and its predecessor the Molonglo Cross, have been at the forefront of research in this area over several decades. Using the Molonglo Cross, \citet{Hunstead72} was the first to observe low frequency variability at 408~MHz.  In an archival survey of calibrator measurements, \citet{gaensler2000most} found that one-third of the bright point sources at 843~MHz were variable, and a weak Galactic latitude dependence was demonstrated, indicating that interstellar scintillation was at least partly responsible. The MOST has been used to discover several hundred pulsars in a number of surveys \citep{Manchester85}, and a survey for short-duration transients has been performed by \citet{Amy89} but with a null result. As a follow-up instrument, the MOST was the first telescope to detect prompt radio emission from SN1987A (\citeauthor{Turtle87} \citeyear{Turtle87}), and it has been used to monitor a number of Galactic accreting systems \citep{Hannikainen98}, a brightening supernova remnant \citep{Murphy08} and a magnetar flare \citep{Gaensler05}.

The MOST archive is a unique resource which is able to address the limitations of sky coverage, sensitivity and cadence that have accompanied other blind surveys. In addition, the experience gained from analysing wide field-of-view images provides an opportunity to develop techniques suitable for upcoming wide-field transient and variability surveys such as the Variables And Slow Transients (VAST) survey for the Australian Square Kilometre Array Pathfinder (ASKAP) \citep{Chatterjee10} and the LOFAR Transients Key Project \citep{Fender08}.

In $\S$2 we describe the MOST and its image archive. In  $\S$3 we describe our method of extracting light curves from this archive and in $\S$4 we present the results of applying our method, including quality checks and selected sources. In  $\S$ 5 we discuss our results, \change{and in $\S$6 we discuss some noteworthy sources. We draw our conclusions in $\S$7.}

\section{Observations with the Molonglo Observatory Synthesis Telescope}

The Molonglo Observatory Synthesis Telescope (MOST) is located near Canberra, Australia and was constructed by modification of the East-West arm of the former One-Mile Mills Cross telescope. The MOST is an east-west synthesis array comprising two cylindrical paraboloids each of dimension 778~m $\times$ 12~m separated by 15~m. Radio waves are received by a line feed system of 7744 circular dipoles. The telescope is steered in the North-South axis by mechanical rotation of the paraboloids about their long axis, and in the East-West axis by phasing the feed elements along the arms. By tracking the field over 12~h, a full synthesis image can be formed. The near-continuous UV coverage from 15~m to 1.6~km results in good response to complex structure and low sidelobe levels. Technical specifications are shown in Table~\ref{tab:mostspecs}. The MOST has been described in detail by \citet{Mills81} and \citet{robertson1991themost}.

Since 1986 the MOST has observed a single field for a 12~h synthesis almost every week night and often during the day on weekends. At the beginning and end of each 12~h synthesis, a set of up to 8 different calibrator sources is observed. Calibrator measurements are discarded if they do not pass a number of checks, and the remaining measurements are averaged to obtain gain and pointing solutions for the beginning and end of the observation. The gain and pointing solutions are linearly interpolated over the synthesis time between two calibrator observations. The full list of calibrators is described by \citet{CambellWillson94}. Known variable calibrators were removed from the list following the analysis of \citet{gaensler2000most}.

Our present analysis is performed on the final images processed according the procedure described by \citet{Green1999MGPS1} and \citet{Bock99}.

\begin{table}
\centering
\caption{Technical specifications of the MOST. The Declination range is given for a fully synthesised (12~h) image. The sensitivity and field sizes are mode-dependent and shown in Table~\protect \ref{tab:imagetypes}.}
\label{tab:mostspecs}
\begin{tabular}{lr}
\hline
Parameter & Value \\
\hline
Centre Frequency & 843~MHz \\
Bandwidth & 3~MHz \\
Polarisation & Right Hand Circular (IEEE) \\
Declination range for full synthesis & $-30 \degree$ to $-90\degree$ \\
Synthesised beam & $43 \times 43 \csc | \delta |$ arcsec$^2$ \\
Restoring beam & $45 \times 45 \csc | \delta |$ arcsec$^2$  \\ 
Area per pixel &  $11 \times 11 \csc | \delta | $ arcsec$^2$ \\
Dynamic Range (typical) & 100:1 \\
\hline
\end{tabular}
\end{table}

\subsection{Observing modes}
\label{sec:modes}
\begin{figure*}
\centering
\includegraphics[width=\textwidth]{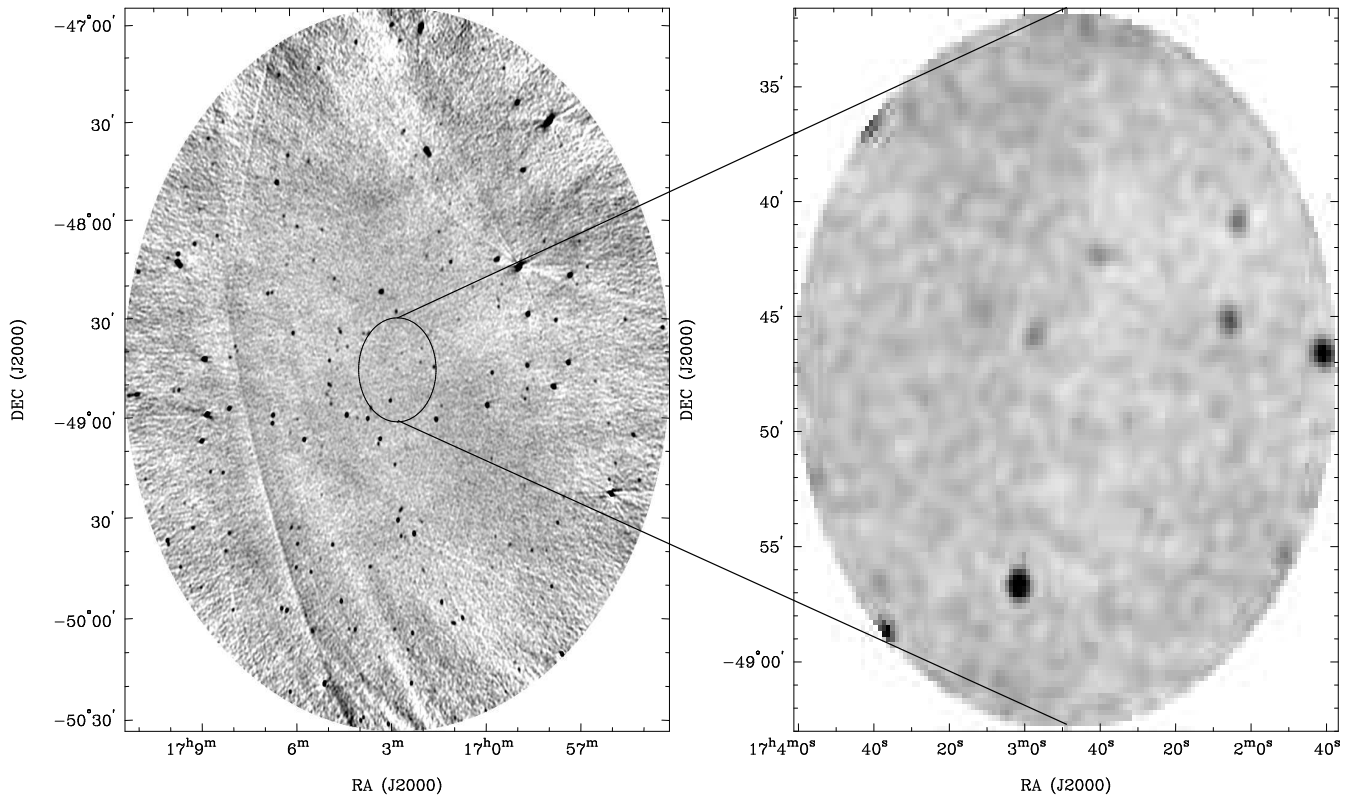}
\caption{Examples of MOST images from the MGPS surveys. Left: An `I' field from MGPS-2, Right: A `B' field from MGPS-1. Both images are centred on $l=333.9\degree$, $b=-4.4\degree$. The grey scale is the same for each image ranging from $-5 \unit{mJy~beam^{-1}}$ to $+15 \unit{mJy~beam^{-1}}$. The resolution of each image is the same $(45 \times 61 )$~arcsec, and the measured background RMS values are $1.5$ and $0.85\unit{mJy~beam^{-1}}$ respectively. The `B' field overlaps the `I' field in the region shown. A number of typical artefacts are visible, including radial spokes from bright sources, regions of negative emission, grating rings and increased noise towards the edge of the `I' field.}
\label{fig:exampleImages}
\end{figure*}

The MOST is capable of observing in a number of different modes depending on the desired signal-to-noise ratio and field size, as summarised in Table~\ref{tab:imagetypes}. The MOST field of view is elliptical with the major axis aligned in the North-South direction and size $a = a_0 \csc{|\delta|}$, where $\delta$ is the declination of the field centre and $a_0$ is the mode-specific minor axis size. The area of an image is approximately equal to $\pi (a_0/2)^2 \csc|\delta_i|$ and the spatial resolution of each image is the same for each mode at $45 \times 45 \csc|\delta|$~arcsec. Examples of images observed in two different modes are shown in Fig.~\ref{fig:exampleImages}.

All MOST images are primary-beam corrected, which results in increased noise towards the edges of images. The increased noise is most pronounced in the wide-field `I' images, which are larger than the half-power point of the primary beam. Therefore, we define an ellipse-shaped available area, parameterised by the minor axis shown in Table~\ref{tab:imagetypes}, outside which we ignore any detections and measurements.

We define an image as `usable' if it passes the visual inspection and post-facto calibration procedure described in Section \ref{sec:post_facto_calibration}.

\begin{table*}
\centering
\caption{MOST observing mode, image parameters (see Section \ref{sec:modes}) and number of images in the archive. The columns are: the mode code, typical Root Mean Square (RMS) of the noise in the image, full and available minor axes, total number of images in the archive and equivalent area (assuming the available minor axis) and number and equivalent area of images for which visual inspection and post-facto calibration were successful (see Section \protect \ref{sec:post_facto_calibration}). The `H'  and `I' modes were only available since the 1993 wide-field upgrade described by \protect \citet{Large94}.}
\label{tab:imagetypes}
\begin{tabular}{|c|c|c|c|r|r|r|r|}
\hline
&    					& \multicolumn{2}{c}{Minor Axis ($a_0$)} &  \multicolumn{2}{c}{Total Images} & \multicolumn{2}{c}{Usable Images} \\
& Typical RMS         & Full & Available  &  Count & Area & Count & Area\\
Mode & (mJy beam$^{-1})$                   & (deg) & (deg)      &             & (deg$^2$) &        &  (deg$^2$) \\
\hline
B & 0.8 & 0.20 & 0.19 & 928 & 220.3 & 3 & 0.4 \\
C & 1.1 & 0.39 & 0.37 & 85 & 141.2 & 8 & 4.3 \\
D & 1.3 & 0.58 & 0.52 & 2292 & 3045.8 & 872 & 1028.4 \\
H & 1.8 & 1.18 & 1.00 & 6 & 29.5 & 0 & 0.0 \\
I & 2.0 & 1.36 & 1.02 & 3915 & 18156.4 & 2128 & 9266.1 \\
\hline
\end{tabular}
\end{table*}

\subsection{Archive coverage}
The primary survey conducted with MOST was the Sydney University Molonglo Sky Survey (SUMSS; \citeauthor{Mauch2003sumss} \citeyear{Mauch2003sumss}), which covers the sky south of $ -30\degree$ excluding the Galactic plane ($|b| < 10\degree$), and was carried out between mid-1997 and 2006. The MOST was also used to conduct two Molonglo Galactic Plane Surveys: MGPS-1 \citep{Green1999MGPS1}, covering the Galactic plane from $245\degree \le l \le 355\degree $ and $|b| \le 1.5\degree$, and MGPS-2 \citep{murphy2007sem} covering $245\degree < l < 365\degree $ and $|b|< 10 \degree$. There has also  been a wide range of directed observations and monitoring programs. Here we analyse 7227 images, observed between 1986 December 18 and 2008 August 28.

\begin{figure*}
\centering
\includegraphics[width=0.49\textwidth]{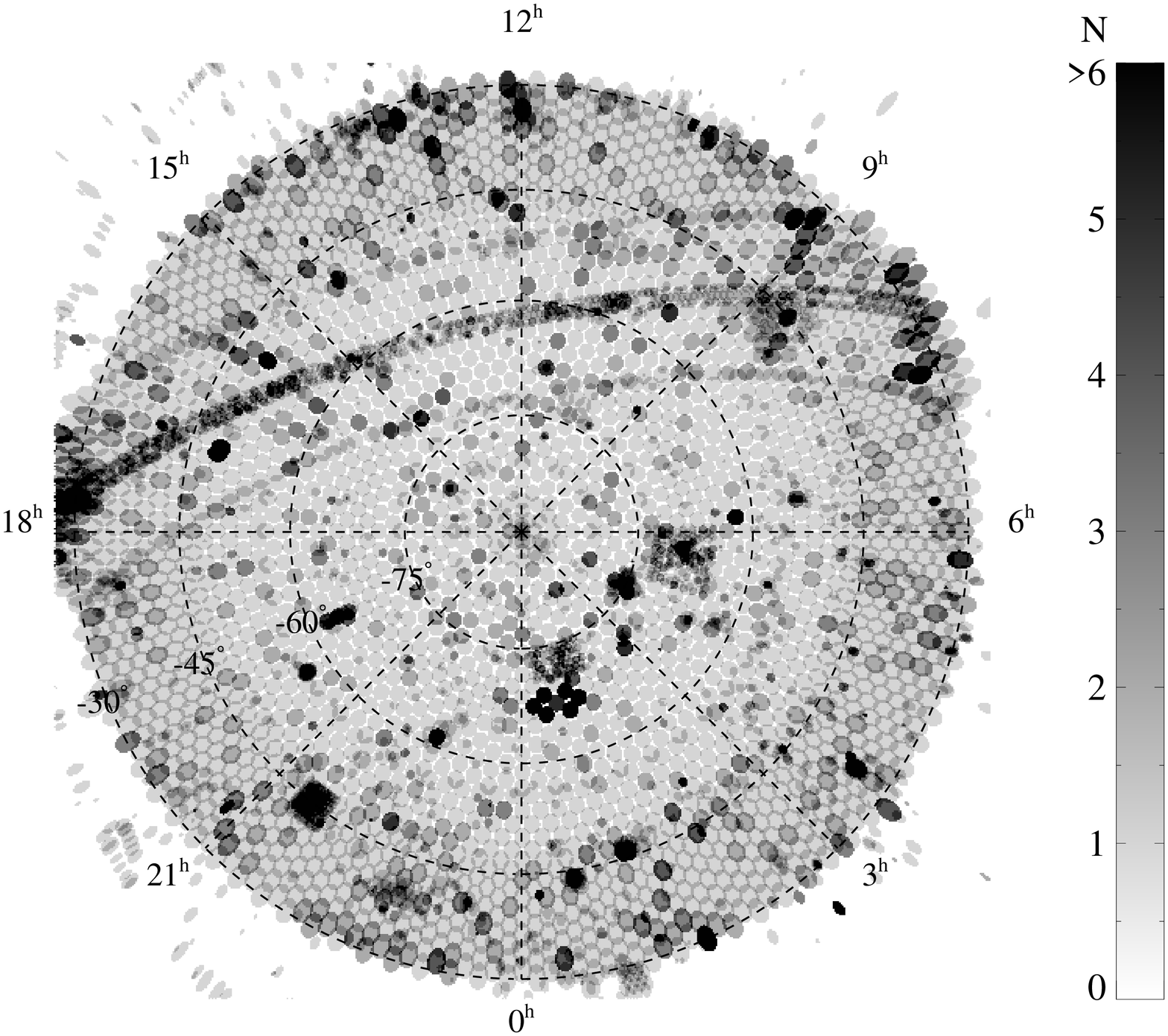}
\includegraphics[width=0.49\textwidth]{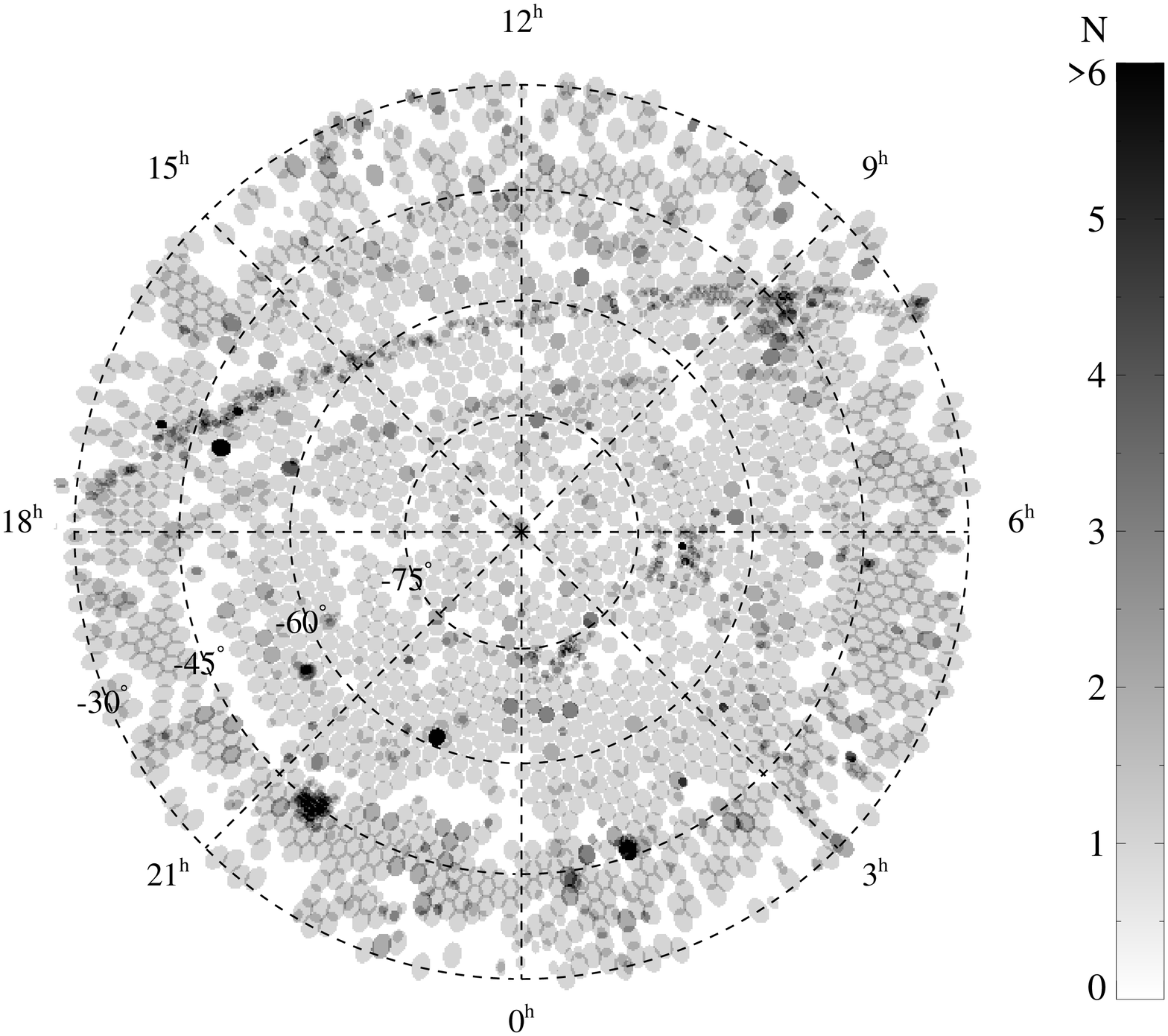}
\caption{Coverage of the MOST archive in a Lambert equal area projection sampled on 0.1\degree grid and centred on the South Celestial Pole. Left: the entire archive. Right: usable images after visual inspection and post-facto calibration (see Section \ref{sec:post_facto_calibration}). Some 4200 of the 7200 images were not useable and appear only in the left panel. The colour axis is number of epochs with light grey being no images, dark grey being one image (unusable for detecting transients and variability) and colours yellow to blue indicating 2 or more images. The shape of each image is the available size shown in Table~\ref{tab:imagetypes}, not the full image size. The SUMSS survey covers the sky south of $-30\degree$ and the Galactic plane surveys: MGPS-1 and MGPS-2 can be clearly seen as a band from the left-middle to top-right of both images. Repeated fields include the Hubble Deep Field South (52 usable epochs), SN1987A (13 usable epochs),  GX339-4 (13 usable epochs) and GRO1655-40 (15 usable epochs).}
\label{fig:coverageplot}
\end{figure*}

A plot of the total and usable sky coverage of the archive is shown in Fig.~\ref{fig:coverageplot}. All of the sky south of $-30 \degree$ has been covered at least once due to the SUMSS survey, and the Galactic plane has been observed twice with the MGPS-1 and MGPS-2 surveys. Some regions have been covered multiple times due to monitoring programs (e.g. the microquasar GX339$-$4 has been observed 14 times) or deep imaging (e.g. the Hubble Deep Field South, which has 52 usable epochs). A number of images that were excluded from further analysis due to poor quality discovered during visual inspection, or difficulty in obtaining a post-facto calibration, resulting in the usable coverage being significantly less than the total coverage.

For the purposes of comparing survey areas, we define the two-epoch equivalent coverage area as:

\begin{equation}
\label{eq:A2epoch}
A_{\mathrm{two-epoch}} = \sum_{i=1}^{N_{\mathrm{pos}}}{A_i \left( N_{\mathrm{epoch, i}} - 1 \right )},
\end{equation}

\noindent where $N_{\mathrm{pos}}$ is the total number of independent directions on the sky measured in the survey, $A_i$ is the area of a particular direction of interest, and $N_{\rm epoch, i}$ is the number of epochs a particular direction has been measured.

\begin{figure}
\centering
\includegraphics[width=1.0\linewidth]{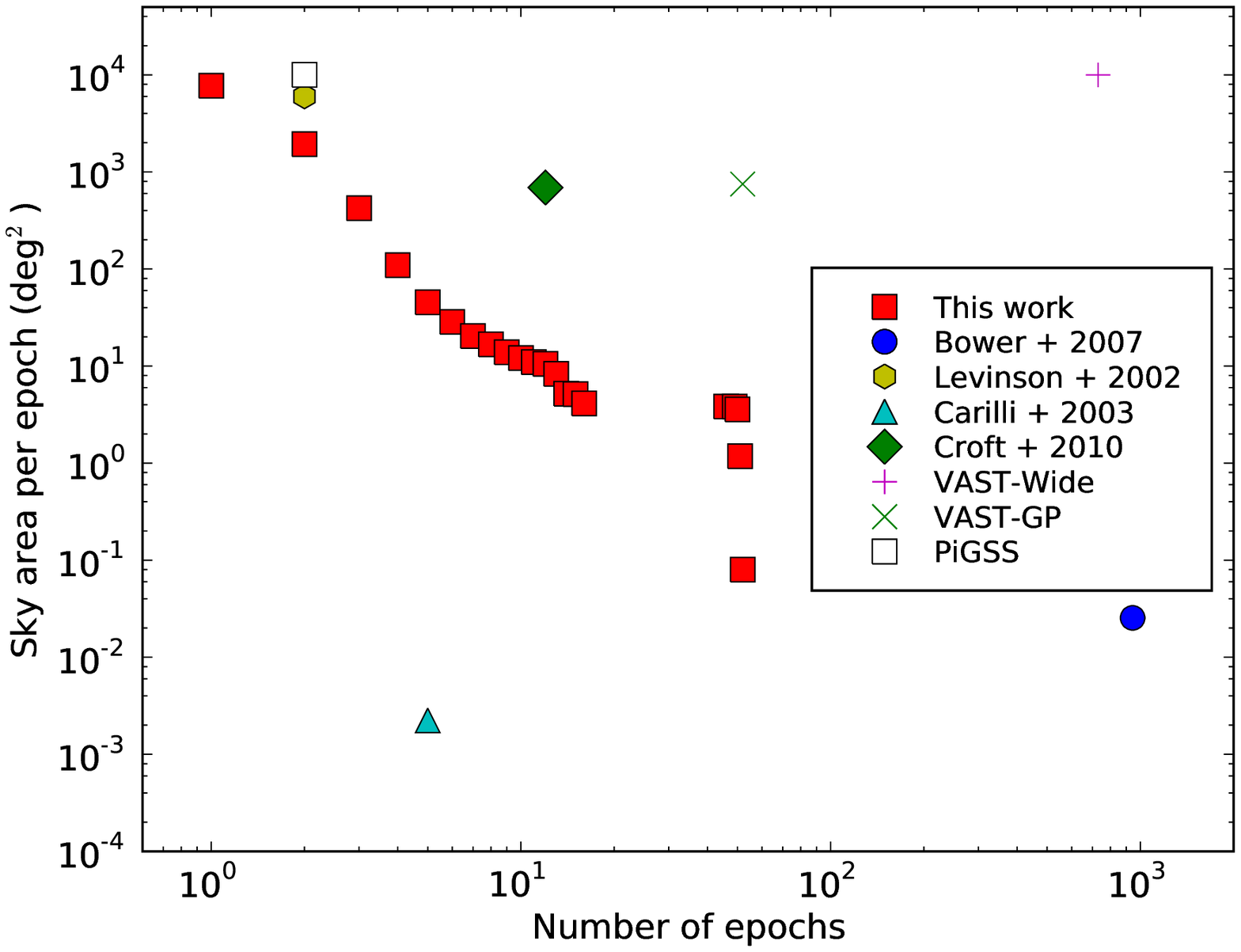}

\caption{Area of epoch vs. number of epochs for blind radio transient surveys. At the extremes, \protect \citet{bower2007sta}  covered $0.03 \unit{deg^2}$ with 844 epochs, \protect \citet{levinson2002} covered $5990 \unit{deg^2}$ with 2 epochs, \change{\citet{carilli2003vsrs} covered a $0.002 \unit{deg^2}$ with 5 epochs and \citet{Croft10} covered $690 \unit{deg^2}$ with 12 epochs}. This work, shown in red squares, spans some of the intervening parameter space. VAST-Wide and VAST-GP are the proposed wide-field and Galactic plane surveys of the Variables And Slow Transients (VAST) project, planned for ASKAP, and the Pi GHz Sky Survey (PiGSS) \citep{Bower10} is planned for the \changeb{Allen} Telescope Array.}
\label{fig:coveragehistogram}
\end{figure}

\begin{figure}
\centering
\includegraphics[width=1.0\linewidth]{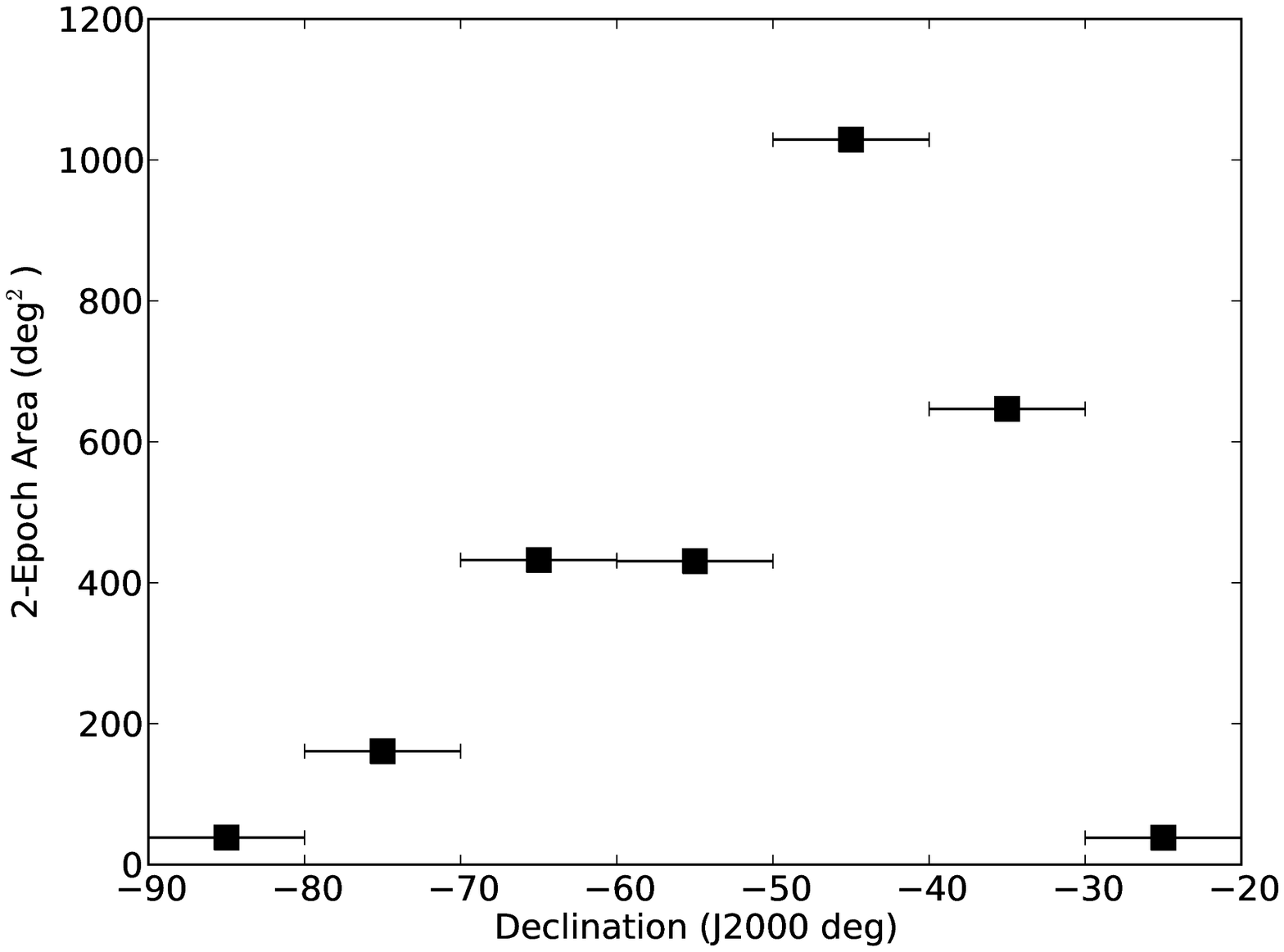}
\includegraphics[width=1.0\linewidth]{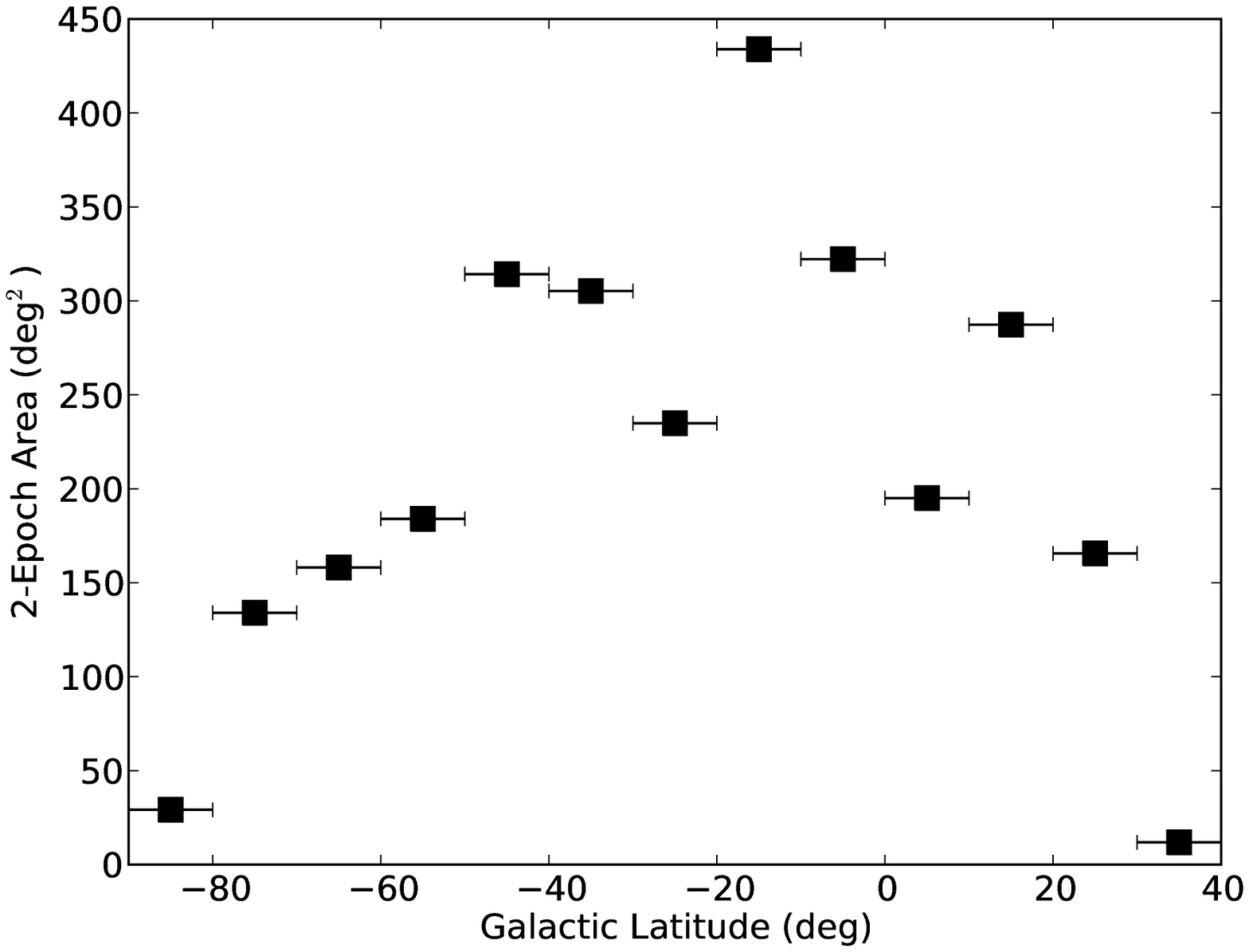}

\caption{Two epoch coverage vs. declination (above) and vs. Galactic latitude (below). The horizontal error bars denote the bin widths of 10\degree.}
\label{fig:coverage_bins}
\end{figure}

A comparison of the two-epoch equivalent coverage area of the MOST archive with other blind radio variability surveys is shown in Fig.~\ref{fig:coveragehistogram} and Table~\ref{tab:2epoch_vs_survey}. The MOST archive covers $\sim 10^4 \unit{deg^2}$ once, and $\sim 2 \times 10^{3}\unit{deg^2}$ twice. The MOST archive has a maximum of 52 epochs on a small ($\sim 0.1 \unit{deg^2}$) patch of sky. At the large-epoch extreme, \citet{bower2007sta} covered a very small ($0.024 \unit{deg^2}$) patch of sky 844 times. Plots of the two-epoch equivalent coverage area versus declination and Galactic latitude of the MOST archive are shown in Fig.~\ref{fig:coverage_bins}.

\begin{table}
\centering
\caption{Two-epoch coverage area ($A_{\rm two-epoch}$) for a range of recent and proposed blind radio variability surveys. VAST-Wide and VAST-GP are the wide-field and Galactic plane surveys of the Variables And Slow Transients (VAST) project, planned for the Australian Square Kilometre Array Pathfinder (ASKAP).}
\label{tab:2epoch_vs_survey}
\begin{tabular}{lr}
\hline
Survey &$A_{two-epoch} (\unit{deg^2}$) \\
\hline
\citet{carilli2003vsrs} & 0.0008 \\
\citet{bower2007sta} & 28 \\
This Work & 2776 \\
\citet{levinson2002} & 5990 \\
\citet{Croft10} & 6900 \\
PiGSS (proposed) \citep{Bower10} & 10000 \\
VAST-GP (proposed) & 38250 \\
VAST-Wide (proposed) & 7290000 \\
\hline
\end{tabular}
\end{table}

\begin{figure}
\centering
\includegraphics[width=\linewidth]{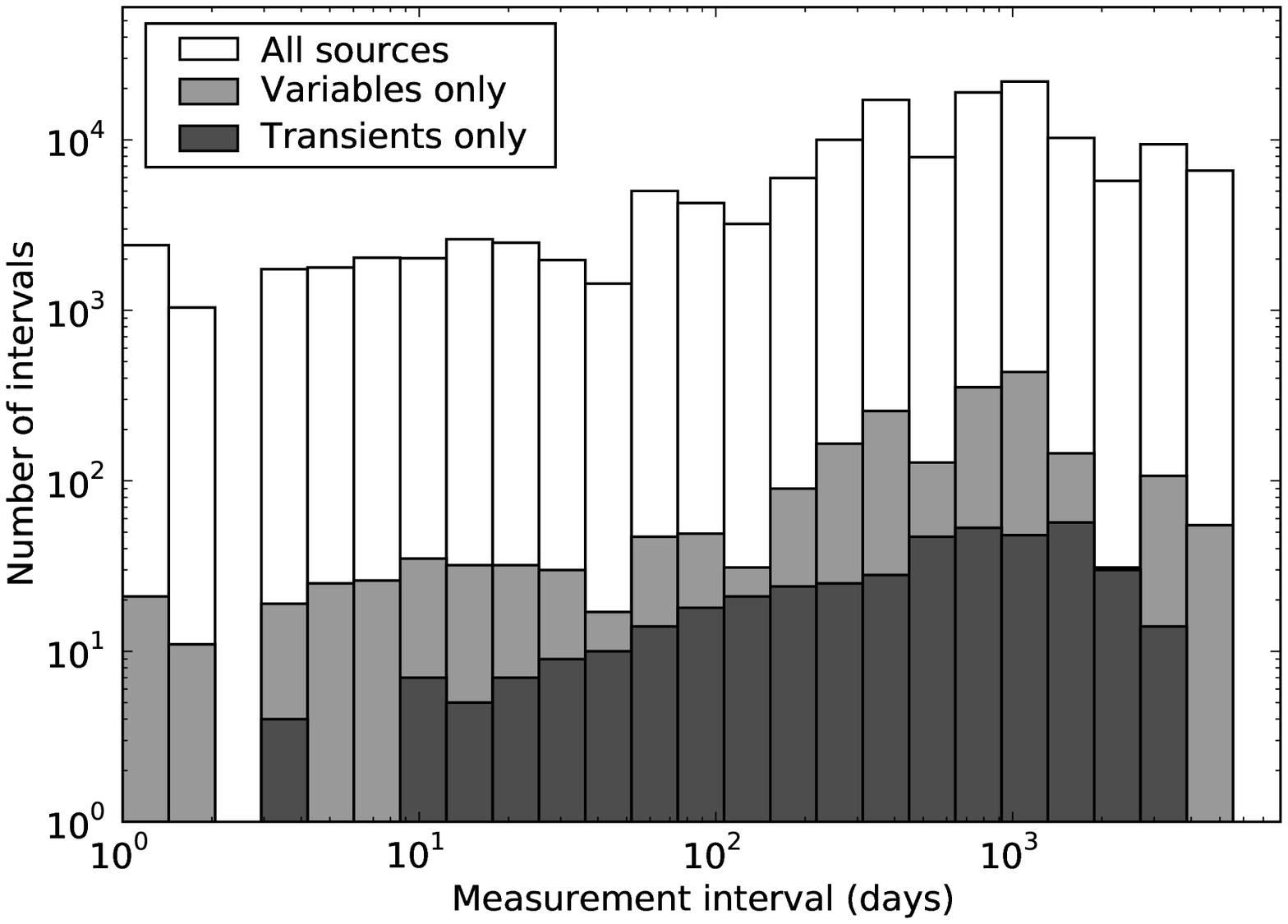}
\caption{\change{Histogram of the number of non-redundant pairs of measurements in the light curves of all sources in the sample, the variable sources only, and the transient sources only, binned by the interval between the measurements in days. }}
\label{fig:intervals_histogram}
\end{figure}

\change{The interval between epochs covers a wide range. Figure \ref{fig:intervals_histogram} shows a histogram of the number of intervals between non-redundant pairs \changeb{of} measurements of each source as a function of the interval in days. At the two extremes, Figure \ref{fig:intervals_histogram} shows that there are over two thousand pairs of measurements separated by one day, and the same number of measurement pairs separated by over 20 years. The archive is most sensitive to changes in flux density on the time-scale of around one thousand days, where it contains over ten thousand pairs of measurements. The results for transient and variable radio sources are discussed in Section \ref{sec:variability_time_scales}.}

\subsection{Image artefacts and systematic errors}

\subsubsection{Visually obvious artefacts}
MOST images suffer from a number of image artefacts that make accurate automatic point source detection and fitting difficult \citep{Mauch2003sumss}. In particular, some images are contaminated by extended emission (especially in the Galactic plane), with negative bowls or radial spokes around bright sources, grating rings due to the telescope geometry and bands across images due to solar and terrestrial interference (see, for example Fig.~\ref{fig:exampleImages}). To mitigate the effects of these artefacts, we discard clearly affected images by visual inspection, considered only point source fits with small formal errors, and measured the RMS and mean in the region around each detection to correct for localised offsets and noise. Extended emission was not explicitly flagged but point sources with small formal errors of fit were rarely found close to regions of extended emission.

\subsubsection{{\sc clean} artefacts}

\begin{figure}
\centering
\includegraphics[width=\linewidth,angle=270]{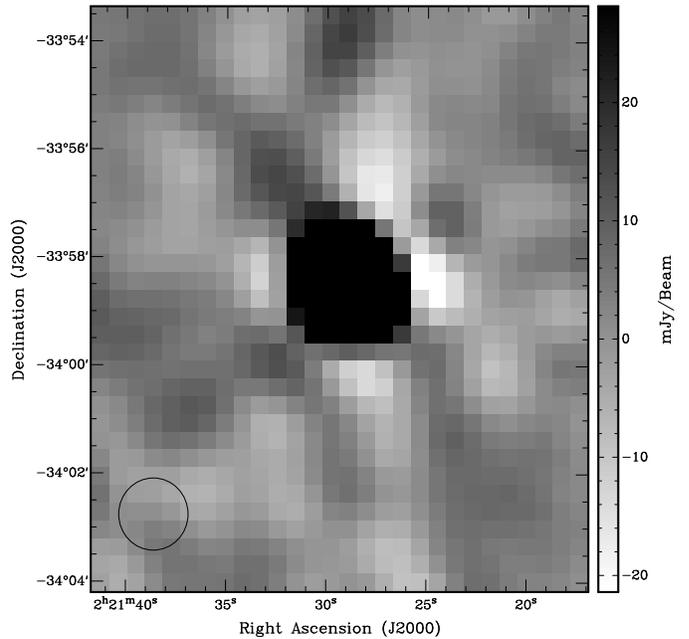}
\caption{Example of poor dirty beam modelling producing negative regions around SUMSS J022129$-$335825 after {\sc clean}ing. Fitting a 2D Gaussian in a 10x10 pixel region centred on the source results in an integrated flux density of $394.0 \unit{mJy}$ which is less than peak intensity of $449.3 \pm 16.6 \unit{mJy~beam^{-1}}$. The size of the fitted Gaussian is consistent with the size of the synthesised beam (shown in the bottom left). The large difference in integrated and peak flux densities is due to poor dirty beam modelling.}
\label{fig:bowlproblem}
\end{figure}

As with any interferometer, poor dirty beam modelling  can result in negative flux around sources after {\sc clean}ing (see \citeauthor{Green1999MGPS1} \citeyear{Green1999MGPS1} for a discussion). An example of this effect is shown in Fig.~\ref{fig:bowlproblem}. When performing automatic source fitting, the negative surface brightness around the edge of the source reduces the integrated flux density to a value significantly less than the peak flux density, which should not occur if {\sc clean} is using the correct beam. In a sample of 6767 otherwise good point source fits in the range $0\unit{h} < \unit{RA} < 10\unit{h}$, only 45 per cent had peak and integrated flux density measurements in formal agreement, and more than half of these had a peak value that significantly exceeded the integrated flux density. Assuming that variable and transient sources must be point sources at the MOST resolution, integrated flux densities were not used in the succeeding analysis, with peak values used throughout.

\subsubsection{Variable calibrators}
\citet{gaensler2000most} found that, between 1984 and 1996, 18 of the 55 Molonglo calibrators were variable on characteristic time-scales between 300 and 3000 days. The Molonglo calibration procedure averages the measurements of up to 8 calibrator sources, so variation of a single source is expected to have a relatively small effect on the flux density scale calibration. The characteristic time-scale was much longer than the synthesis time so we expect no variability on that time-scale. In 1996, the variable sources were removed from the calibrator list and the post-1996 flux density calibration (which included SUMSS and MGPS-2) can be considered more reliable. In any case, the flux density scale for every image was calibrated against SUMSS/MGPS-2 using the post-facto calibration procedure (see $\S$\ref{sec:post_facto_calibration}), so any residual error in the flux density scale should be removed by this technique.

\subsubsection{Radio frequency interference}
\change{ Observations performed after 2007 January at the Molonglo site have been subject to interference from strong local sources of Radio Frequency Interference (RFI)}. The interference affects almost all wide-field `I' images observed after 2007 January, but `B' fields are much less affected. No other modes were used during this period. The strongest RFI is usually excised during the reduction process, which results in incomplete coverage of the UV plane and higher sidelobe levels. Remaining low-level RFI resulted in significantly increased image noise levels. Occasionally, the interference can manifest itself as gain variations across the image, which is difficult to identify with visual inspection alone. Images affected in this way were detected and removed by the post-facto calibration procedure (see $\S$\ref{sec:post_facto_calibration}).

\subsubsection{Missing calibration values}
The standard MOST image reduction pipeline assumes default values for calibration parameters if none are supplied. Inspection of image headers revealed that over 600 images had been reduced with default calibration values, rather than the measured values. We were able to re-reduce these images with the same calibration measurements used to produce the results of \citet{gaensler2000most}.

In spite of this effort, we found large variations in the flux density scale on some occasions, which we attribute to the variable calibrators, wrongly applied calibration and interference during calibrator observations. Rather than discarding images that appeared good apart from flux density scale, we included them in our analysis and calibrated with the post-facto calibration procedure (see $\S$\ref{sec:post_facto_calibration}).

\subsubsection{Resolution effects}
\change{
Many radio transient and variability surveys, such as that performed by \citet{levinson2002}, have suffered from the potential for mismatched resolutions to cause systematic errors in the analysis. Mismatched resolutions can cause a number of systematic effects. In particular, higher resolutions can resolve a large source into multiple components, as well as resolve out the extended flux, resulting in a lower flux density measurement.

The MOST synthesised beam is fixed in the east-west direction, but elongates in the north-south direction with the declination of the image centre. The difference in resolution is largest when a source is imaged at the southern tip of an image centred on the declination limit ($\delta_{\rm max} = -30\degree$), and the northern edge of an image centred to the south, with declination $\delta_2 \simeq \delta_{\rm max} - 2 a_{0, \rm{max}} \csc | \delta_{\rm max} |$. The largest minor axis occurs for the `I' images, with $a_{0, \rm{max}} = 1.36\degree$ (see Table~\ref{tab:imagetypes}). The difference in resolution in the worst case is therefore $45 (\csc |\delta_{\rm max}| - \csc |\delta_2 |) \simeq 9.2$~arcsec in the north-south direction, \changeb{which corresponds to a maximum difference in resolution of overlapping images of approximately 20 per cent.}

By way of comparison, \citet{levinson2002} compared NVSS and FIRST surveys, whose resolutions differed by a factor of 5 (45 to 5~arcsec). We suspect resolution difference affects a very small fraction of our measurements as only those sources in the narrow band of overlap between north-south adjacent images at northern declinations are affected, and that multiple match and extended flux effects will be small as the change in resolution is only 20 per cent.
}
\section{Light curve extraction}

There are two broad categories of techniques that can be applied to a set of astronomical images to detect variable and transient emission: image comparison and catalogue comparison. Image comparison, as used for example by \citet{carilli2003vsrs},  involves comparing an image of interest with a reference image, either by subtraction or division. Source finding can then be performed on the difference or quotient image. Catalogue comparison makes source catalogues by performing source finding on each image and forming a light curve by measuring the flux density separately on each image. The light curve is then subjected to a statistical analysis. \citet{bower2007sta} used a cataloguing approach.

An image comparison approach is not feasible for the MOST archive since MOST image artefacts change as a function of pointing centre and observing mode, which would result in a large number of false detections when a reference image is compared with a candidate image. Instead we use a catalogue-based approach as discussed in the following sections.

\subsection{Source detection}
\label{sec:source_detection}
Source detection was performed using the {\sc sfind} task \citep{2002AJ....123.1086H} from the {\sc miriad} package \citep{miriad.userguide}. {\sc sfind} uses the False Discovery Rate method for controlling the fraction of false positives when performing multiple hypothesis testing. It accepts two parameters: the size of the square box over which it should measure the background RMS surface brightness, and the percentage of false pixels to accept when applying the False Discovery Method. We chose a box size of $50 \times 50$ pixels and a false pixel rate of 10 per cent. {\sc sfind} automatically produces Gaussian source fits and background RMS for each detected source. We used only the fitted positions.  We separately measured the flux densities as described in $\S$\ref{sec:measurement} below as we found the fitted flux densities and background RMS from {\sc sfind} to be occasionally unreliable.

\subsection{Source association}
\label{sec:association}
An initial catalogue was created by merging the SUMSS catalogue\footnote{Version 2.1 dated 2008 Mar 11\\ \url{http://www.physics.usyd.edu.au/sifa/Main/SUMSS}} and the MGPS-2 catalogue\footnote{Version dated 2007 Aug 15 \\ \url{http://www.physics.usyd.edu.au/sifa/Main/MGPS2}} and source detection was performed independently on each image as described in $\S$ \ref{sec:source_detection}. We associated each detection with a catalogued source if the positions of the detected and catalogued sources were coincident within 10\arcsec. If there was no previously known source within 10\arcsec, a new source was added to the catalogue.

\subsection{Measurement}
\label{sec:measurement}

Once a source has been detected, we measured flux density at the position of the source on all images in the archive for which this position is in the field of view. We do this for a number of reasons:

\begin{enumerate}
\item To obtain fitted flux densities at all epochs.
\item To a obtain upper limits on flux density for epochs in which a source is not present.
\item To obtain measurements of the surface brightness in a region around a source, which is useful for quantifying the RMS error of the measurement (for error estimates) and correcting for a non-zero mean level.
\end{enumerate}

To obtain a post-facto calibration solution for an image, we require accurate flux densities of bright point sources. To measure flux density for bright sources we fit a Gaussian source to a $10 \times 10$ pixel square box on the detected position. For weak sources ($<5\sigma$), Gaussian source fitting occasionally produces large errors, partly due to the large number of free parameters. Therefore, to measure flux density of weaker sources, we also performed a 2D parabolic fit to a $3 \times 3$ pixel region centred on the source position. To quantify a non-detection, we measured the peak pixel in a $3 \times 3$ pixel square region centred on the source position, and to measure the local background we measured the RMS and mean flux density in an annular region of inside radius 66\arcsec and outside radius 99\arcsec (6 and 9 pixels in radius, respectively).  We did not mask out other bright sources in the annular region, so measurements of the local background can be biased in densely populated areas.

\subsection{Post-facto calibration}
\label{sec:post_facto_calibration}
In our initial investigations, we found a large number of sources that appeared to have varying flux densities, which we found was due to variations in the gain of individual images.

In order to reduce the number of false variability detections due to this effect, we implemented a post-facto calibration scheme by comparing the Gaussian-fitted peak flux density measurements of the bright sources on each image, against the SUMSS  flux densities measured by \citet{Mauch2003sumss} or the MGPS-2 flux densities measured by \citet{murphy2007sem}.

\begin{table}
\centering
\caption{Threshold values of fitted formal errors used to select point sources for detection and post-facto calibration.}
\begin{tabular}{@{} lr @{}}
\hline
Parameter & Formal error threshold \\
\hline
Major axis & $<15 \arcsec$ \\
Minor axis & $<12 \arcsec$ \\
Position in RA & $<5 \arcsec$ \\
Position in Dec & $<6 \arcsec$ \\
Peak Value & $<10 \unit{mJy~beam^{-1}}$ \\
\hline
\end{tabular}
\label{tab:fiterrors}
\end{table}

For each image, we performed a 2-stage fit for the gain. First, we selected all point sources that had formal fit errors within the bounds shown in Table \ref{tab:fiterrors}, a peak flux density above 40~mJy, and a counterpart in SUMSS or MGPS-2. We then performed a linear  least squares fit to the relationship between peak fitted flux density measurements and catalogued flux densities, weighted by the inverse of the quoted peak flux density measurement error ($\sigma$) from the catalogue. The fit was constrained to go through zero. We use the fitted Gaussian measurements because it provides a more accurate flux density measurement than the 2D parabolic fit for the bright sources that we used for the calibration and is more sensitive to image errors, thus enabling the rejection of images containing subtle image errors that affect a large fraction of sources. All points that deviated from the fit by more than $3\sigma$ were removed from the set and the fit was performed again on the remaining points. The gradient of this fit was used as the gain calibration factor, and was applied to all measurements derived from the image in question.

In addition to producing a gain estimate, this procedure also enables us to remove poor quality images from the remaining analysis by measuring the goodness-of-fit, both before and after the deviant points were removed. As a measure of goodness-of-fit we adopted the reduced $\chi^2$, defined as

\begin{equation}
\tilde{\chi}^2_{\mathrm{cal}} = \frac{\chi^2_{\mathrm{cal}}}{N_{\mathrm{dof}}} = \frac{\chi^2_{\mathrm{cal}}}{N_{\mathrm{cal}} - 1}
\end{equation}

\noindent where $N_{\mathrm{dof}}$ is the number of degrees of freedom in the fit and $N_{\mathrm{cal}}$ is the number of sources being used for the calibration.

We discarded images for which the following conditions were all met:

\begin{itemize}
\item The initial $\tilde{\chi}^2_{\mathrm{cal}}$ was greater than 30.
\item More than 30 per cent of the sources were more than $3\sigma$ from the fit.
\item \changeb{The $\tilde{\chi}^2$ was greater than 5 after removing the sources which were $3\sigma$ from the fit.}
\end{itemize}

The post-facto calibration procedure assumes that the majority of sources in the sky are static over time-scales of several years and variations in flux density are not spatially correlated over the MOST field of view. As such, refractive scintillation caused by a nearby cloud of ionised gas covering a significant fraction of the MOST field of view will be removed by the calibration and rendered undetectable. We consider this possibility physically unlikely.

\begin{figure}
\centering
\includegraphics[width=\linewidth]{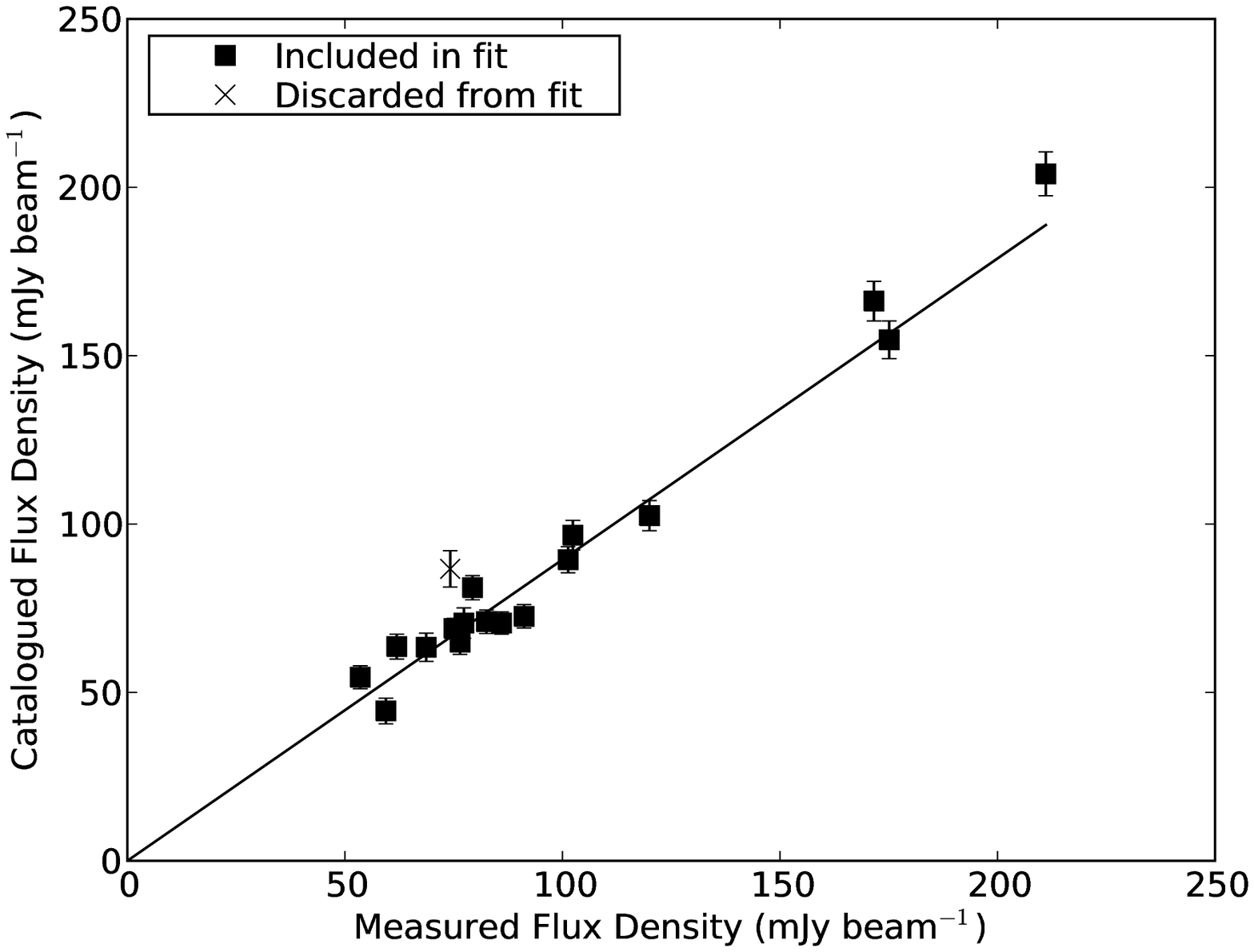}
\includegraphics[width=\linewidth]{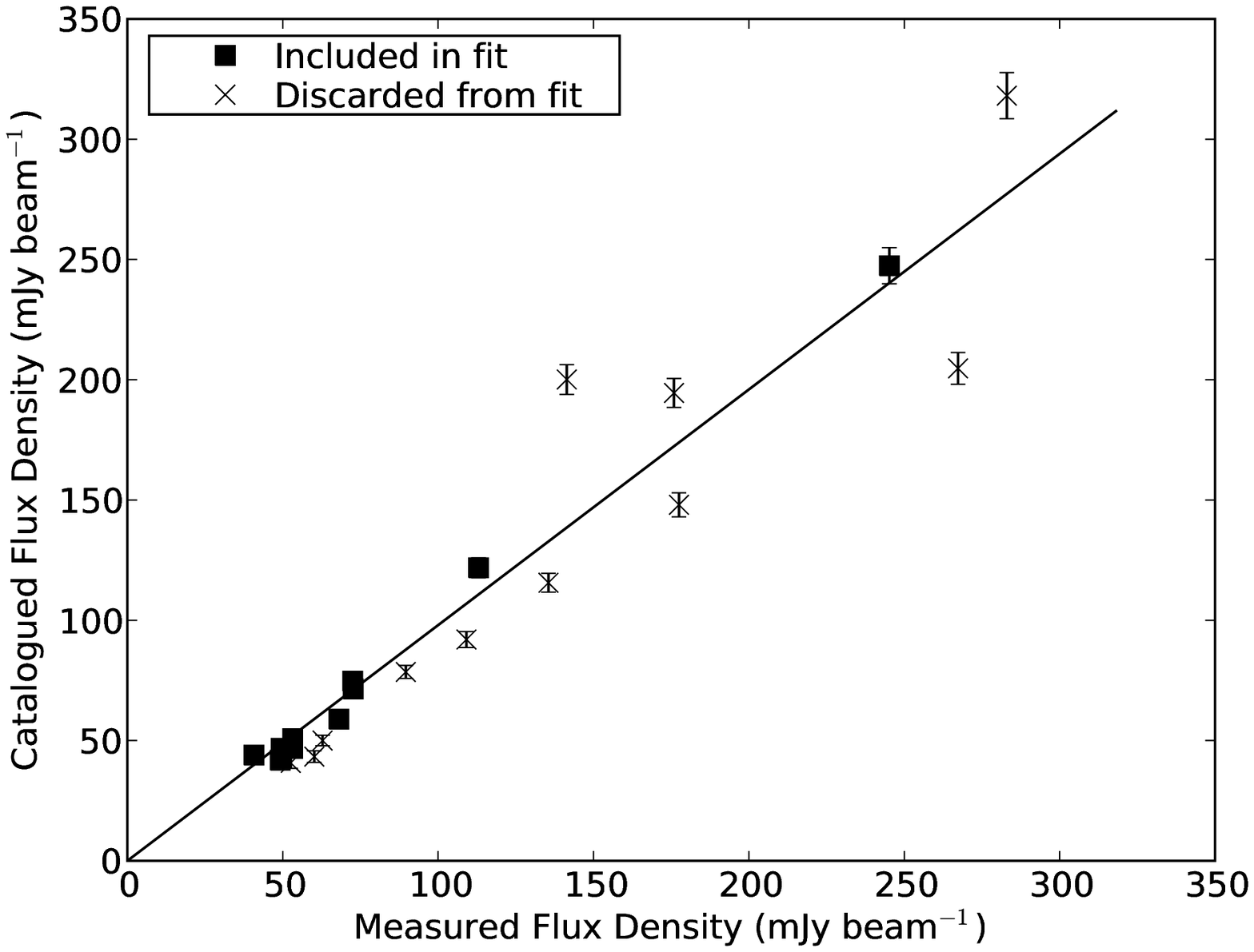}
\caption{Examples of post-facto calibration of two images. Catalogued flux densities (from SUMSS/MGPS-2) are plotted against the peak flux density measured with a Gaussian fit on a 10x10 pixel box centred on each source position. The solid line is the line of best fit after removing the sources more than 3$\sigma$ from the initial best fit line including all points (where $\sigma$ is the quoted error from SUMSS/MGPS-2). The squares are the points included in the fit and the crosses are the sources removed. Top panel: An `I' image (observed 2000 March 25) with good fit and gain of 0.89. Bottom panel: An `I' image (observed 1999 November 9) with a poor fit due to large discrepancies between measured and catalogued flux densities for many sources, and gain of 0.98.}
\label{fig:post_facto_examples}
\end{figure}

\begin{figure}
\centering
\includegraphics[width=\columnwidth]{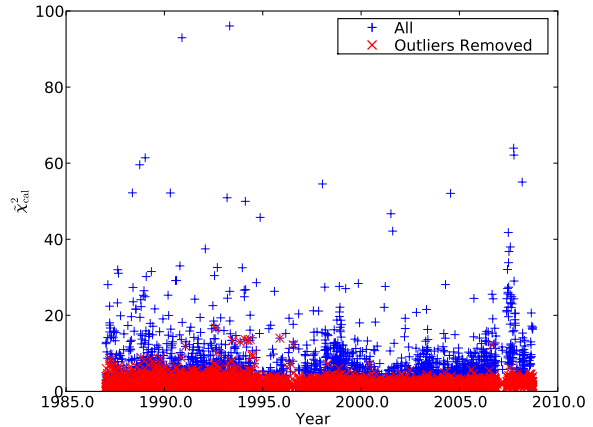}
\caption{$\tilde{\chi}^2_{\mathrm{cal}}$ of the post-facto calibration fit versus image observation date, before and after  $3 \sigma$ outliers were removed. Note the significant number of images with large value of reduced $\tilde{\chi}^2_{\mathrm{cal}}$ (before outlier removal) in 2007 and 2008, when terrestrial interference was affecting the telescope.}
\label{fig:reduced_chi2_vs_time}
\end{figure}

Examples of the post-facto calibration procedure applied to two images are shown in Fig. \ref{fig:post_facto_examples}. A plot of the $\tilde{\chi}^2_{\mathrm{cal}}$ vs time is shown in Fig. \ref{fig:reduced_chi2_vs_time}, from which we were able to detect the interference that began affecting the telescope in 2007 by a group of images with large $\tilde{\chi}^2_{\mathrm{cal}}$. This increased our confidence in the usefulness of the post-facto calibration procedure and, in particular, in its ability to recognise poor quality images.

\begin{figure}
\centering
\includegraphics[width=\columnwidth]{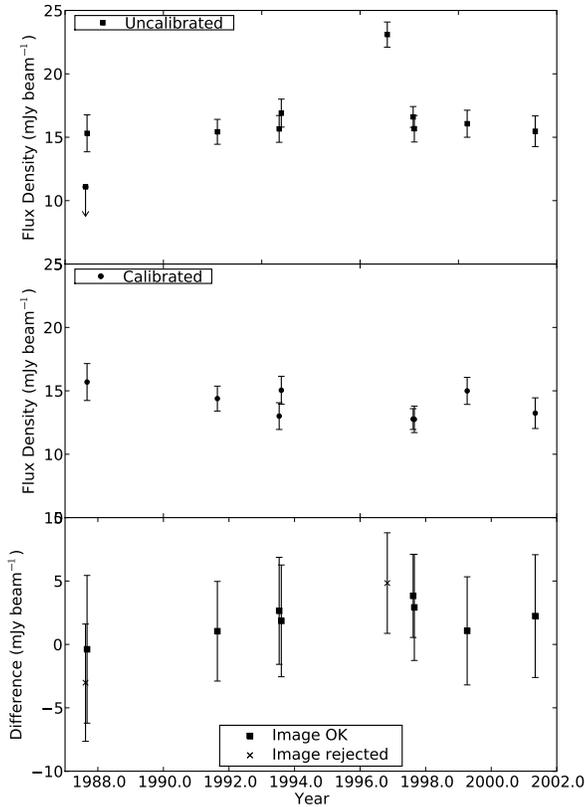}
\caption{Light curves illustrating the effect of post-facto calibration on SUMSS J201139$-$561858. Top panel: Pre-calibrated light curve, centre panel: calibrated light curve, bottom panel: difference between the calibrated and uncalibrated light curves. The uncalibrated light curve has a modulation index (see Equation \ref{eq:modindex}) of 0.181 compared to that of the calibrated light curve of 0.0842. Two images were discarded from the light curve by the post-facto calibration. }
\label{fig:before_after_cal}
\end{figure}

An example of the post-facto calibration procedure applied to the light curve of a single source is shown in Fig.~\ref{fig:before_after_cal}. The procedure removes the most discrepant points and corrects the remainder, which results in less scatter in the light curves.

\begin{figure}
\centering
\includegraphics[width=\columnwidth]{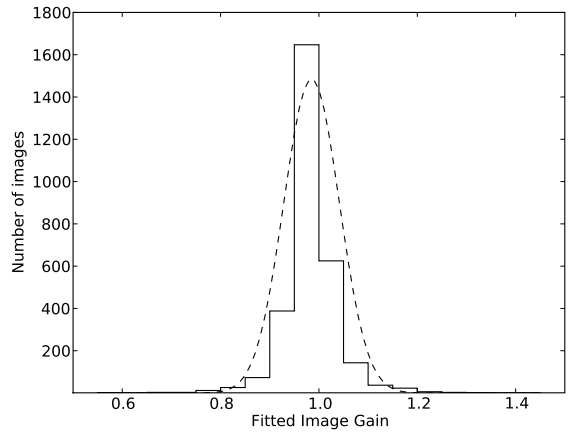}
\caption{Histogram of post-facto calibrated gains for  2~999 usable images. The dashed line is a Gaussian distribution fit to the gains, with a mean gain of 0.985 and standard deviation of 0.056.}
\label{fig:fitted_gain_histogram}
\end{figure}

\subsection{Light curve characterisation}
\label{sec:light_curve_characterisation}

\subsubsection{Variability Classification}
Astronomical sources exhibit a wide variety of light curve shapes. For example, interstellar scintillation produces  random variations around a mean, whereas stellar flares are episodic with short duty cycles and flux densities significantly above a quiescent level.

Determining whether a given light curve matches any of a wide variety of light curve shapes, and quantifying the fidelity of the match, are beyond the scope of this work. For simplicity, we adopt the classification measure of \citet{kesteven1977.2.7GHzvariability} to measure random variability in the light curve, namely the $\chi^2$ of the residuals of a weighted fit to a line of constant flux density. It is computed according to the following expressions:

\begin{equation}
\chi^2_{\mathrm{lc}}  =   \sum_{i=1}^N \frac{(S_i - \bar{S})^2}{\sigma_i^2} \label{eq:chi2}
\end{equation}

\noindent where:

\begin{eqnarray}
\label{eq:bars}
\bar{S} &  = & \frac{\sum_{i=1}^N S_i/\sigma_i^2}{\sum_{i=1}^N 1/\sigma_i^2}, \\
\label{eq:sigmai} \sigma_i &=& \sqrt{ (g \sigma_{bg,i})^2 +(0.05 S_i)^2}, \\
S_i &=& g ( S^{raw}_i - \rho_i ),
\end{eqnarray}

\noindent and \changeb{$\bar{S}$ is the weighted mean flux density, $\sigma_i$ is the estimated flux density error,} $S^{raw}_i$ is the flux density measured on the image with the 2D parabolic fit, $g$ is the gain coefficient calculated by the post-facto gain calibration and $\rho_i$ and $\sigma_{bg,i}$ are respectively the mean and RMS surface brightness of an annular region around the source. The weighting factor of $1/\sigma_i^2$ captures both the noise level in the surrounding region and the absolute flux density calibration error, from the MGPS-2 and SUMSS catalogues, assumed to be 5 per cent. For a source with a peak flux density at the detection limit of $14\unit{mJy~beam^{-1}}$, $\sigma_i$ is dominated by the background RMS term and has a typical value of $\sigma_i \simeq 3 \unit{mJy}$. For a source with flux density of $100\unit{mJy~beam^{-1}}$ the flux density calibration term dominates and results in a typical error of $\sigma_i \simeq 6 \unit{mJy}$.

To determine whether a source is variable or not, we work from the null hypothesis that all sources are static with added Gaussian measurement error. Applying the measured $\chi^2_{\mathrm{lc}}$ to the analytic $\chi^2$ cumulative distribution function (CDF) yields the probability of rejecting the null hypothesis, i.e. the probability that the source is genuinely variable, which we write as $P(\chi^2_{\mathrm{lc}})$. Working with probabilities rather than $\chi^2$ values also enables us to compare the statistics of light curves with differing number of measurements, which is absorbed in the CDF. Those light curves whose probability exceeds a certain threshold can be classified as variable.

Past investigations such as those of \citet{kesteven1977.2.7GHzvariability}, \citet{gregory1986radiopatrol} and \citet{seilstad1983.10.8GHz} applied a probability threshold of 99 per cent to derive a set of variable sources. In the case of \citet{gregory1986radiopatrol}, the 99 per cent threshold left 4 per cent of the sources classified as variable. Of these sources, 1 per cent of the total, or 25 per cent of the classified variables must be false positives, simply from the choice of a low threshold.

As we have almost 30~000 light curves, a 25 per cent false positive rate is too high. To arrive at a more rigourous threshold we define a confidence measure that is the excess sources over expected number given the null hypothesis, expressed as a percentage. Specifically, we use:

\begin{eqnarray}
C(T)= \frac{N_{\mathrm{actual}}(T) - N_{\mathrm{noise only}}(T)}{N_{\mathrm{actual}}(T)} \label{eq:ct}
\end{eqnarray}

\noindent where $N_{\mathrm{actual}}(T)$ is the number of sources with light curves whose $P(\chi^2_{\mathrm{lc}})$ is greater than $T$, $N_{\mathrm{noise only}}(T) = N_{\mathrm{sources}}\left(1 - P(\chi^2_{\mathrm{lc}} > T)\right)$ is the number of sources expected if there were only noise in the sample, and $N_{\mathrm{sources}}$ is the total number of sources in the sample. \change{The confidence measure $C(T)$ in Equation \ref{eq:ct} represents the fraction of genuinely variable sources contained in the set of sources with $P(\chi^2_{\rm lc}) > T$.}

\subsubsection{Variability measures}
\label{sec:varmeas}

Previous work has adopted a number of different metrics to quantify variability. To enable comparison with previous work, we adopt two different metrics, the first is the fractional variability defined as:

\begin{equation}
\label{eq:v}
\mathcal{V} =  \frac{S_{\mathrm{max}} - S_{\mathrm{min}}}{S_{\mathrm{max}} + S_{\mathrm{min}}}.
\end{equation}

\noindent where $S_{\mathrm{max}}$ and  $S_{\mathrm{min}}$ are respectively the maximum and minimum flux densities. This measure is most suited to characterising sources with few measurements or flaring behaviour and was used by  \citet{gregory1986radiopatrol}.

The second metric is the modulation index, defined as:
\begin{equation}
\label{eq:modindex}
m=\sigma/\bar{S}
\end{equation}

\noindent where $\sigma$ is the standard deviation of the light curve and $\bar{S}$ is the weighted mean defined in Equation \ref{eq:bars}. This metric is more suited to measuring random variability in data sets with a large number of measurements and was used by \citet{gaensler2000most}.

\subsubsection{Detections}
A measurement must satisfy a number of criteria in order to be considered an acceptable measurement of a point source rather than a result of an image artefact.

The analysis of the SUMSS catalogue by \citet{Mauch2003sumss} showed that the SUMSS survey was complete above 15~mJy and we found a number of convincing detections above 14~mJy. Gaussian fits can have major and minor axes significantly different from the synthesised beam, either because a source is partially resolved, or because of a strong image artefact. Restricting the major and minor axes of the fit to less than 20 per cent larger than the synthesised beam ensures the fit is a point source and reduces the number of false detections due to artefacts.

Therefore, we define a `detection' as a Gaussian fit to a source that satisfies all of the following criteria:

\begin{itemize}
\item Fit errors are below the thresholds in Table \ref{tab:fiterrors},
\item Fit position is less than $5\arcsec$ from the catalogued position (c.f. $\S$ \ref{sec:association}),
\item Peak flux density $\geq14$~mJy,
\item Ratio of fit major axis to synthesised beam major axis is $< 1.2$,
\item Ratio of fit minor axis to synthesised beam minor axis is $< 1.2$,
\item Image passes the post-facto calibration.
\end{itemize}

\subsubsection{Detection signal-to-noise ratio}
To find transient sources, it is particularly important to measure the significance of a given detection or non-detection. We define the detection signal-to-noise ratio as $d = S_{\mathrm{cenmax}}/\sigma_{bg}$ where $S_{\rm cenmax}$ is the maximum surface brightness in a $3 \times 3$ pixel box centred on the source position, and $\sigma_{\mathrm{bg}}$ is the RMS surface brightness (in $\unit{mJy}$) in an elliptical region centred on the source position.

\subsection{List of variable and transient sources}
\label{sec:criteria}
We computed all metrics described in $\S$\ref{sec:light_curve_characterisation} for all sources and produced lists of variable sources, and another list of transient sources. The list of variable sources was produced by finding all sources with:

\begin{itemize}
\item Light curve probability exceeding a threshold that yields a confidence metric greater than 90 per cent, i.e. $P(\chi^2_{\mathrm{lc}}) > T_0$ where $C(T_0) > 90 \unit{per cent}$.
\item A `detection' on every image.
\end{itemize}

A variable source, therefore, has at least two detections and no non-detections. The list of transient sources was produced by finding all sources with:

\begin{itemize}
\item Minimum detection signal-to-noise ratio $d_{min} < 5.0$
\item Maximum detection signal-to-noise ratio $d_{max} > 6.0$
\item Flux density at the maximum detection signal-to-noise ratio epoch (measured with a parabolic fit) $>14~\unit{mJy~beam^{-1}}$
\item $P(\chi^2_{\mathrm{lc}}) > 99\%$ computed with $S_{\rm cenmax}$ as the flux density value.
\end{itemize}

A transient source, therefore, has at least one detection and at least one non-detection. We use the parabolic fit, as this was less prone to error at low signal-to-noise ratio. We compute the $\chi^2_{\mathrm{lc}}$ for the transient list using the $S_{\rm cenmax}$ rather than the 2D parabolic fitted flux density as this is the most robust measure of the upper limit to the flux density, and we apply the error as per Equation \ref{eq:sigmai}. We reduce the threshold on the $P(\chi^2_{\mathrm{lc}})$ with respect to the variable sources, because the flux density of the non-detections is not known.

\section{Results}

Throughout this section we assume a Hubble constant of $H_0 = 72 \unit{~km~s^{-1} Mpc^{-1}}$ and define the spectral index, $\alpha$, according to $S \propto \nu ^{\alpha}$ where $S$ is the flux density and $\nu$ is the observed frequency.

\change{When referring to a radio source, we use the popular name (e.g. SN1987A) if one exists. Otherwise, if the source appears in either the MGPS2 or SUMSS catalogues we use the MGPS2 or SUMSS naming convention, or if not, we use the source coordinates in the form `JHHMMSS-DDMMSS' without prefix.}

\subsection{Data Quality}
\subsubsection{Number of usable images}
Of the 7227 images in the archive, 2999 were retained after visual inspection and post-facto calibration. The area of an equivalent two-epoch survey (see Equation \ref{eq:A2epoch})  is $2775.7 \unit{deg^{2}}$.

\subsubsection{Number of measurements}
\begin{figure}
\centering
\includegraphics[width=\linewidth]{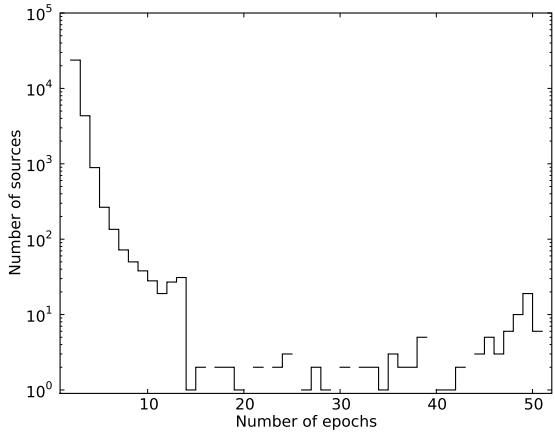}
\caption{Histogram of the number of epochs measured per source in the variability and transient analysis. At the two extremes, \changeb{ there are 26000 sources with two measurements and 11 sources with 51 measurements.} Gaps in the histogram indicate no sources have that number of epochs.}
\label{fig:nepochs_histogram}
\end{figure}

There are 29730 sources that have light curves with measurements at two or more epochs. A histogram of the number of epochs measured on each source is shown in Fig.~\ref{fig:nepochs_histogram}.

\subsubsection{Analysis of the population of light curves}

\begin{figure*}
\centering
\includegraphics[height=21cm]{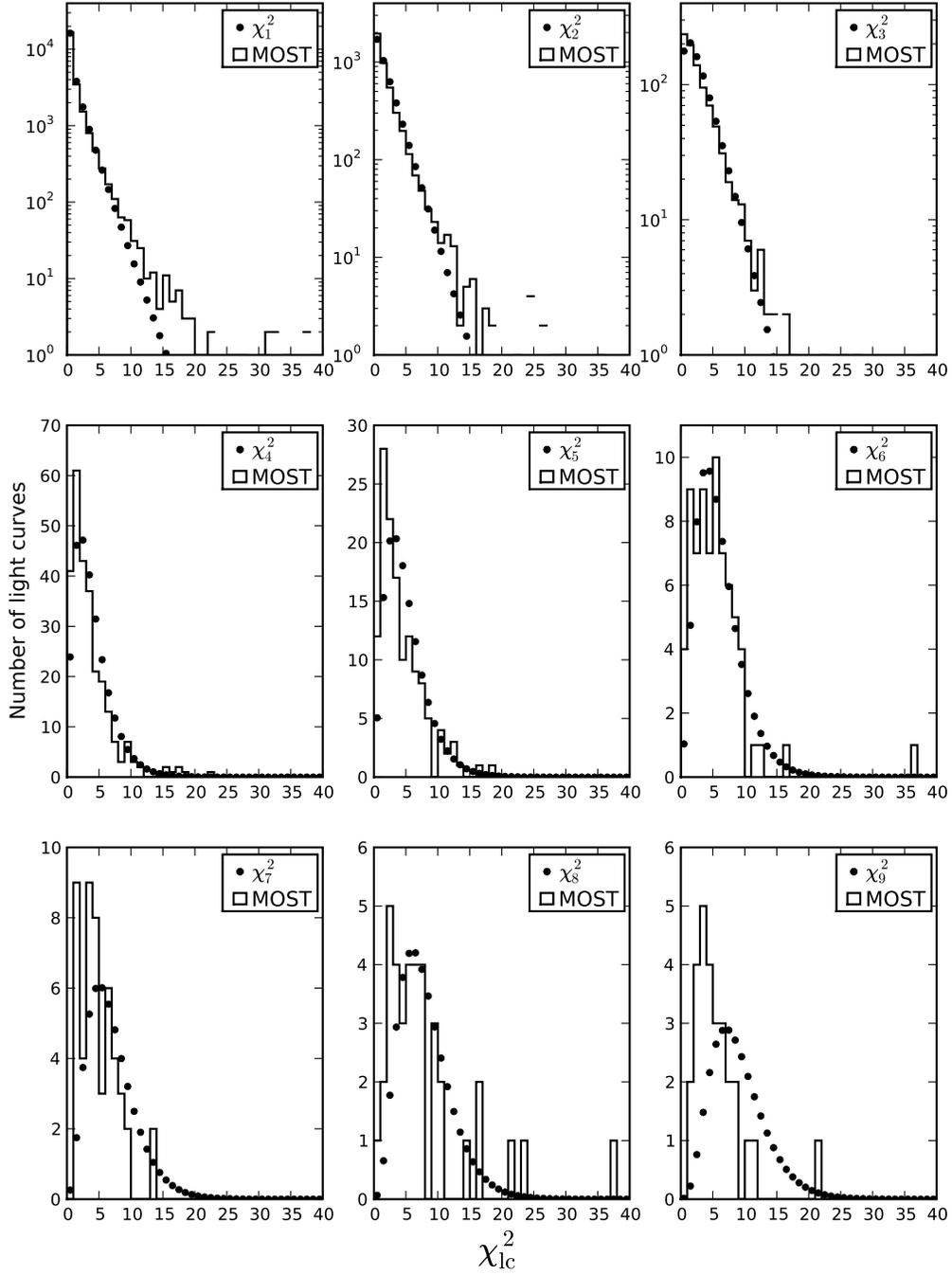}
\caption{Histograms of $\chi_{\rm lc}^2$ values for light curves with between 2 and 10 measurements ($N_{meas}$). Also plotted as  filled circles is the theoretical $\chi_{N_{\mathrm{meas}}-1}^2$ probability distribution with $N_{\mathrm{dof}} = N_{\mathrm{meas}}-1$ degrees of freedom binned as for the measurements.}
\label{chi2_pdfs}
\end{figure*}

To check that the flux density error estimate is consistent over the population of light curves, we compare the measured and theoretical distributions of the residuals from a line of constant flux density. As described in Equation \ref{eq:chi2}, the residuals of the best-fitting light curve of constant flux density follow a $\chi^2$ distribution with $N_{\mathrm{meas}}-1$ degrees of freedom. Fig.~\ref{chi2_pdfs} shows the histograms of the $\chi_{\rm lc}^2$ of the residuals for light curves in our variability sample. The theoretical and actual curves agree reasonably well, with the slightly larger number of sources at small values of $\chi_{\rm lc}^2$ indicating that the flux density errors have likely been marginally overestimated.

\begin{figure}
\centering
\includegraphics[width=\linewidth]{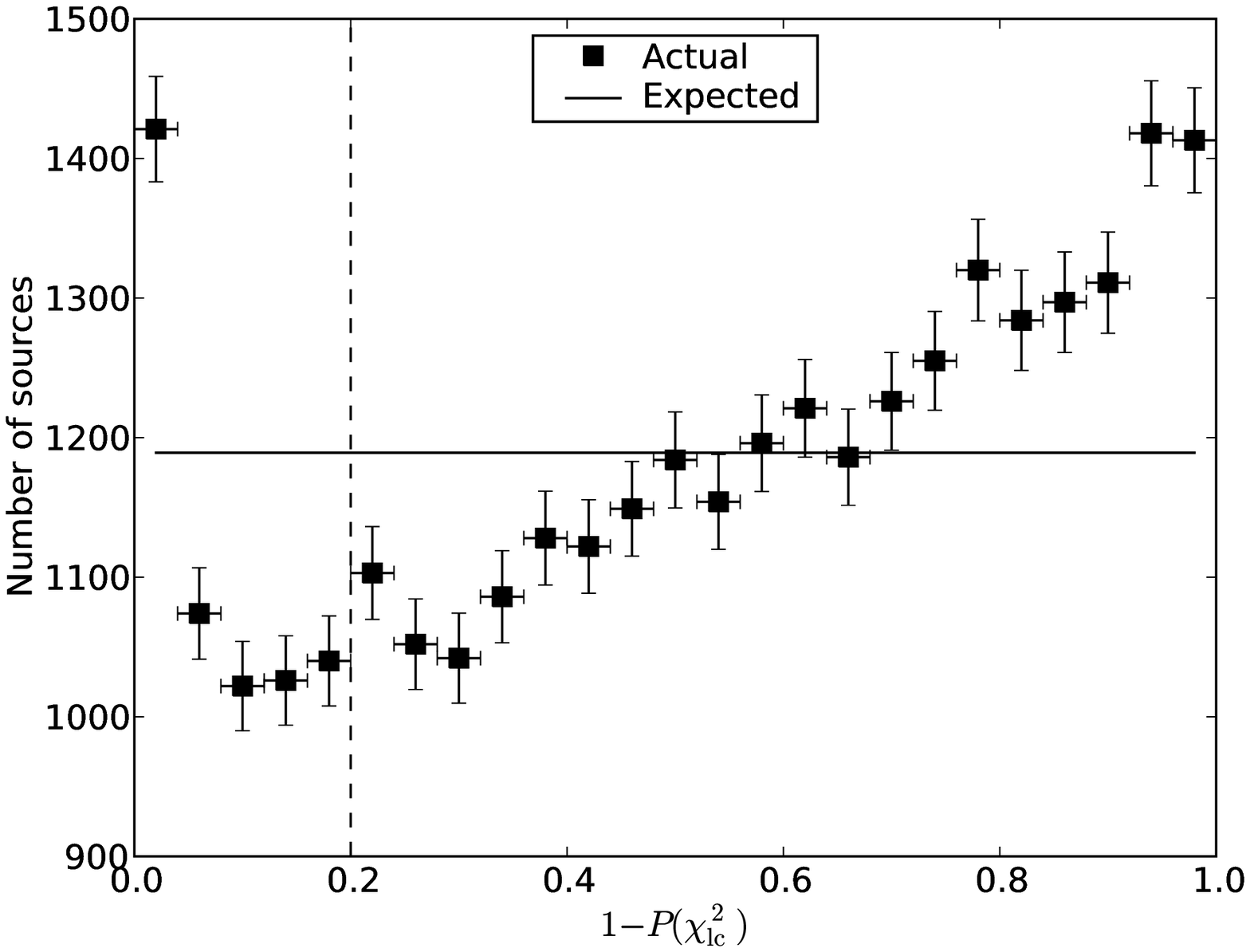}
\includegraphics[width=\linewidth]{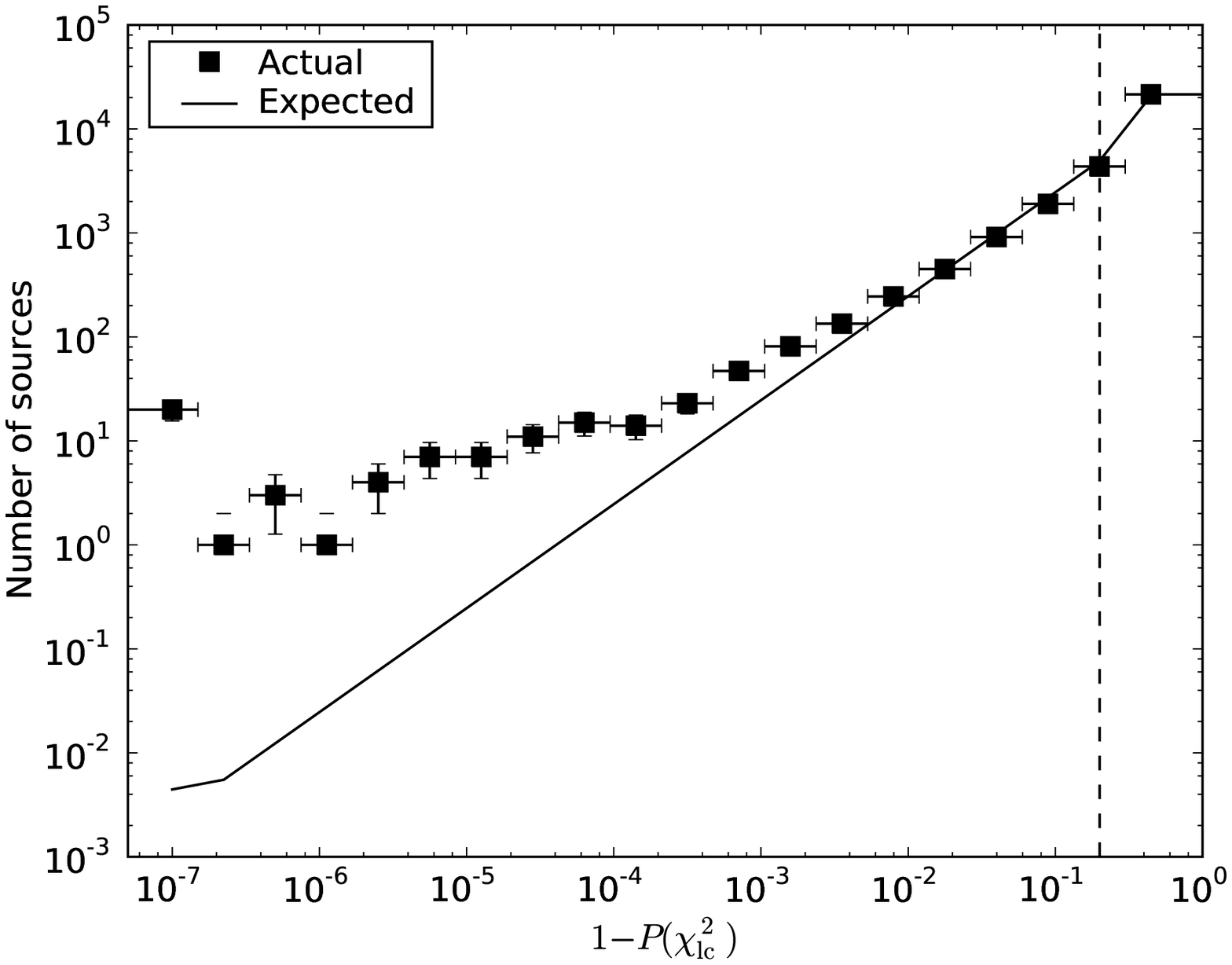}
\caption{Top panel: Number of sources vs $1 - P(\chi_{\rm lc}^2)$ on a linear scale. The y error bars are the Poisson errors and the x error bars indicate the bin widths. The solid line depicts the expected number of sources in each probability bin if the scatter in the light curves was due to Gaussian noise. The point at the far left indicates a population of variable sources and the excess over the solid line indicates that there are approximately 350 genuinely variable sources in our sample. The increasing general trend from left to right indicates there is a deficiency of moderately variable and excess of static sources, which agrees with Fig.~\ref{chi2_pdfs}. The region to the left of the dashed vertical line at $1 - P(\chi_{\rm lc}^2)=0.2$ indicates the baseline number of sources used to infer the actual number of variable sources (see $\S$\ref{sec:sourcecounts}).
Bottom panel: \changeb{The same data but re-binned on a log scale to illustrate how the number of sources departs from the expected number for small values of $1 - P(\chi_{\rm lc}^2)$}.}
\label{fig:nsources_vs_threshold}
\end{figure}

In order to compare distributions of sources with differing numbers of measurements, we can compute the probability of a given $\chi_{\rm lc}^2$ value occurring by chance by using the analytic $\chi^2$ cumulative distribution function with the appropriate degrees of freedom. If the variation in the light curves is attributed entirely to Gaussian noise, the histogram of the computed probabilities should be uniform and independent of the degrees of freedom. Fig.~\ref{fig:nsources_vs_threshold} depicts the expected and actual probability histograms. There is a general trend in the actual histogram showing an excess of sources with high $1-P(\chi_{\rm lc}^2)$, which we attribute to overestimated flux density errors leading to smaller measured $\chi_{\rm lc}^2$ values. Because a large fraction of sources have a small number of measurements (see Fig.~ \ref{fig:nepochs_histogram}), at this degree of freedom a small $\chi_{\rm lc}^2$ corresponds to a higher value of $1-P(\chi_{\rm lc}^2)$ which in turn leads to the observed excess. In addition to the general trend, there is a single bin at the lowest $1-P(\chi_{\rm lc}^2)$, which we attribute to a population of genuinely variable sources. The genuinely variable sources are clearly seen as an excess of sources, departing from the theoretical curve as the probability is reduced.

\begin{figure}
\centering
\includegraphics[width=\linewidth]{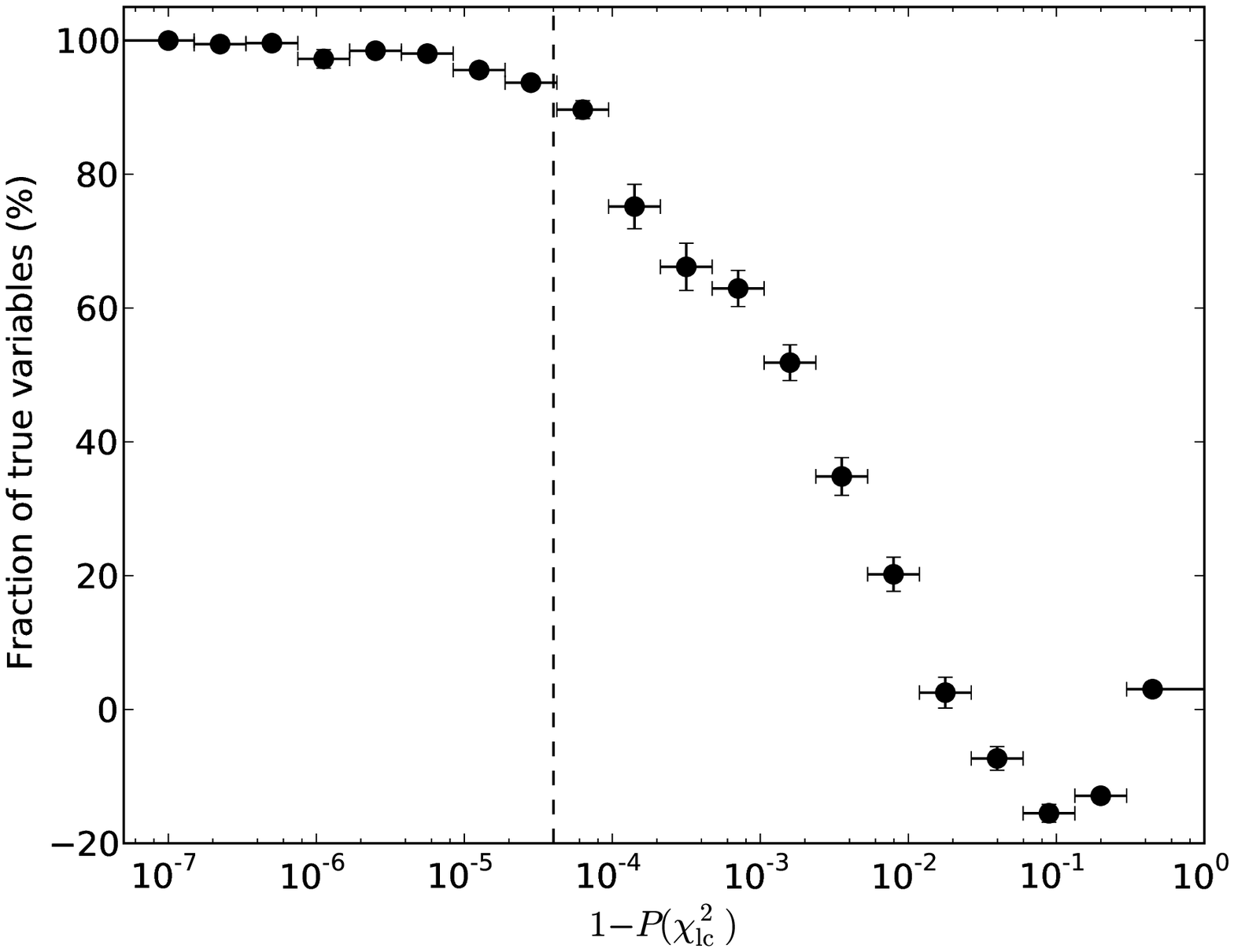}
\caption{Confidence measure ($C(T)$) vs $1 - P(\chi_{\rm lc}^2)$ on a log scale. To achieve a fraction of genuine variable sources of  90 per cent requires $1 - P(\chi_{\rm lc}^2) = 4 \times 10^{-5}$. The y errors are the Poisson errors propagated through the definition of the confidence metric, and the the x errors indicate the bin width. The confidence is negative at large values of $1 - P(\chi_{\rm lc}^2)$ due to a deficiency of moderately variable sources at large values of $1 - P(\chi_{\rm lc}^2)$ (see Fig.~\protect\ref{fig:nsources_vs_threshold}.). The dashed vertical line indicates the adopted threshold of $1-P(\chi_{\rm lc}^2) = 4 \times 10^{-5} = T_0$, to the left of which the confidence of finding a genuinely variable source is greater than 90 per cent.}
\label{fig:confidence_vs_logthreshold}
\end{figure}

Fig.~\ref{fig:confidence_vs_logthreshold} shows the confidence measure $C(T)$ vs $1-P(\chi_{\rm lc}^2)$ on a log scale. From this plot we deduce that applying a threshold of $T_0 = 1-P(\chi_{\rm lc}^2) = 4 \times 10^{-5}$ ensures the fraction of spurious variable sources in the list of candidate variables is less than 10 per cent.

\subsubsection{Primary beam effects}
\begin{figure}
\centering
\includegraphics[width=\columnwidth]{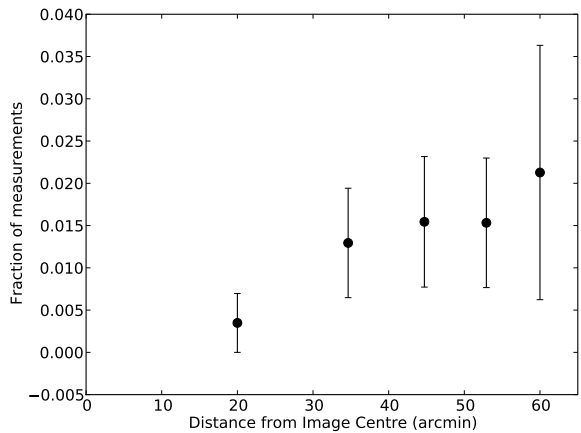}
\caption{Plot of the fraction of flux density measurements that deviate from $\bar{S}$ by more than 20 per cent, binned by distance from the image centre. Only images centred on the HDFS were considered, and 2 known variable sources were removed. The error bars are Poisson errors}.
\label{fig:deviant_fraction_vs_distance}

\end{figure}

In the MOST archive, a significant fraction of the overlapping coverage is at the edges of adjacent images. Errors in primary beam correction, which increase with radial distance from the pointing centre, could therefore introduce false variability. To investigate this effect, we took the light curves of all sources appearing in images centred on $\alpha = $22h30m00s, $\delta=$-61d06m36s (J2000), in the Hubble Deep Field South (HDFS) region. We calculated the fraction of measurements that deviated by more than 20 per cent from $\bar{S}$ (see Equation \ref{eq:bars}), binned by radial distance from the pointing centre, as shown in Fig.~\ref{fig:deviant_fraction_vs_distance}. We found no strong dependence with distance and therefore conclude that the primary beam correction is accurate, and that the adopted available minor axis of the primary beam (see Table~\ref{tab:imagetypes} and $\S$ \ref{sec:modes}), outside which we ignore measurements, is acceptable.

\subsubsection{Declination effects}
\begin{figure}
\centering
\includegraphics[width=\linewidth]{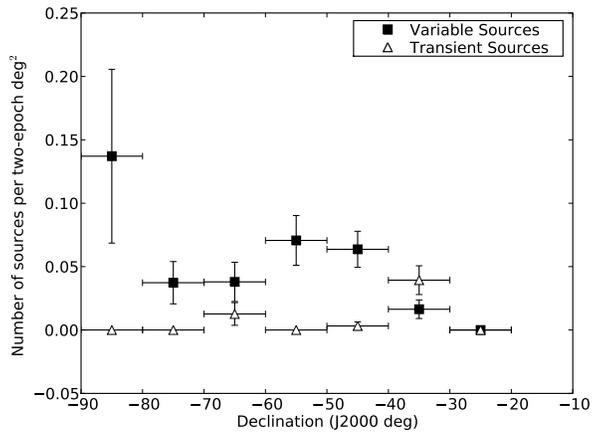}
\caption{Fraction of variable and transient sources vs declination.}
\label{fig:frac_variables_vs_dec}
\end{figure}

The sensitivity and resolution of MOST images is strongly declination dependent, with noise and beam size increasing towards Northerly declinations. Therefore, to ensure our sample is not contaminated by declination dependent systematic effects, we plot in Fig.~\ref{fig:frac_variables_vs_dec} the fraction of transient and variable sources as a function of declination. We find no declination dependence.

\subsection{Variable sources}
Table~\ref{tab:variables} lists the 53 sources that we classify as variable according to the criteria described in $\S$\ref{sec:criteria}. The Table also lists radio and optical counterparts where available. Light curves for each of these sources are shown in Appendix \ref{app:var_light_curves}.

\begin{landscape}
\begin{table}
\centering
\caption{53 radio sources classified as variable with a confidence greater than 90\%. The columns are: the source name, right ascension and declination (J2000), Galactic longitude and latitude, number of accepted measurements, probability of $\chi_{\rm lc}^2$ of the residuals to a line of constant flux density, maximum, weighted mean and minimum flux density at 843~MHz, modulation index, $\mathcal{V}$ value, NVSS flux density at 1.4~GHz (if covered by NVSS area), SuperCOSMOS classification: D = point source detected, N = no source detected, C = more than 1 point source in the error box, E = extended according to SuperCOSMOS catalogue, G = clearly resolved galaxy, the SuperCOSMOS optical magnitudes in blue (B), red (R) and infra-red (I), and the B$-$R colour.}
\label{tab:variables}
\scriptsize

\begin{tabular}{lllrrcrrrrrrrrrrrrrr|}
\hline
Source name & $\alpha$ & $\delta$ & $l$ & $b$ & $N_{\mathrm{meas}}$ & $1 - P(\chi_{\rm lc}^2)$ & $S_{\rm max}$ & $\bar{S}$  & $S_{\rm min}$ & $m$ & $\mathcal{V}$ & NVSS & Supercos & B  & R & I  & B$-$R\\
& (h m s) & (d m s) & (deg) & (deg) & & & \multicolumn{3}{c}{$(\unit{mJy~beam^{-1}})$ } & & &  $\unit{mJy~beam^{-1}}$ & & (mag) & (mag) & (mag)  & (mag)  \\
\hline
SUMSS J000716$-$793206 & 00 07 16 & -79 32 00 & 305.4 & -37.4 & 3 & $5.0 \times 10^{-6}$ & 29.4 & 22.9 & 18.8 & 0.2 & 0.22 & - & N & --- & --- & --- & --- \\
SUMSS J001859$-$811133 & 00 18 58 & -81 11 32 & 304.5 & -35.8 & 2 & $2.4 \times 10^{-5}$ & 114.9 & 94.9 & 83.8 & 0.2 & 0.16 & - & D &     20.93 &     20.46 & --- &       .48 \\
SUMSS J003722$-$421508 & 00 37 22 & -42 15 12 & 312.8 & -74.6 & 2 & $6.5 \times 10^{-6}$ & 25.3 & 19.2 & 15.2 & 0.4 & 0.25 & - & D &     20.54 &     20.15 & --- &       .38 \\
SUMSS J004417$-$375258 & 00 44 17 & -37 52 57 & 310.4 & -79.1 & 2 & $2.4 \times 10^{-14}$ & 43.5 & 23.8 & 18.7 & 0.6 & 0.40 &    18.1 & E &     20.30 &     19.14 &     18.91 &      1.16 \\
SUMSS J010426$-$400412 & 01 04 26 & -40 04 12 & 292.0 & -76.8 & 2 & $2.3 \times 10^{-8}$ & 62.6 & 44.6 & 37.7 & 0.4 & 0.25 & - & D &     20.00 &     19.15 &     18.92 &       .85 \\
SUMSS J011019$-$455112 & 01 10 20 & -45 51 16 & 292.8 & -70.9 & 7 & $2.3 \times 10^{-6}$ & 20.9 & 16.0 & 10.7 & 0.2 & 0.32 & - & G & --- & --- & --- & --- \\
SUMSS J011136$-$394340 & 01 11 36 & -39 43 40 & 285.8 & -76.7 & 2 & $1.0 \times 10^{-9}$ & 44.5 & 27.9 & 23.5 & 0.4 & 0.31 &    21.5 & E &     19.32 &     18.35 &     18.77 &       .97 \\
SUMSS J011839$-$452657 & 01 18 40 & -45 26 59 & 288.2 & -70.9 & 9 & $9.4 \times 10^{-6}$ & 45.7 & 37.4 & 32.0 & 0.1 & 0.18 & - & N & --- & --- & --- & --- \\
SUMSS J012615$-$561946 & 01 26 15 & -56 19 47 & 293.2 & -60.1 & 2 & $9.5 \times 10^{-6}$ & 14.1 & 9.3 & 6.4 & 0.5 & 0.38 & - & C & --- &     20.50 &     18.35 & --- \\
SUMSS J013013$-$742024 & 01 30 15 & -74 20 24 & 299.4 & -42.5 & 4 & $1.0 \times 10^{-5}$ & 29.9 & 24.2 & 20.9 & 0.2 & 0.18 & - & N & --- & --- & --- & --- \\
SUMSS J021036$-$405154 & 02 10 36 & -40 51 55 & 258.1 & -68.7 & 2 & $8.8 \times 10^{-6}$ & 40.4 & 32.6 & 26.1 & 0.3 & 0.22 & - & D &     21.28 &     19.96 & --- &      1.32 \\
SUMSS J022223$-$802356 & 02 22 23 & -80 23 59 & 298.4 & -35.9 & 2 & $2.4 \times 10^{-7}$ & 56.7 & 41.1 & 35.0 & 0.3 & 0.24 & - & D &     19.95 &     19.70 & --- &       .24 \\
SUMSS J031201$-$523432 & 03 12 00 & -52 34 32 & 266.9 & -53.6 & 2 & $1.4 \times 10^{-5}$ & 30.8 & 24.7 & 18.9 & 0.3 & 0.24 & - & N & --- & --- & --- & --- \\
SUMSS J032102$-$405735 & 03 21 02 & -40 57 32 & 247.0 & -56.4 & 2 & $2.9 \times 10^{-8}$ & 39.8 & 27.8 & 19.4 & 0.5 & 0.35 & - & D &     20.21 &     19.64 & --- &       .58 \\
SUMSS J040608$-$540445 & 04 06 08 & -54 04 54 & 263.9 & -45.6 & 2 & $1.6 \times 10^{-8}$ & 19.6 & 12.5 & 9.0 & 0.5 & 0.37 & - & C &     21.71 &     19.33 &     18.49 &      2.38 \\
SUMSS J043624$-$563332 & 04 36 24 & -56 33 31 & 265.7 & -40.8 & 3 & $9.8 \times 10^{-10}$ & 81.4 & 61.5 & 49.6 & 0.2 & 0.24 & - & D &     20.22 &     19.63 & --- &       .58 \\
SUMSS J044508$-$541155 & 04 45 08 & -54 11 53 & 262.4 & -40.0 & 2 & $1.3 \times 10^{-8}$ & 45.7 & 33.1 & 28.2 & 0.3 & 0.24 & - & D &     20.71 &     19.99 & --- &       .72 \\
SUMSS J044624$-$391548 & 04 46 24 & -39 15 45 & 242.8 & -40.3 & 2 & $2.5 \times 10^{-5}$ & 47.1 & 37.8 & 32.2 & 0.3 & 0.19 &    44.2 & D &     18.41 &     17.64 &     17.21 &       .77 \\
SUMSS J044829$-$401714 & 04 48 28 & -40 17 12 & 244.2 & -39.9 & 2 & $3.4 \times 10^{-9}$ & 58.9 & 43.8 & 35.4 & 0.4 & 0.25 & - & D &     19.44 &     19.39 &     19.15 &       .05 \\
SUMSS J045530$-$720442 & 04 55 30 & -72 04 41 & 283.9 & -34.5 & 3 & $< 1 \times 10^{-14}$ & 38.5 & 17.0 & 11.9 & 0.6 & 0.53 & - & C &     21.50 &     21.68 & --- &      -.18 \\
SUMSS J050230$-$651321 & 05 02 30 & -65 13 22 & 275.6 & -35.8 & 4 & $2.3 \times 10^{-9}$ & 40.4 & 31.4 & 25.5 & 0.2 & 0.23 & - & C &     19.47 &     18.42 & --- &      1.05 \\
SUMSS J061952$-$643917 & 06 19 52 & -64 39 17 & 274.3 & -27.7 & 4 & $2.3 \times 10^{-8}$ & 49.5 & 33.5 & 28.5 & 0.3 & 0.27 & - & E &     21.57 &     19.41 &     19.32 &      2.16 \\
SUMSS J062724$-$405328 & 06 27 24 & -40 53 28 & 248.9 & -21.6 & 2 & $2.6 \times 10^{-5}$ & 24.0 & 16.9 & 13.7 & 0.4 & 0.27 & - & N & --- & --- & --- & --- \\
SUMSS J064951$-$620221 & 06 49 51 & -62 02 21 & 272.1 & -23.9 & 2 & $2.4 \times 10^{-6}$ & 46.2 & 36.4 & 30.1 & 0.3 & 0.21 & - & D &     21.52 &     20.84 & --- &       .68 \\
MGPS2 J082437$-$463505 & 08 24 38 & -46 35 09 & 263.6 & -5.1 & 2 & $3.3 \times 10^{-5}$ & 44.3 & 33.6 & 29.6 & 0.3 & 0.20 & - & N & --- & --- & --- & --- \\
MGPS2 J083322$-$444141 & 08 33 22 & -44 41 39 & 263.0 & -2.8 & 3 & $2.0 \times 10^{-8}$ & 102.4 & 83.3 & 64.6 & 0.2 & 0.23 & - & C & --- &     21.03 & --- & --- \\
MGPS2 J083343$-$431958 & 08 33 43 & -43 19 58 & 261.9 & -1.9 & 2 & $3.8 \times 10^{-6}$ & 26.2 & 18.3 & 12.4 & 0.5 & 0.36 & - & N & --- & --- & --- & --- \\
MGPS2 J083420$-$385401 & 08 34 20 & -38 53 57 & 258.4 & +0.8 & 3 & $3.8 \times 10^{-6}$ & 66.5 & 48.8 & 40.8 & 0.3 & 0.24 &    33.7 & N & --- & --- & --- & --- \\
MGPS2 J083524$-$482342 & 08 35 24 & -48 23 42 & 266.1 & -4.7 & 3 & $4.6 \times 10^{-7}$ & 51.8 & 36.9 & 31.4 & 0.3 & 0.24 & - & C &     22.12 & --- &     19.50 & --- \\
MGPS2 J083555$-$480638 & 08 35 54 & -48 06 38 & 266.0 & -4.5 & 3 & $4.6 \times 10^{-6}$ & 41.6 & 31.1 & 26.9 & 0.2 & 0.21 & - & C &     19.94 &     18.88 &     18.31 &      1.06 \\
MGPS2 J090857$-$523612 & 09 08 57 & -52 36 13 & 272.8 & -3.3 & 3 & $5.7 \times 10^{-7}$ & 41.7 & 31.1 & 25.6 & 0.3 & 0.24 & - & C & --- & --- & --- & --- \\
MGPS2 J091052$-$473743 & 09 10 52 & -47 37 41 & 269.4 & +0.3 & 2 & $< 1 \times 10^{-14}$ & 34.3 & 15.6 & 10.8 & 0.7 & 0.52 & - & N & --- & --- & --- & --- \\
MGPS2 J095141$-$550630 & 09 51 41 & -55 06 31 & 279.1 & -0.8 & 2 & $3.0 \times 10^{-5}$ & 42.4 & 33.3 & 29.4 & 0.3 & 0.18 & - & N & --- & --- & --- & --- \\
SUMSS J100222$-$423135 & 10 02 22 & -42 31 33 & 272.7 & +10.2 & 2 & $2.1 \times 10^{-5}$ & 25.0 & 16.7 & 13.6 & 0.4 & 0.30 & - & N & --- & --- & --- & --- \\
SUMSS J101048$-$805633 & 10 10 47 & -80 56 32 & 296.7 & -20.1 & 2 & $3.7 \times 10^{-5}$ & 19.5 & 13.4 & 11.0 & 0.4 & 0.28 & - & N & --- & --- & --- & --- \\
MGPS2 J104530$-$505429 & 10 45 30 & -50 54 30 & 283.5 & +7.2 & 2 & $9.6 \times 10^{-10}$ & 94.7 & 68.2 & 57.3 & 0.3 & 0.25 & - & C & --- & --- & --- & --- \\
SUMSS J113907$-$462506 & 11 39 06 & -46 25 06 & 290.2 & +14.7 & 2 & $< 1 \times 10^{-14}$ & 35.7 & 18.3 & 14.3 & 0.6 & 0.43 & - & C &     21.34 &     19.97 &     19.48 &      1.37 \\
SUMSS J120525$-$470714 & 12 05 26 & -47 07 10 & 294.9 & +15.0 & 3 & $1.6 \times 10^{-5}$ & 29.0 & 22.1 & 16.4 & 0.3 & 0.28 & - & N & --- & --- & --- & --- \\
MGPS2 J161340$-$640022 & 16 13 40 & -64 00 22 & 323.0 & -9.3 & 2 & $9.3 \times 10^{-6}$ & 34.9 & 26.9 & 23.0 & 0.3 & 0.21 & - & C &     20.09 &     18.70 &     18.36 &      1.39 \\
MGPS2 J170932$-$441750 & 17 09 32 & -44 17 49 & 343.2 & -2.5 & 3 & $2.1 \times 10^{-6}$ & 36.1 & 30.0 & 21.4 & 0.3 & 0.26 & - & C &     20.48 &     19.07 &     18.52 &      1.41 \\
MGPS2 J171300$-$541538 & 17 12 60 & -54 15 39 & 335.4 & -8.8 & 2 & $1.7 \times 10^{-6}$ & 33.4 & 26.9 & 20.0 & 0.4 & 0.25 & - & C &     20.76 &     20.16 & --- &       .60 \\
SUMSS J173100$-$533721 & 17 31 00 & -53 37 21 & 337.5 & -10.7 & 2 & $3.2 \times 10^{-5}$ & 55.8 & 46.0 & 39.3 & 0.2 & 0.17 & - & C &     20.85 &     19.73 &     19.13 &      1.12 \\
SUMSS J175605$-$464450 & 17 56 06 & -46 44 49 & 345.6 & -10.7 & 2 & $1.1 \times 10^{-8}$ & 34.1 & 22.3 & 16.8 & 0.5 & 0.34 & - & C &     16.71 &     15.56 &     15.00 &      1.14 \\
SUMSS J183056$-$464744 & 18 30 56 & -46 47 43 & 348.2 & -16.1 & 3 & $5.2 \times 10^{-6}$ & 16.4 & 12.1 & 8.3 & 0.3 & 0.33 & - & C & --- & --- & --- & --- \\
SUMSS J190344$-$505108 & 19 03 44 & -50 51 08 & 346.0 & -22.6 & 3 & $1.5 \times 10^{-10}$ & 44.9 & 33.3 & 23.1 & 0.3 & 0.32 & - & C &     19.73 &     19.25 &     19.48 &       .48 \\
SUMSS J200936$-$554236 & 20 09 35 & -55 42 31 & 342.2 & -33.0 & 4 & $4.2 \times 10^{-6}$ & 21.2 & 16.8 & 12.2 & 0.3 & 0.27 & - & G & --- & --- & --- & --- \\
SUMSS J201524$-$395949 & 20 15 24 & -39 59 48 & 0.9 & -32.5 & 2 & $2.0 \times 10^{-5}$ & 31.5 & 23.3 & 16.4 & 0.4 & 0.31 &     8.5 & G & --- & --- & --- & --- \\
SUMSS J202536$-$813918 & 20 25 36 & -81 39 19 & 311.8 & -30.2 & 2 & $1.3 \times 10^{-7}$ & 62.1 & 47.8 & 39.7 & 0.3 & 0.22 & - & D &     20.26 &     19.88 & --- &       .39 \\
SUMSS J211009$-$772600 & 21 10 09 & -77 25 58 & 315.4 & -33.7 & 2 & $1.3 \times 10^{-10}$ & 39.0 & 25.9 & 20.6 & 0.4 & 0.31 & - & N & --- & --- & --- & --- \\
SUMSS J214235$-$724745 & 21 42 35 & -72 47 40 & 318.9 & -37.9 & 3 & $1.8 \times 10^{-6}$ & 14.3 & 9.8 & 7.8 & 0.3 & 0.30 & - & N & --- & --- & --- & --- \\
SUMSS J223017$-$614701 & 22 30 16 & -61 47 03 & 327.1 & -48.2 & 50 & $3.0 \times 10^{-5}$ & 58.4 & 47.0 & 37.9 & 0.1 & 0.21 & - & D &     21.14 &     19.16 &     18.02 &      1.98 \\
SUMSS J223225$-$615308 & 22 32 24 & -61 53 13 & 326.8 & -48.3 & 33 & $< 1 \times 10^{-14}$ & 257.7 & 221.1 & 167.5 & 0.1 & 0.21 & - & N & --- & --- & --- & --- \\
SUMSS J225444$-$500528 & 22 54 44 & -50 05 30 & 339.1 & -58.0 & 2 & $3.5 \times 10^{-7}$ & 24.5 & 17.0 & 12.7 & 0.4 & 0.31 & - & E &     20.92 &     20.46 & --- &       .46 \\
\hline
\end{tabular}
\end{table}
\end{landscape}

\begin{landscape}
\begin{table}
\centering
\caption{15 radio sources classified as transient. The columns are: the source name, right ascension and declination (J2000), Galactic longitude and latitude, number of detections, number of accepted observations, probability of $\chi_{\rm lc}^2$ of the residuals to a line of constant flux density, maximum flux density at 843~MHz in $\unit{mJy~beam^{-1}}$, the maximum and minimum detection significance, NVSS flux density at 1.4~GHz in $\unit{mJy~beam^{-1}}$ (if covered by NVSS area, ND if not detected), SuperCOSMOS classification: D = point source detected, N = no source detected, C = more than 1 point source in the error box, E = extended according to SuperCOSMOS catalogue, G = clearly resolved galaxy,  the SuperCOSMOS optical magnitudes in blue (B), red (R) and infra-red (I), and the B$-$R colour. More information on these sources, including identifications, can be found in Table~\protect \ref{tab:transients_comments}. }
\label{tab:transients}
\scriptsize
\centering
\begin{tabular}{lllrrrrrrrrrrrrrrrr|}
\hline
Source name & $\alpha$ & $\delta$ & $l$ & $b$ & $N_{det}$ & $N_{\rm obs}$ & $1 - P(\chi_{\rm lc}^2)$& $S_{\rm max}$ & $d_{\rm max}$ & $d_{\rm min}$ & NVSS & Supercos & B &  R & I &  B$-$R\\
& (h m s) & (d m s) & (deg) & (deg) & & & & ($\unit{mJy~beam^{-1}}$) &  & & ($\unit{mJy~beam^{-1}}$) & & (mag) & (mag) & (mag) & (mag) \\
\hline
SN1987A & 05 35 28 & -69 16 07 & 279.7 & -31.9 & 2 & 13 & $6.5 \times 10^{-13}$ & 151.8 & 13.5 & -0.4 & - & C &     18.16 &     13.25 &     12.65 &      4.91 \\
SUMSS J055712$-$381106 & 05 57 12 & -38 11 05 & 244.3 & -26.5 & 1 & 2 & $2.5 \times 10^{-3}$ & 16.7 & 7.1 & 2.9 &     5.2 & E &     17.89 &     16.80 &     16.18 &      1.10 \\
J060938$-$333508 & 06 09 38 & -33 35 08 & 240.3 & -22.7 & 1 & 3 & $1.4 \times 10^{-6}$ & 21.3 & 6.6 & 2.3 & ND & G & --- & --- & --- & --- \\
J061051$-$342404 & 06 10 50 & -34 24 04 & 241.2 & -22.8 & 2 & 4 & $3.4 \times 10^{-3}$ & 14.3 & 8.3 & 2.2 & ND & N & --- & --- & --- & --- \\
SUMSS J062636$-$425807 & 06 26 36 & -42 58 07 & 251.0 & -22.4 & 1 & 2 & $7.0 \times 10^{-4}$ & 19.3 & 15.0 & 2.9 & - & D &     19.39 &     18.18 &     16.81 &      1.20 \\
J062716$-$371736 & 06 27 16 & -37 17 36 & 245.3 & -20.5 & 2 & 3 & $3.1 \times 10^{-3}$ & 14.9 & 6.6 & 2.9 &     2.7 & N & --- & --- & --- & --- \\
J064149$-$371706 & 06 41 49 & -37 17 06 & 246.3 & -17.8 & 1 & 2 & $4.4 \times 10^{-4}$ & 14.4 & 6.1 & 2.4 & ND & N & --- & --- & --- & --- \\
SUMSS J102641$-$333615 & 10 26 40 & -33 36 14 & 271.3 & +20.2 & 1 & 2 & $9.8 \times 10^{-3}$ & 17.2 & 7.3 & 2.7 & ND & N & --- & --- & --- & --- \\
SUMSS J112610$-$330216 & 11 26 09 & -33 02 13 & 283.0 & +26.5 & 1 & 2 & $4.6 \times 10^{-3}$ & 16.6 & 7.8 & 2.8 &     7.1 & N & --- & --- & --- & --- \\
Nova Muscae 1991 & 11 26 27 & -68 40 31 & 295.3 & -7.1 & 1 & 2 & $< 1 \times 10^{-14}$ & 140.4 & 67.8 & 0.1 & - & C &     21.57 & --- &     18.72 & --- \\
J121032$-$381439 & 12 10 32 & -38 14 39 & 294.2 & +23.9 & 1 & 2 & $7.4 \times 10^{-3}$ & 15.0 & 9.5 & 2.8 &     9.6 & N & --- & --- & --- & --- \\
J135304$-$363726 & 13 53 04 & -36 37 26 & 316.5 & +24.6 & 1 & 3 & $1.0 \times 10^{-3}$ & 16.3 & 6.6 & 2.9 &     5.1 & D & --- &     20.71 & --- & --- \\
J153613$-$332915 & 15 36 13 & -33 29 15 & 338.2 & +17.9 & 1 & 2 & $1.5 \times 10^{-3}$ & 18.4 & 6.2 & 2.3 &     7.7 & N & --- & --- & --- & --- \\
GRO 1655$-$40 & 16 54 00 & -39 50 44 & 345.0 & +2.5 & 2 & 9 & $< 1 \times 10^{-14}$ & 6504.2 & 17.4 & -0.8 & ND & E &     18.58 &     15.26 &     13.96 &      3.32 \\
SUMSS J224152$-$300823 & 22 41 52 & -30 08 23 & 19.1 & -61.5 & 1 & 3 & $9.6 \times 10^{-3}$ & 15.6 & 9.4 & 0.6 & ND & N & --- & --- & --- & --- \\
\hline
\end{tabular}
\end{table}
\end{landscape}

\subsection{Transient sources}
Table~\ref{tab:transients} lists 15 sources we classify as transients by the criteria described in $\S$\ref{sec:criteria}. Eight sources that satisfied these criteria are not shown in the table, because visual inspection showed incorrect classification due to image artefacts, such as nulls causing non-detections or false detections from grating rings.

\subsection{Optical Counterparts}
\begin{figure}
\centering
\includegraphics[width=\columnwidth]{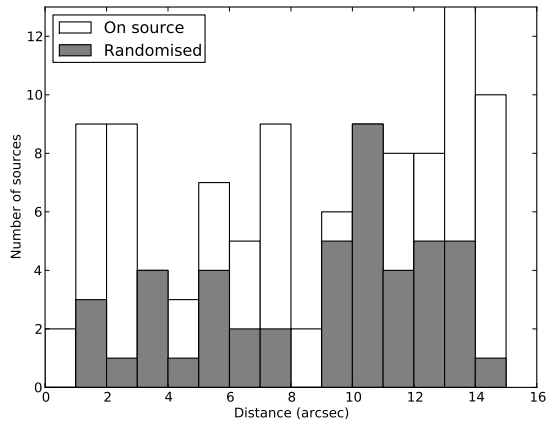}
\caption{Histogram of the number of SuperCOSMOS B sources as a function of radius for randomly chosen fields, and for uncrowded MOST transient and variable sources.}.
\label{fig:optical_matches}

\end{figure}

\change{The deepest optical survey of the area of sky covered by our survey is the SuperCOSMOS Sky Survey \citep{supercos1}, which was produced by digitising photographic plates from the UK Schmidt Telescope. The SuperCOSMOS Sky Survey has image and catalogue data in three colours with limits down to $B \sim22$,  $R \sim20$ and  $I \sim19$. To determine an optical match for each radio source, we visually inspected overlays of the MOST data sources with the SuperCOSMOS B image, and searched the SuperCOSMOS catalogues.

From visual inspection of the overlay plots we were immediately able to identify four of the 68 transient and variable radio sources have radio positions directly on, or within 10~arcsec of the centre of a resolved optical galaxy. These sources that have an optical classification of `G' in Table~\ref{tab:variables} and Table~\ref{tab:transients}. To determine the probability of random false association, we visually inspected images of 68 random positions on the sky and found that no positions fell near resolved optical galaxies of similar size. We conclude that the radio sources that are positionally coincident with resolved optical galaxies are very likely to be physically associated.

Many of our radio sources are found in optically crowded regions in the Magellanic Clouds. The positional uncertainty for MOST sources is $< 5$~arcsec \citep{Mauch2003sumss}; therefore, we consider any region with 2 or more optical matches within 5~arcsec to be crowded. Such sources have an optical classification of `C' in Table~\ref{tab:variables} and Table~\ref{tab:transients}, and optical magnitudes shown in the tables are those of the nearest candidate to the known radio position.

We consider any radio source with a single optical counterpart within 5~arcsec to be a detection. These sources have an optical classification of `D' if the match is classified as point-like (mean class is 1) in the SuperCOSMOS catalogue, or `E' if the match is extended in the SuperCOSMOS catalogue (mean class is 2). To measure the probability of a false match we compare the number of optical sources contained within a circle centred on the radio source position, and on a random position in the sky (Figure \ref{fig:optical_matches}) over a range of radii. We find that, within 5~arcsec, there are 9 optical matches for the random positions, and 27 optical matches for the radio source positions, corresponding to a false match probability at the 5~arcsec match radius of 33 per cent.

The remaining radio sources have no point-like optical counterparts within 5~arcsec to within  $B \sim22$,  $R \sim20$ and  $I \sim19$. These sources have an optical classification of `N' in Table~\ref{tab:variables} and Table~\ref{tab:transients}.

} 

\subsection{Gamma Ray Burst Counterparts}
\change{
Some of our transient and variable sources could be related to GRB afterglows. In the standard picture, radio afterglow at 843~MHz would be detected several weeks after a GRB event and would fade over several months.

We searched the master catalogue of the \emph{Swift} GRB satellite that has been operational since 2005 Jan \footnote{\url{http://www.swift.ac.uk/swift_live/arnie5.php}}, for GRBs within 10~arcmin\ of all transient and variable radio sources. We found no GRBs within the position uncertainties of any transient or variable radio sources.

We also searched the \citet{Stern01} catalogue for GRBs detected by the BATSE instrument aboard the \emph{Compton Gamma-Ray Observatory}, which was operational between 1991 and 2000, for our transient sources only.  For each transient source, between 5 and 10 counterparts were identified within positional uncertainties (which are typically tens of degrees). No GRBs were detected in the weeks before the first detection of a transient radio source with a typical GRB afterglow light curve.

\change{The vast majority of GRBs are detected at $z > 0.1$ \citep{Jakobsson06} with the remainder being detected in gamma rays at much larger distances. All of our transient sources with resolved optical counterparts are at $z <0.1$. Hence, if our transient sources had a gamma-ray trigger, we would expect them to be detected if a GRB satellite had been operational at the time.}

We note that a number of transient sources with radio light curves consistent with GRB afterglows (i.e. J060938$-$333508, J062636$-$425807, J064149$-$371706 and SUMSS J102641$-$333615), would not have had a GRB detection in either \emph{Swift} BAT or BATSE catalogues as neither instrument was fully operational when the first MOST detection was made.

} 

\section{Discussion}

\subsection{Overall statistics}
Our sample comprises light curves for 29730 sources, with a two-epoch equivalent sky coverage (c.f. Equation \ref{eq:A2epoch}) of $2775.7 \unit{deg^{2}}$. The shortest interval between accepted measurements of a single source is 24~h and the longest interval between accepted measurements is 7301 days ($\simeq 20 \unit{years}$).

For a source to be in our sample, it must appear in at least two images for which the exact two-epoch coverage applies, rather than the equivalent coverage. The two-epoch (exact) coverage is approximately $2000\unit{deg^2}$ (c.f. Figure \ref{fig:coveragehistogram}) resulting in a source density of $15\unit{deg^{-2}}$. The SUMSS source density is  $31.6\unit{deg^{-2}}$ meaning the overall completeness of our sample is 47 per cent.

\begin{figure}
\centering
\includegraphics[width=\linewidth]{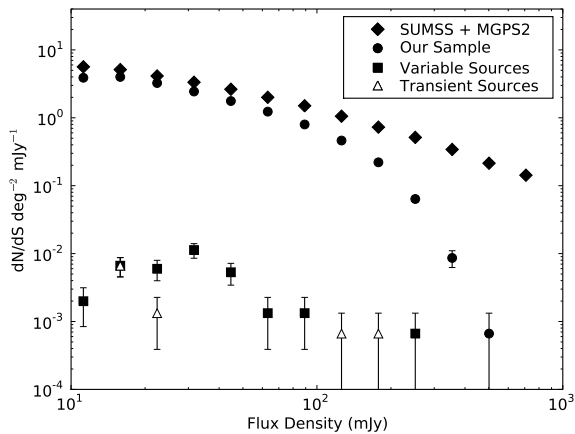}
\caption{Log N-Log S plot for the combined SUMSS and MGPS-2 catalogue, all sources used for our variability analysis, and the sources classified as variable and transient. The error bars are Poisson errors. The sources available for variability analysis are incomplete above 100$\unit{mJy~beam^{-1}}$, as strong sources have poor fits in MOST images due to artefacts.}
\label{fig:logn_logs}
\end{figure}

A Log N-Log S plot for the SUMSS and MGPS-2 catalogues, our whole sample, the transient sources and the variable sources is shown in Fig.~\ref{fig:logn_logs}.  At low flux densities the catalogue is incomplete due to variations in the noise floor and at high flux densities it is due to artefacts around bright sources. In general the low completeness is due to the strict requirements on goodness of fit we imposed in order to remove false positives due to artefacts.

\begin{figure}
\centering
\includegraphics[width=\linewidth]{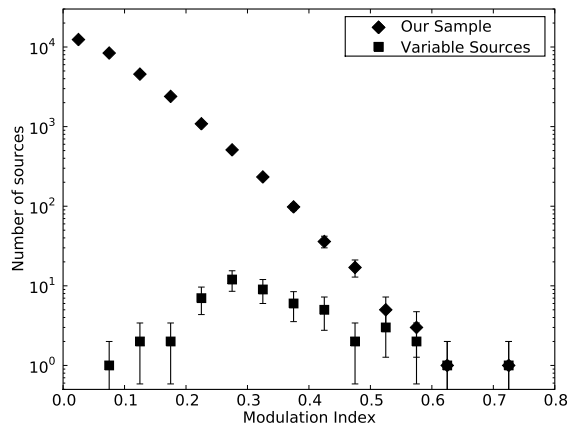}
\caption{Histogram of modulation indices  ($m=\sigma/\bar{S}$, see $\S$\ref{sec:varmeas}) for all sources, including variable sources.}
\label{fig:modindex_histogram}
\end{figure}

\begin{figure}
\centering
\includegraphics[width=\linewidth]{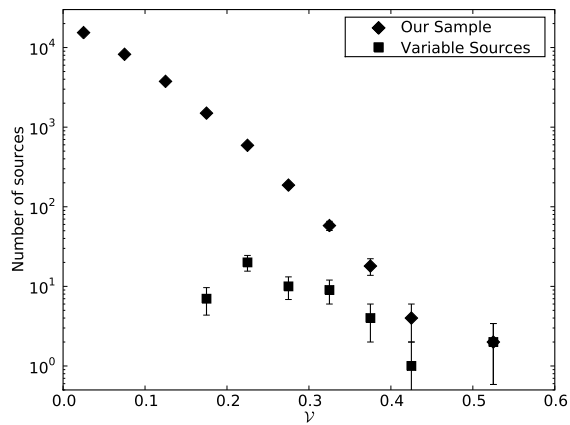}
\caption{Histogram of fractional variability (see Equation~\ref{eq:v}) for all sources and variable sources.}
\label{fig:v_histogram}
\end{figure}

Fig.~\ref{fig:modindex_histogram} and \ref{fig:v_histogram} show histograms of modulation index, and $\mathcal{V}$ for all sources in our sample.

\subsection{Variability time-scales}
\label{sec:variability_time_scales}
\begin{figure}
\centering
\includegraphics[width=\linewidth]{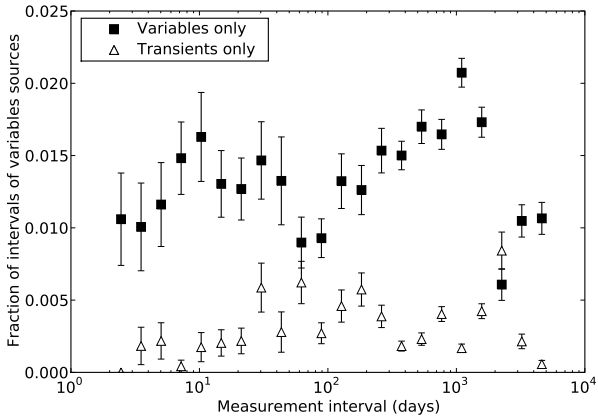}
\caption{\change{The ratio of the number of non-redundant pairs of intervals in the light curves for for variable and transient sources, to the total number of intervals for all sources, as a function of the interval in deays. The y errors are the Poisson errors.}}
\label{fig:intervals_ratio}
\end{figure}

\change{
If there is a sufficient number of measurements on the light curve of a source, a structure function can be formed to make an estimate of the variability time-scale (e.g. \citeauthor{gaensler2000most} \citeyear{gaensler2000most} or \citeauthor{lovell2008MASIV} \citeyear{lovell2008MASIV}). As the majority of our sources have only a small number of measurements (see Fig.~\ref{fig:nepochs_histogram}), structure functions cannot be used. To determine if there is a preferred variability time-scale for our sources, we compare the distribution of intervals between measurements of our variable sources, with that of the whole sample as follows: For each source, we take each non-redundant pair of flux density measurements and bin by the time interval between these measurements, i.e. if there are $N_{\rm meas}$ measurements on the light curve, there are $N_{\rm meas}(N_{\rm meas} -1)/2$ intervals. We then form a histogram comprised of all sources, and the subset of variable sources in Table~\ref{tab:variables} only. If the variable sources have a preferred time-scale, this should be evident in the variable sources constituting a larger fraction of their measurement intervals at a particular time-scale. Fig.~\ref{fig:intervals_histogram} depicts the histograms of intervals described above.

Fig.~\ref{fig:intervals_ratio} shows the ratio of the number intervals belonging to the variable and transient sources to the number of intervals belonging to the whole sample peaks. The ratio peaks at around 1000 days and then falls rapidly by 2000 days. This suggests that any blind survey for variable radio sources will detect the majority of variable sources in the first 1000 to 2000 days.

}

\subsection{Source counts for variable sources}
The number of variable sources in our sample can be determined in several ways. The source list in Table~\ref{tab:variables} contains 55 sources and we are confident that at least 90 per-cent of them are truly variable, which implies a source count of 0.020 sources per two-epoch square degree, but this is not the whole story. Fig.~\ref{fig:nsources_vs_threshold} shows there is a clear excess of sources with $P(\chi_{\rm lc}^2)< 0.1$. While we cannot determine whether a particular source's variation is attributable to measurement errors or true variation, we can compare the number of apparently variable sources to the expected number if only Gaussian measurement errors were present. If we conservatively take the `expected' line as the number of non-varying sources, we arrive at an excess of 200 sources, corresponding to a lower limit on the the two-epoch areal density of variable sources at 843~MHz of $0.072 \unit{deg^{-2}}$. This would appear to be an underestimate, as the actual number of sources with $P(\chi_{\rm lc}^2)< 0.2$ is substantially less than the expected number, due to over-estimated errors. If we extrapolate the actual number of sources from $P(\chi_{\rm lc}^2)= 0.2$, we arrive an an excess of approximately 350 sources with $P(\chi_{\rm lc}^2)< 0.2$ at 843~MHz with flux densities between 15 and 100$\unit{mJy~beam^{-1}}$, which corresponds to a two-epoch areal density of variable sources at 843~MHz of $0.126 \unit{deg^{-2}}$. After accounting for completeness of 47 per cent, the areal density rises to $0.268 \unit{deg^{-2}}$ in 2 epochs.

A definitive statement about the number of variable sources requires a caveat on the time-scale of the variability. Our data are not regularly sampled and so we can only make broad statements about variability on time-scales greater than 1 day and less than about 10 years.

\subsection{Interpretation as interstellar scintillation}
At 843~MHz, our observations are firmly in the strong or refractive regime of interstellar scintillation. In these conditions \citet{Rickett86} derives the maximum rate of change of flux of sources with a brightness of $T_{12} \times 10^{12}\unit{K}$  at latitudes $|b| \geq 10\degree$ as

\begin{equation}
\label{eq:smax}
(dS/dt)_{max} < S_{rms}/\tau \sim 30T_{12}(\sin b)^{0.5} \unit{~ mJy ~day^{-1}}
\end{equation}

\noindent giving a range of $(dS/dt)_{max}$ for a source with a brightness temperature at the inverse Compton limit of between 12.5 and 30~mJy day$^{-1}$. As our data are sparsely sampled in the time domain, this limit is easily satisfied for all of our sources.

\begin{figure}
\centering
\includegraphics[width=\linewidth]{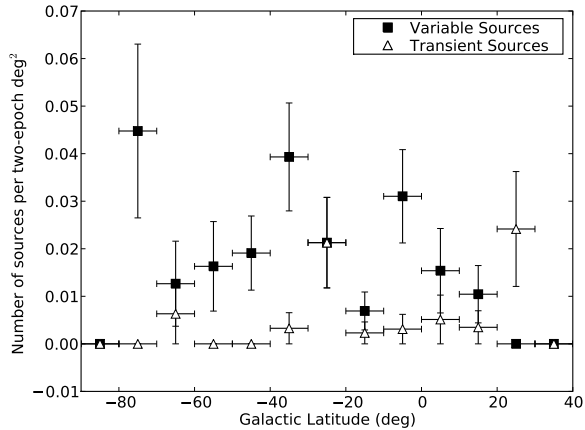}
\caption{Fraction of variable and transient sources vs Galactic latitude.}
\label{fig:frac_variables_vs_gallat}
\end{figure}

\change{The number of sources that scintillate is expected to increase as the line of sight to the source becomes closer to the Galactic plane, and is more likely to pass through a scintillating screen of material. In Figure \ref{fig:frac_variables_vs_gallat} we plot the fraction of sources that we defined as variable and transient, binned by Galactic latitude. The plot suggests, at best, a mild dependence on Galactic latitude. Nonetheless, we expect that refractive interstellar scintillation of compact radio sources is a likely explanation for a large fraction of our variable sources, as also seen by \citet{gaensler2000most}.}

Some 30 per cent of the variable sources have no optical counterpart. In the absence of more data, we cannot positively identify these sources.

\subsection{Source counts for transient sources}
Transient emission on characteristic time-scales of less than the full synthesis time is unlikely to be detected by our technique, because a significant change in flux density would result in a distorted synthesised beam for that source, which would result in large errors in a Gaussian fit and ultimately rejection from the analysis. Therefore, we will omit from our discussion source classes such \changeb{as}  giant pulses from neutron stars \citep{Hankins03} as well as flares from extrasolar planets \citep{Bastian00} and fast extragalactic transients \citep{lorimer2007bmr}, which have time-scales much shorter than 12~h. Also, a source that is constant over 12~h but bright ($>100~\unit{mJy}$), or in a region containing bright sources (such as the Galactic plane), is unlikely to be detected due to image artefacts or large formal errors in source fitting.

\label{sec:sourcecounts}
Table~\ref{tab:transients} shows 15 sources that satisfied our definition of a transient with a flux density above $14 \unit{mJy~beam^{-1}}$, corresponding to an areal density of transient sources at 843~MHz of $6 \times 10^{-3} \unit{deg^{-2}}$ in 2 epochs. After accounting for completeness of 47 per cent, the areal density rises to $1.3 \times 10^{-2} \unit{deg^{-2}}$ in 2 epochs.

\citet{bower2007sta} compute a two-epoch areal density of transient sources of $1.5 \unit{deg^{-2}}$ above $370 \unit{\mu Jy}$ at $5 \unit{GHz}$ and a flux density distribution following $S^{-\gamma}$ with $\gamma=1.5$, which corresponds to an areal density of $2.0 \times 10^{-2} \unit{deg^{-2}}$, assuming a spectral index of -0.7 and flux density limit of $15 \unit{mJy~beam^{-1}}$,  in broad agreement with our findings. We note that there are aspects of our transients which are not the same as those described by \citet{bower2007sta}. In particular, our transients are detected at a frequency an order of magnitude lower, have flux densities an order of magnitude higher, and have upper limits on characteristic time-scales of one or two orders of magnitude longer.

\subsubsection{Known transients}
Our search detected three previously identified radio transients:

\begin{enumerate}
\item GRO $1655-40$ (J165400$-$395044) is a low mass X-ray binary (or microquasar), detected in our survey because of the images taken while monitoring a dramatic radio flare in 1994 \citep{1994IAUC.6052....2D}.

\item Nova Muscae 1991 (J112627$-$684031) is a low mass X-ray binary, detected in our survey from an image taken during the outburst in 1991 \citep{1991IAUC.5181....2K}.

\item SN1987A (J053528$-$691607) was a supernova in the Large Magellanic Cloud. It is detected in our survey due to a regular monitoring campaign \citep{Ball01} during which its flux density has been steadily increasing.

\end{enumerate}

\section{Discussion of selected sources}
\label{sec:selected_sources}
\change{In this section we discuss sources with noteworthy light curves or optical counterparts.}

\subsection{Variable sources with resolved optical counterparts}

\subsubsection{SUMSS J011019$-$455112}
\begin{figure}
\includegraphics[width=\linewidth]{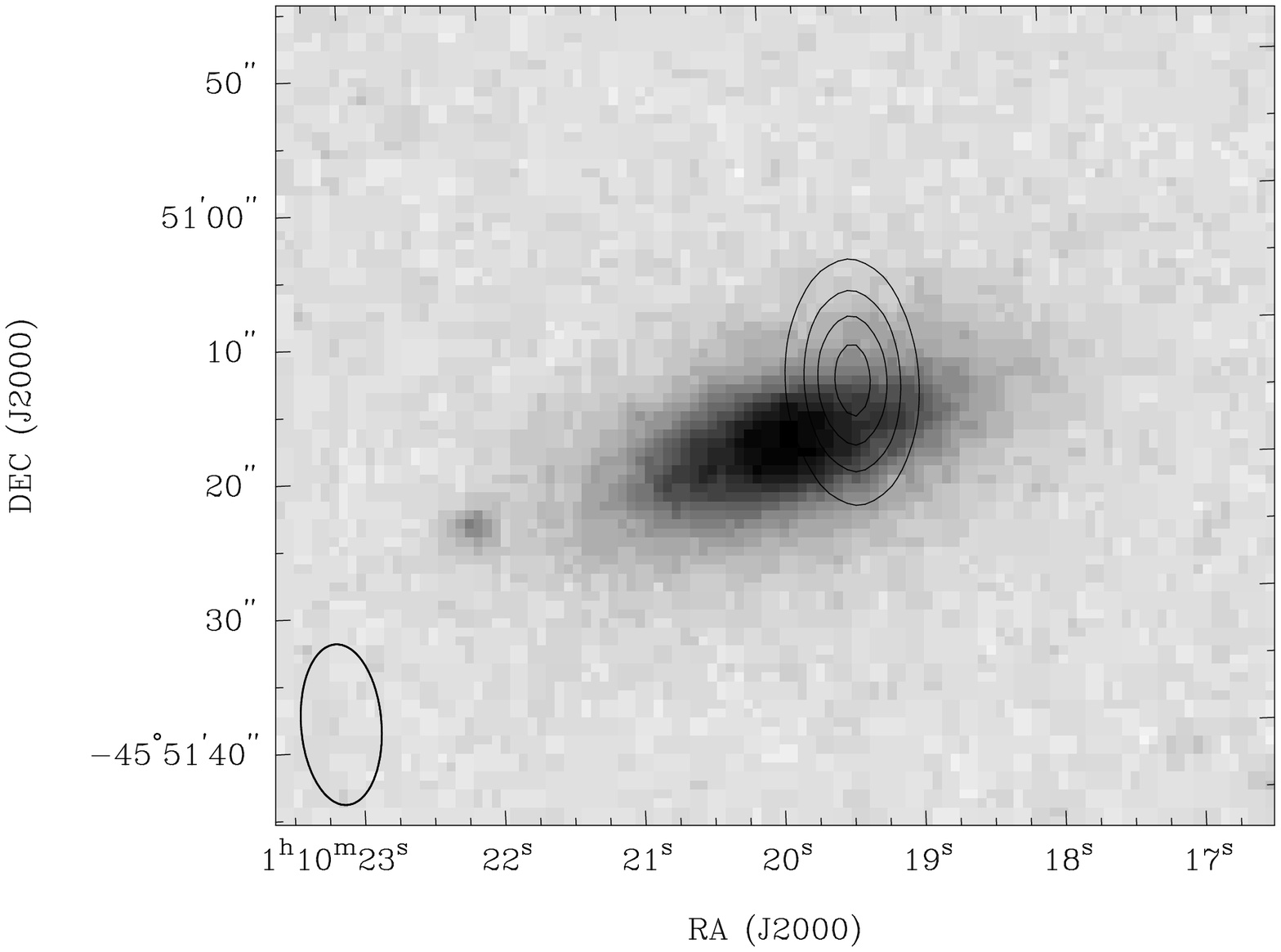}
\includegraphics[width=\linewidth]{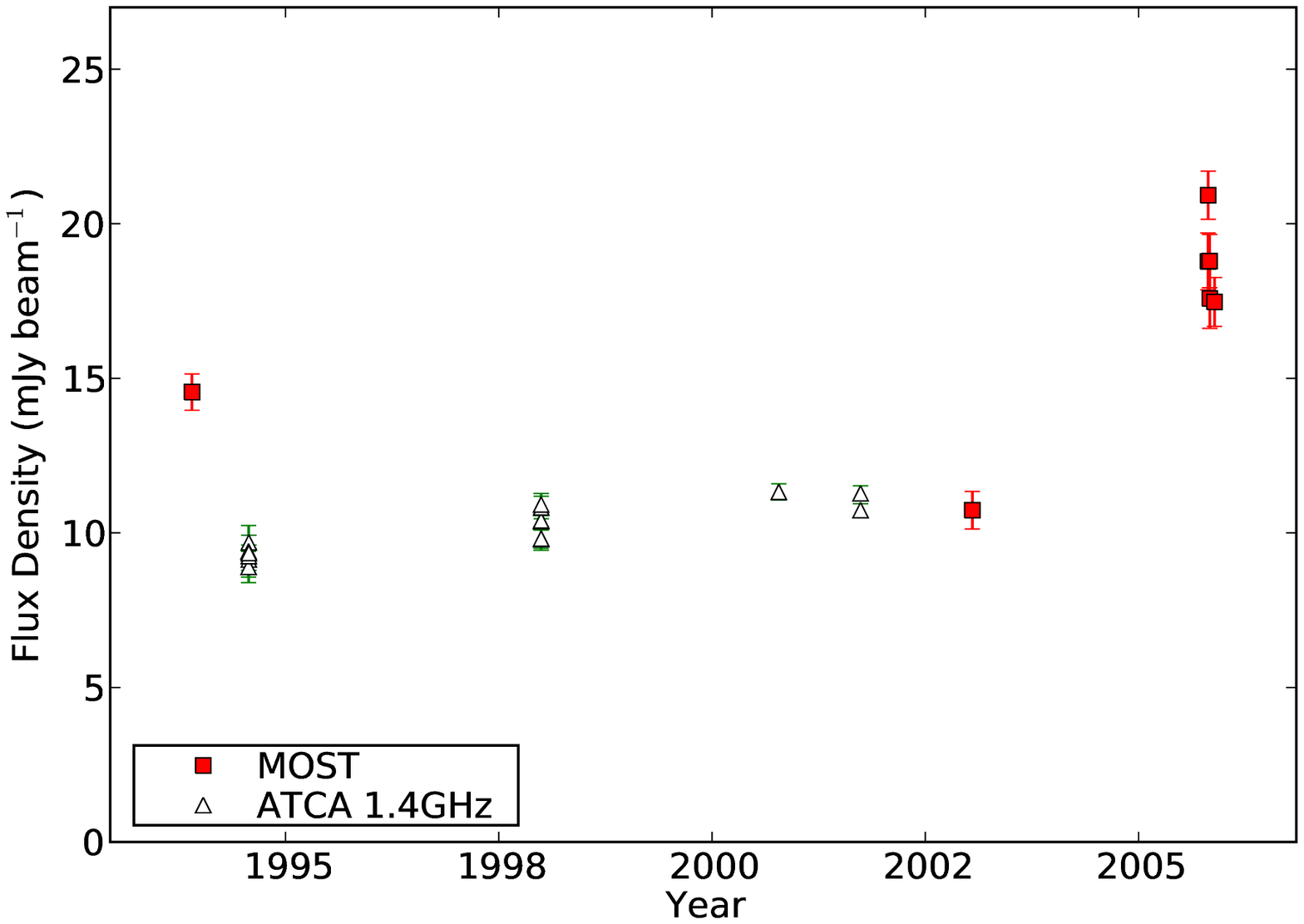}

\caption{Variable source SUMSS J011019$-$455112. Top panel: Radio contours from the Phoenix deep field \protect \citep{hopkins2003phoenix} at 1.4~GHz at 2 to 8 $\unit{mJy~beam^{-1}}$ in steps of 2$\unit{mJy~beam^{-1}}$ overlaid on a SuperCOSMOS B image. Bottom panel: Radio light curves at 1.4~GHz (ATCA archive) and 843~MHz (MOST).}
\label{fig:J011019-455116}
\end{figure}

The MOST light curve of SUMSS J011019$-$455112 (Fig.~\ref{fig:J011019-455116}) shows two decreasing measurements over 10 years, followed two years later by four significantly higher measurements closely spaced in time. \change{We have also obtained flux density and position measurements from archival ATCA observations at 1.4~GHz used for the Phoenix Deep Field survey, which reports the average flux density over all epochs of 10.6~mJy \citep{hopkins2003phoenix}}. The optical overlay of the higher resolution ATCA data shows the source is centred 9.6 arcsec from the centre of the extended 2MASS source 2MASX J01101993$-$455118, an S0 galaxy in the galaxy cluster Abell 2877 at $z=0.023$ \citep{Caldwell1997boe}. Using the highest MOST flux density, the implied isotropic radio luminosity is $L_{\nu} = 2 \times 10^{29} \unit{erg~s^{-1}~Hz^{-1}}$. The spectral index between the mean 1.4~GHz flux density and the minimum and maximum flux densities from MOST is in the range 0 to -1.3 but we note that no measurements are contemporaneous.  \change{The radio source detected by ATCA at 1.4~GHz was a point source at the 12~arcsec resolution of the observations, and had a position consistent with $\alpha=$  1h1m19.442s $\delta=$ $-$45d 51m13.958s (J2000) in all ATCA epochs.}

As the MOST contours are offset from the optical centre of 2MASX J01101993$-$455118, it is unlikely to be associated with AGN activity, so a Radio Supernova (RSN) or GRB afterglow are possible candidates.

Radio Supernovae (RSNe) have typical radio spectral luminosity of $L_{\nu} \simeq 10^{25} \unit{erg~s^{-1}~Hz^{-1}}$, but can reach as high as  $L_{\nu} \simeq 2 \times 10^{27} \unit{erg~s^{-1}~Hz^{-1}}$ \citep{Eck02} and have fading times of 20 years or more. SUMSS J011019$-$455112 has a much higher radio luminosity than even the most luminous known RSN (1979C), \change{so if it is a radio supernova, it is unusually bright.}

The radio luminosity of SUMSS J011019$-$455112 is closer to that of GRB afterglows, which typically have a range of $L_{\nu} \simeq 6 \times10^{28} -  4 \times 10^{31} \unit{erg~s^{-1}~Hz^{-1}}$ \citep{Weiler02}. On the other hand, the 12 year time-scale is typical for a Type II RSN but too long for a GRB afterglow, which typically fade after several weeks. \change{The approximately constant flux density for the MOST and ATCA observations could be explained by pre-existing disc synchrotron emission from the host galaxy, with the flare in the MOST observations being due a flare or stellar explosion overlaid. In such a case, one would expect the disc synchrotron emission to appear resolved at the ATCA resolution and roughly follow the galaxy optical emission, but we find no evidence for resolved emission in the ATCA observations.}

We conclude that SUMSS J011019$-$455112 is associated with an unusual stellar event or explosion in 2MASX J01101993$-$455118.

\subsubsection{SUMSS J201524$-$395949}
\begin{figure}
\includegraphics[width=\linewidth]{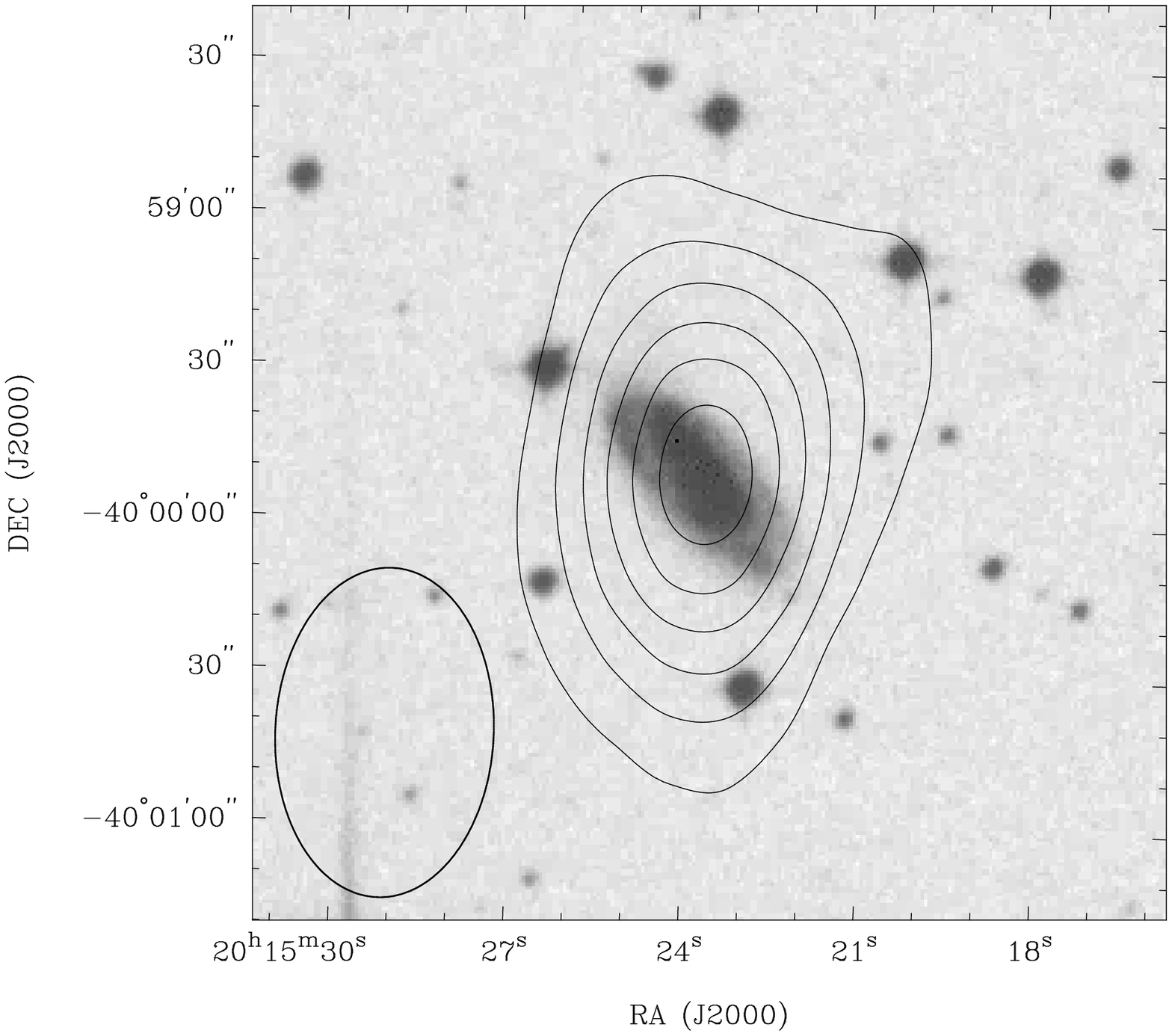}
\includegraphics[width=\linewidth]{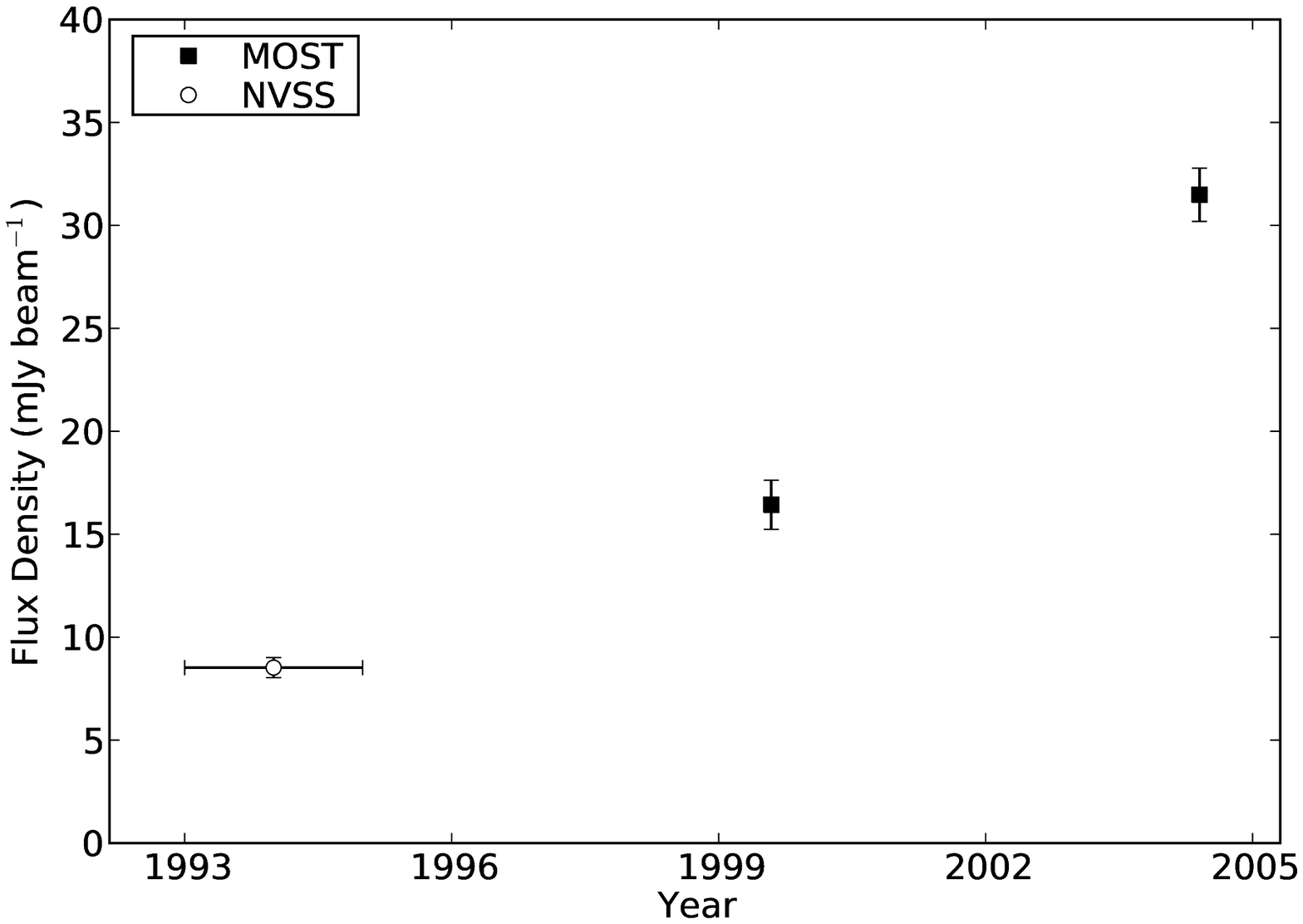}
\caption{Variable J201524$-$395949. Top panel: MOST contours from 5 to $30 \unit{mJy~beam^{-1}}$ in steps of $5 \unit{mJy~beam^{-1}}$for the brightest epoch (2004 May 01) overlaid on a SuperCOSMOS B image. Bottom panel: light curve showing an NVSS detection and a doubling in 843~MHz flux density over 5 years after a 6 year interval.}
\label{fig:J201524-395948}
\end{figure}

The light curve for SUMSS J201524$-$395949 (Fig.~\ref{fig:J201524-395948}) shows an NVSS detection and two MOST detections over 10 years. The MOST flux density doubled over 4 years. The MOST contours are centred on a barred spiral galaxy ESO 340-G06 at redshift $z=0.02$ \citep{6dF} implying, if the radio source is associated with the spiral galaxy, an isotropic radio luminosity of  $L_{\nu} = 3 \times 10^{29} \unit{erg~s^{-1}~Hz^{-1}}$.

The 6dFGS spectrum  \citep{6dF} shows strong Balmer emission lines, indicative of a star-forming galaxy. The \emph{IRAS} $60\unit{\mu m}$ flux is 1~Jy, but the galaxy is not detected at 12 and 25 $\unit{\mu m}$ \citep{IRAS}, suggesting the presence of cold dust. It is rare, but not impossible, for such a star-forming galaxy to contain an AGN.

The radio flux in this case is likely due to a stellar explosion, as widespread star formation or disc synchrotron emission cannot explain the radio variability. Once again, the time-scale of 10~years argues strongly against a GRB afterglow. RSN light curves are known to behave differently at different frequencies, with flux density increase being later and slower at lower frequencies. The increasing flux density over 4 years at 843~MHz is consistent with a Type II RSN but the 6 year interval between MOST and NVSS detections is arguably too long to be due entirely to an RSN, in which case it may be that the RSN occurred after 1996 and the NVSS and first MOST measurement could be attributed to pre-existing disc synchrotron emission or widespread star formation.

\subsubsection{SUMSS J200936$-$554236}
\begin{figure}
\includegraphics[width=\linewidth]{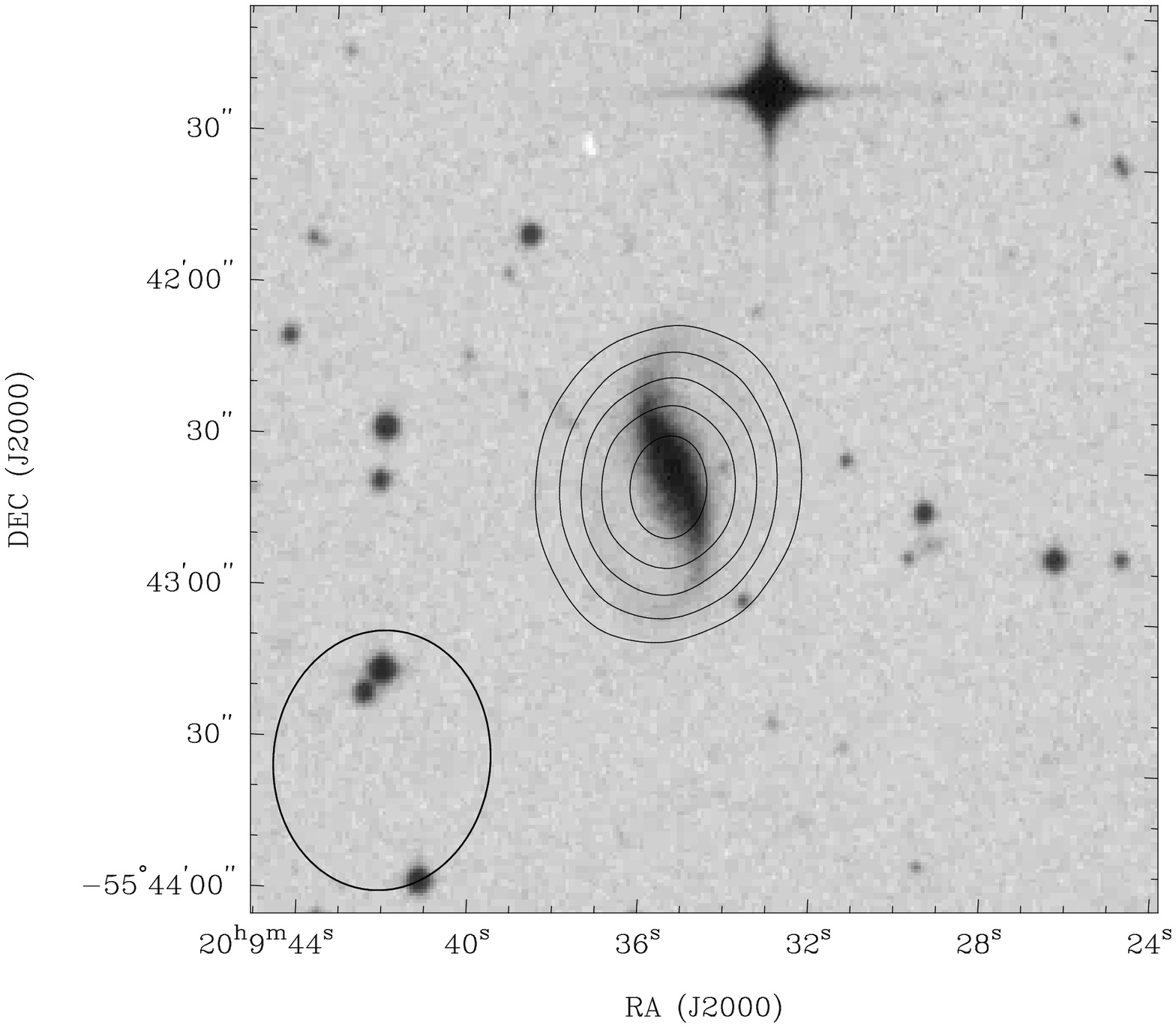}
\includegraphics[width=\linewidth]{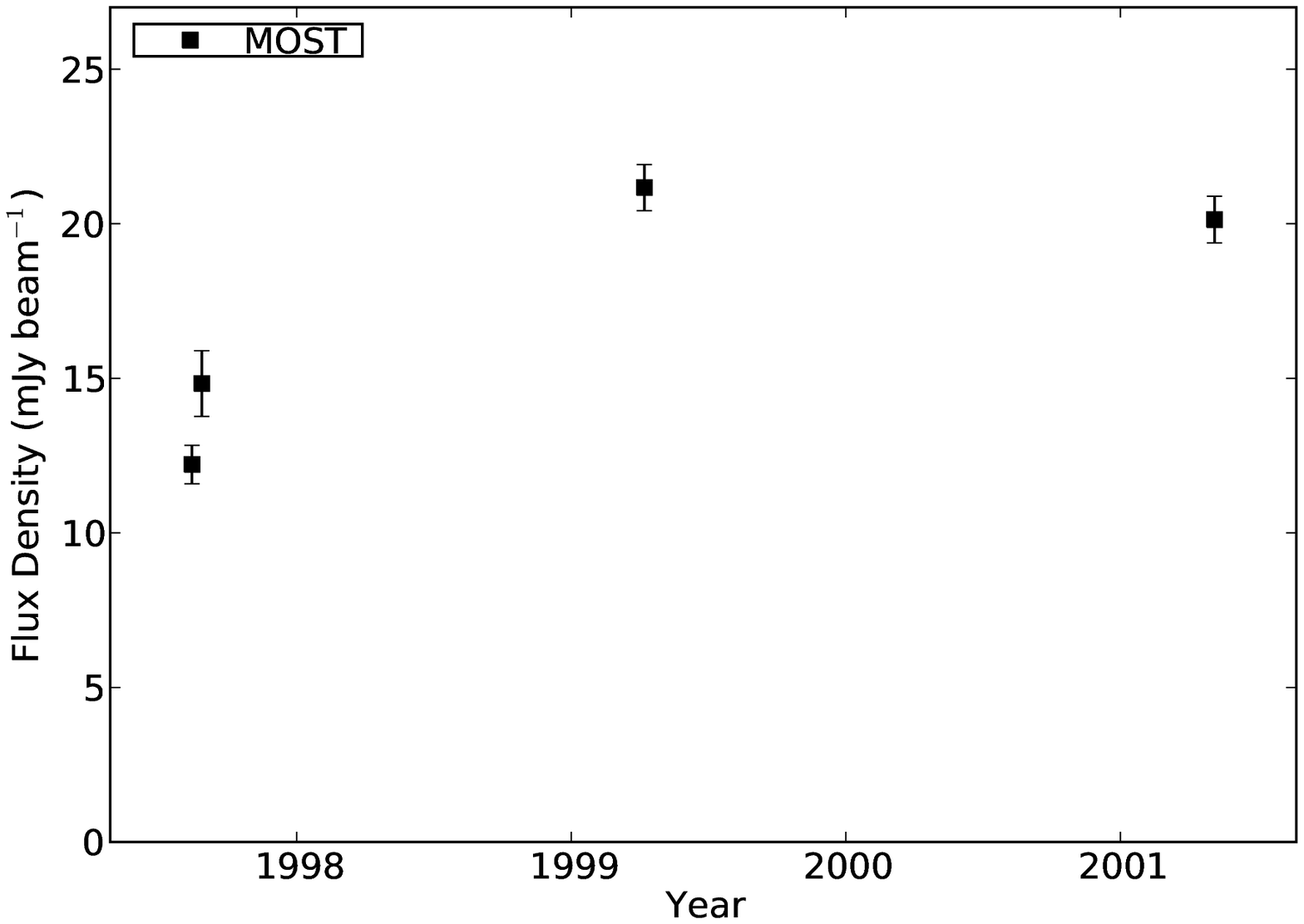}
\caption{Variable SUMSS J200936$-$554236. Top panel: MOST contours at  $-1.5 \unit{mJy~beam^{-1}}$ and 6 to 18$\unit{mJy~beam^{-1}}$ in steps of $3 \unit{mJy~beam^{-1}}$ for the brightest epoch (1999 March 15) overlaid on a SuperCOSMOS B image. Bottom panel: radio light curve indicating a rising and then constant flux density.}
\label{fig:J200935-554231}
\end{figure}

The light curve for  SUMSS J200936$-$554236 (Fig.~\ref{fig:J200935-554231})  shows an increase in flux density over 3 years, and then steady flux density over 2 years. The MOST contours are centred on the barred spiral galaxy IC4957 at redshift $z=0.032$ \citep{1992ApJS...83...29S}. This galaxy is detected across the \emph{IRAS} bands at 0.2~Jy ($25 \unit{\mu m}$), 2.4~Jy ($60 \unit{\mu m}$) and 4.5~Jy ($100 \unit{\mu m}$), implying warm dust. The implied radio luminosity, if the radio source is associated with the spiral galaxy, is  $L_{\nu} = 5 \times 10^{29} \unit{erg~s^{-1}~Hz^{-1}}$.

As the MOST contours are centred on the optical source, AGN activity or ISS of an AGN is a possible explanation but AGNs in barred spiral galaxies are rare \citep{Wilson95}. The warm dust implies star formation, so a stellar event is conceivable. The 3 year time-scale argues strongly against a GRB afterglow interpretation, and the shape and time-scale are consistent with a Type II RSN. The luminosity, once again, is higher than that of any known RSN. We conclude that SUMSS J200936$-$554236 is associated with an unusual stellar event or explosion in IC4957.

\subsection{A compact steep spectrum source: SUMSS J223225$-$615308}
\begin{figure}
\includegraphics[width=\linewidth]{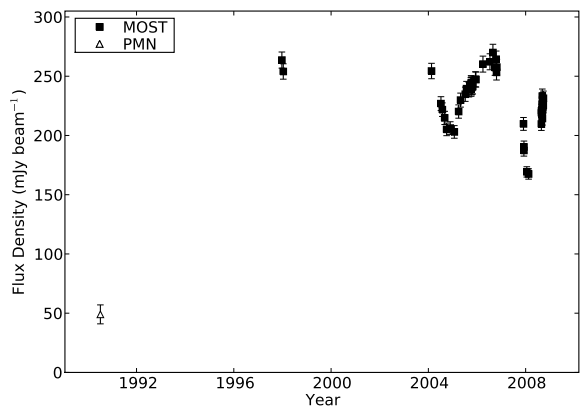}
\caption{Light curve of the candidate compact steep spectrum source: SUMSS J223255$-$615308}
\label{fig:J223255-615313-lightcurve}
\end{figure}

The light curve for SUMSS J223225$-$615308  (Fig.  \ref{fig:J223255-615313-lightcurve}) shows two pronounced dips in flux density over two years and it is the most dramatic of some five variable sources in this field. There are other sources at the same flux density level in this field that do not exhibit variability, ruling out a calibration effect. An overlay of SUMSS J223225$-$615308 and the Parkes-MIT-NRAO (PMN) survey data \citep{Wright94PMN} confirms an association with PMNJ2232$-$6153,  whose reported flux density is $40 \pm 7 \unit{mJy}$ at 4.85~GHz. This flux density together with the range of flux densities shown in the 843~MHz light curve, implies a spectral index between $-0.8$ and $-1.0$, although we note that the PMN and MOST observations are not contemporaneous. If the variation is due to scintillation, this implies a compact source or component, and together with the spectral index implies a classification as a compact steep spectrum source.

\subsection{Transient sources with resolved optical counterparts}
\subsubsection{J060938$-$333508}
\begin{figure}
\includegraphics[width=\linewidth]{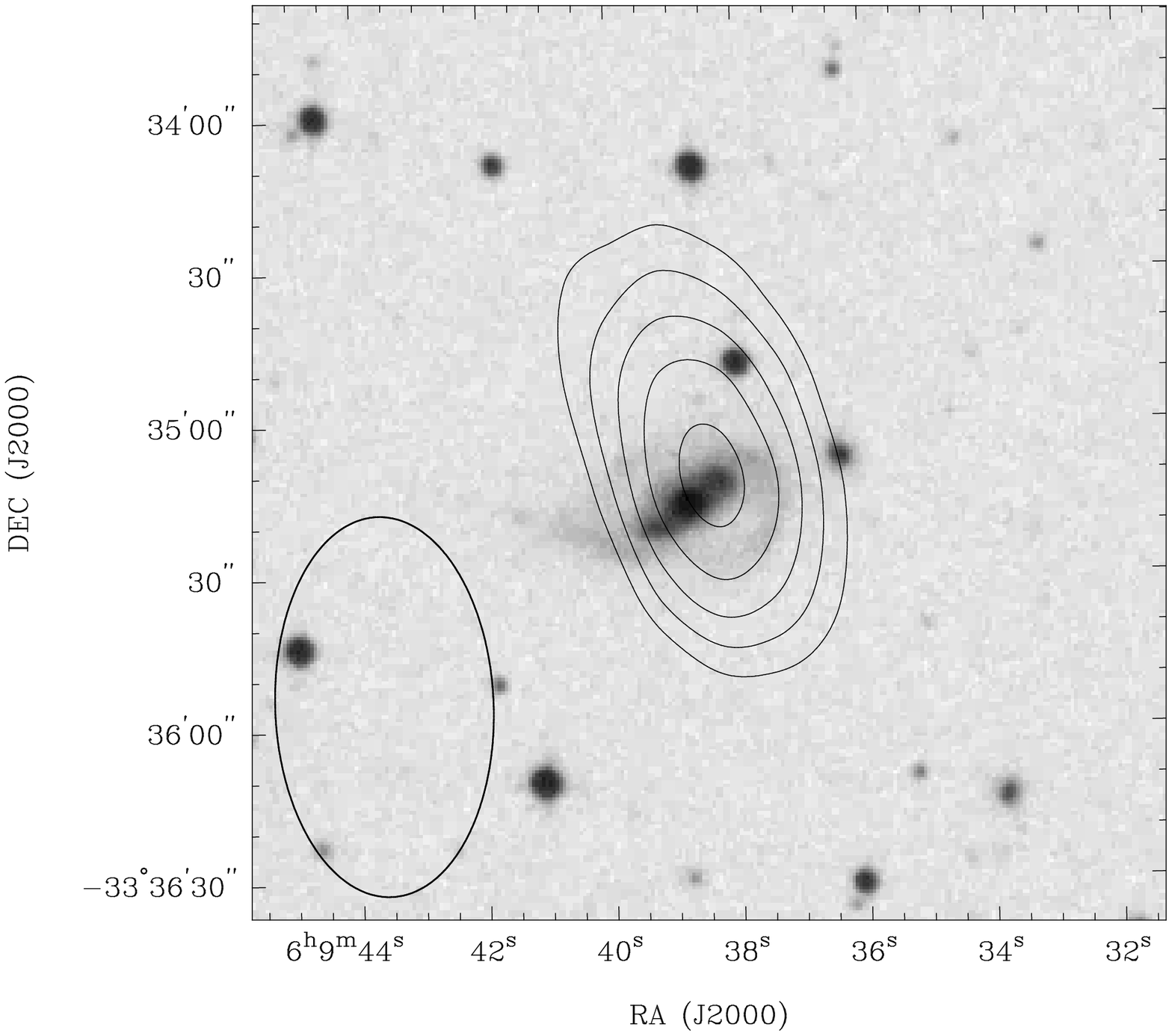}
\includegraphics[width=\linewidth]{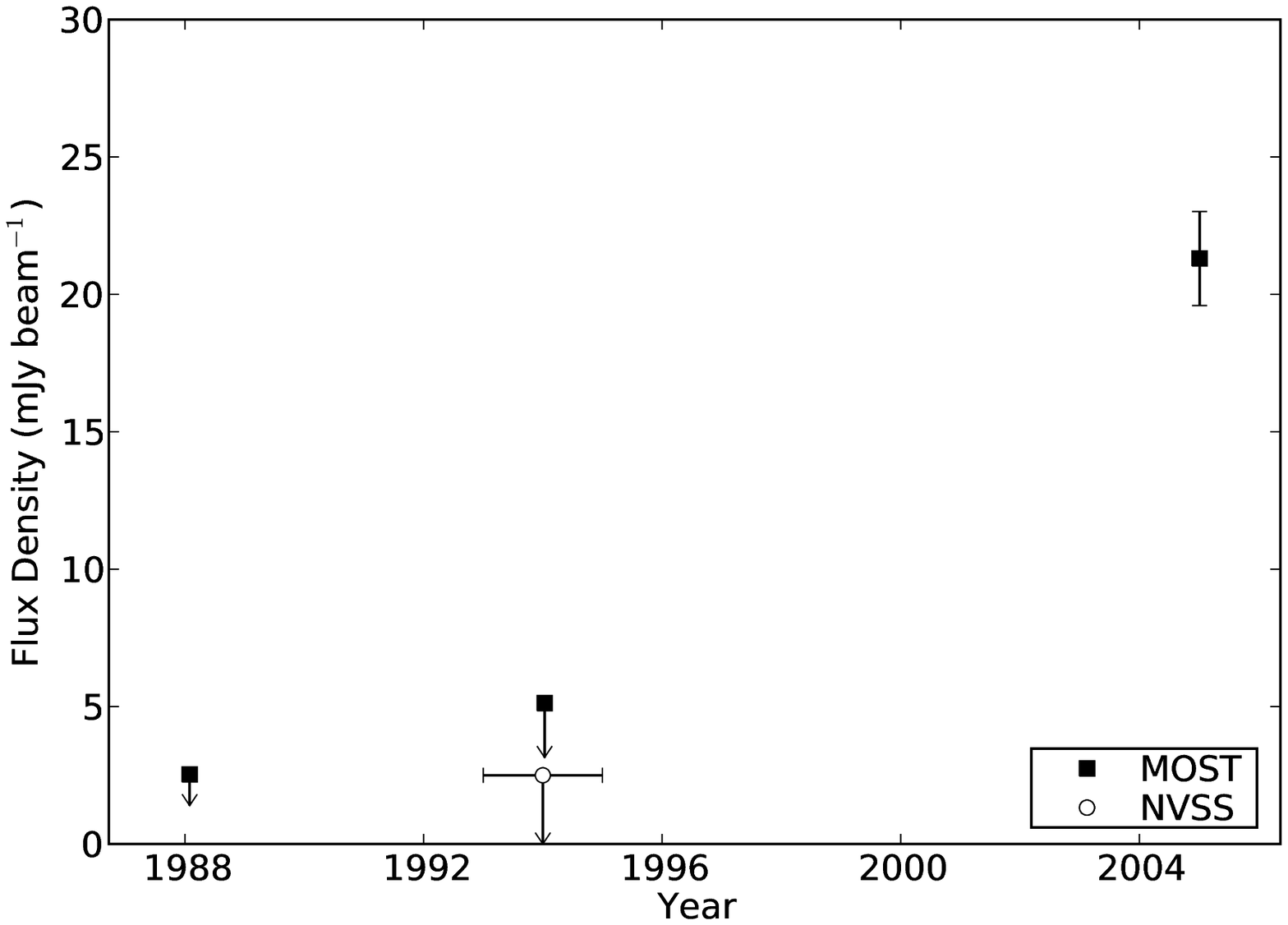}
\caption{Radio transient J060938$-$333508. Top panel: MOST contours from 6 to 18$\unit{mJy~beam^{-1}}$ in steps of 3$\unit{mJy~beam^{-1}}$ for the brightest epoch (2004 December 9) overlaid on a SuperCOSMOS B image. The MOST synthesised beam is drawn at the bottom left. Bottom panel: MOST and NVSS light curves indicating a single detection in 2004 at $21.3 \unit{mJy~beam^{-1}}$ and 3 non-detections. }
\label{fig:J060938-333508}
\end{figure}

The light curve for J060938$-$333508  (Fig.~\ref{fig:J060938-333508}) shows  MOST and NVSS non-detections followed by a single MOST detection. Unlike the vast majority of sources in the MOST archive, the MOST contours appear rotated with respect to the MOST beam possibly indicating a change in flux density over the 12~hr synthesis time. The MOST contours are centred 9 arcsec from the centre of Fairall 1138, a galaxy with spectral type SBab D \citep{Dressler88} and with redshift $z=0.037$ \citep{1998AJ....115..418D}.

Assuming the radio source and galaxy are associated, the inferred isotropic radio luminosity from the brightest epoch (2004 December 9) is $L_{\nu} \simeq 6 \times 10^{29} \unit{erg~s^{-1}~Hz^{-1}}$ at 843~MHz.

The offset from the centre of the optical galaxy, and the fact that spiral galaxies rarely contain an AGN argues against an AGN source for the radio variability. \change{The optical morphology of Fairall 1138 appears quite disturbed and it has a 60 $\mu m$ flux of 345~mJy \citep{Moshir92}, which implies the presence of cold dust. The disturbed morphology, spectral type and presence of dust in Fairall 1138 support the interpretation that it is undergoing star formation, so a stellar explosion, either GRB or RSN are possible explanations.}

The spectral luminosity is very high for a RSN and we are unable to discriminate between Type Ib/c or Type II RSNe by the light curve time-scales, as the time interval between the detection and non-detection epochs is too large.

The spectral luminosity of J060938$-$333508  is within the range of  GRB afterglows.  It is also within the error circle of GRB 940526B, which occurred after the MOST non-detection in 1993, and 10 years before the MOST detection. If J060938$-$333508 is the radio afterglow of GRB 940526B, then the radio detection 10 years after the gamma ray event is unlike known GRB afterglows, which typically peak at 843~MHz a few weeks after the explosion and fade over about 3 years. We consider an association of GRB 940526B and J060938$-$333508 unlikely.

We consider an unusual stellar event in Fairall 1138 as the likely interpretations of this source.

\subsubsection{SUMSS J055712$-$381106}
\begin{figure}
\includegraphics[width=\linewidth]{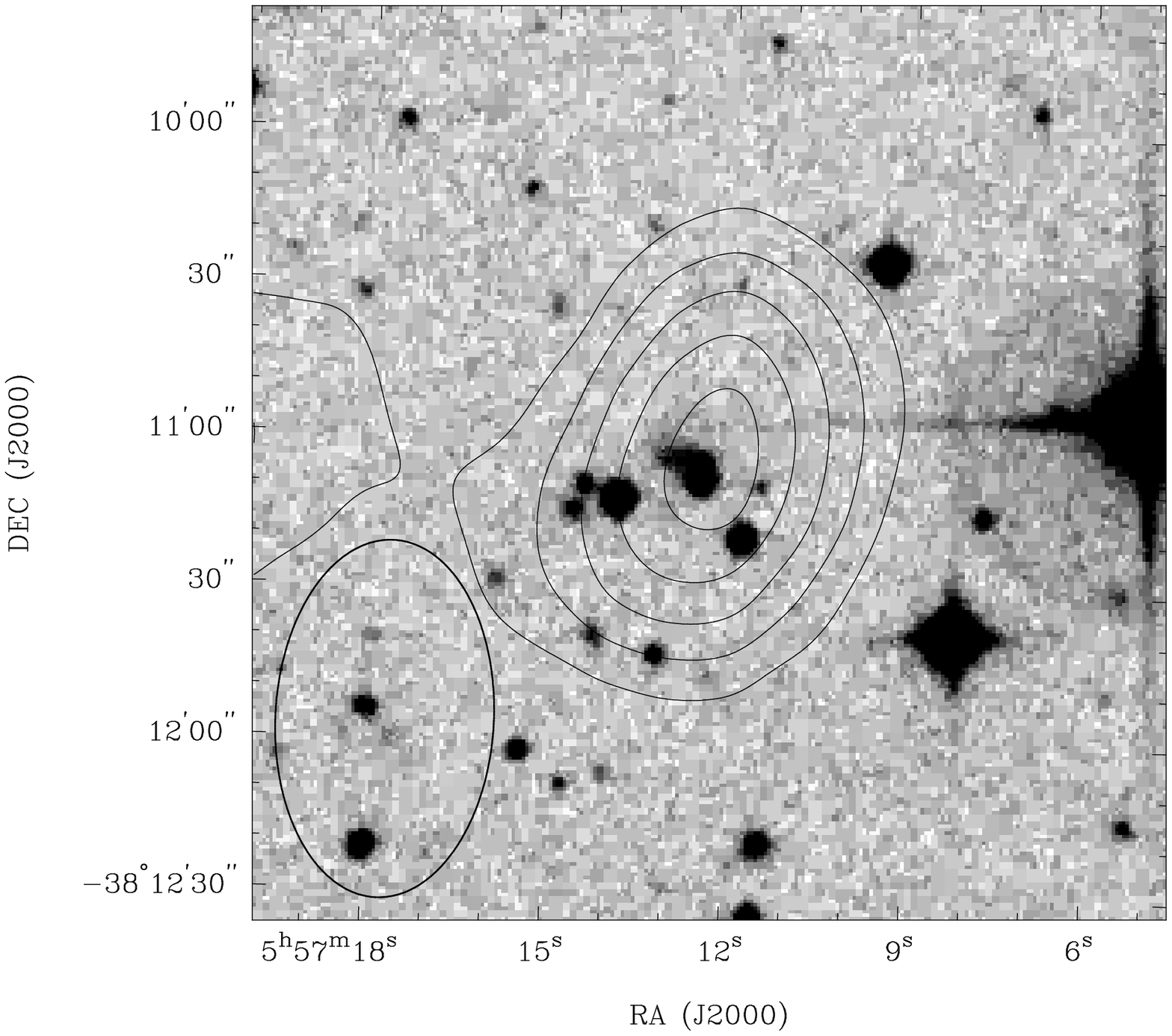}
\includegraphics[width=\linewidth]{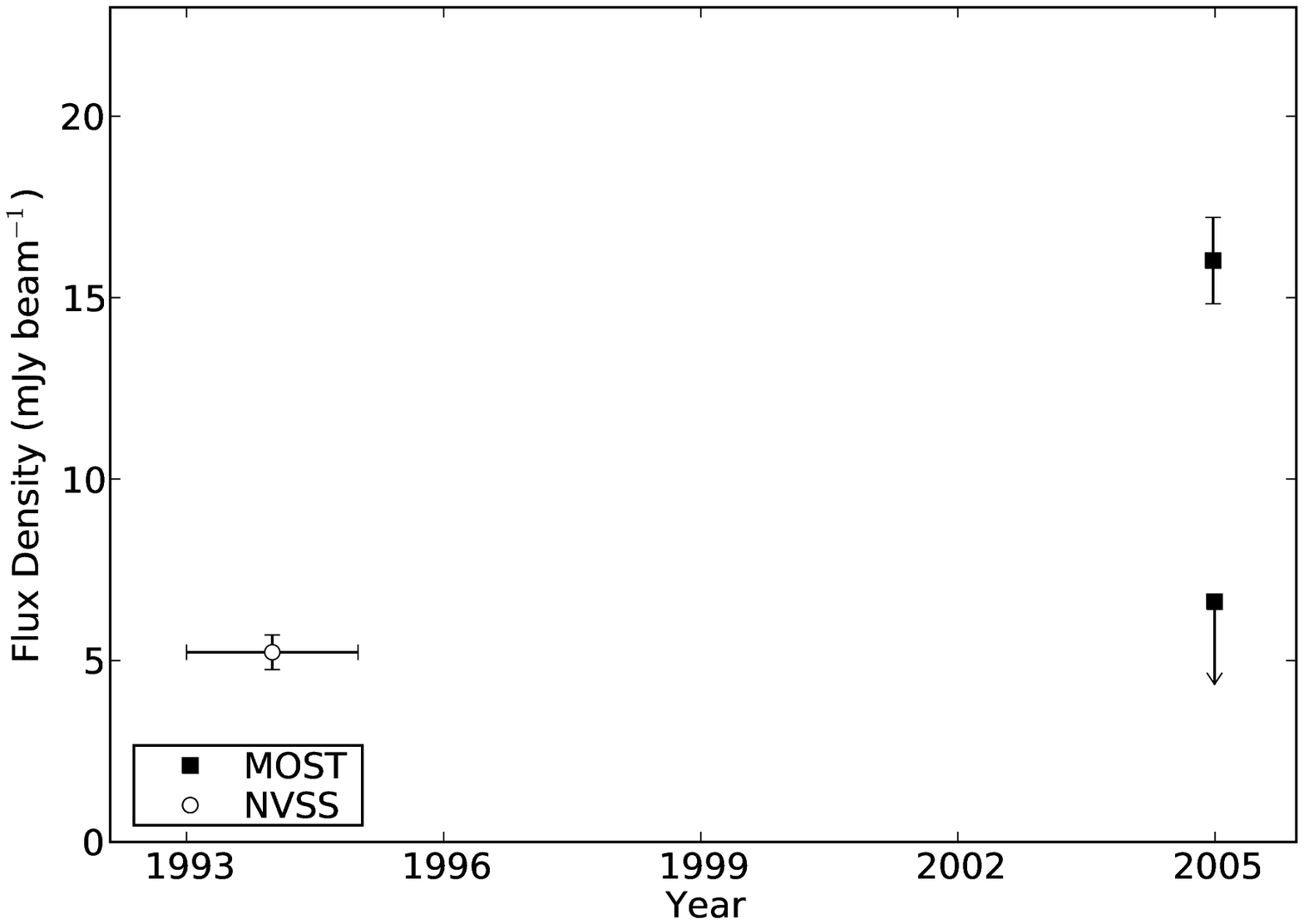}
\caption{Radio transient SUMSS J055712$-$381106. Top panel: MOST contours from 3 to 15$\unit{mJy~beam^{-1}}$ in steps of 3$\unit{mJy~beam^{-1}}$ for the brightest epoch (2004 November 7) overlaid on a SuperCOSMOS B image. Bottom panel: MOST light curve indicating an NVSS and MOST detection, with a non-detection 6 days after the MOST detection.}
\label{fig:J055712-381105}
\end{figure}

The light curve for SUMSS J055712$-$381106  (Fig.~\ref{fig:J055712-381105}) shows an NVSS detection followed by a MOST detection approximately 10 years later and then a non-detection 6 days after that, consistent with either a flaring source or a highly variable source occasionally appearing above our sensitivity limit. The MOST contours appear slightly elongated and rotated with respect to the MOST beam possibly indicating a change in flux density over the 12~hr synthesis time. Flaring or scintillating AGN or flaring radio stars are possible counterparts with these properties.

The SuperCOSMOS B image shows what appears to be a blend of three objects, a star-like object to the south, a faint star-like object immediately to its north, and an extended object to the north-east. The MOST and NVSS radio sources cannot be conclusively associated with any of the three sources in the optical blend\change{which may in fact be a single interacting system. A more accurate position is required for a more definitive association.}

With no clear optical association we cannot positively classify this source, but the time-scales between the NVSS detection and MOST detection and non-detection rule out RSNe and GRB afterglows.

\subsection{Transient sources with point-like optical counterparts}

\subsubsection{SUMSS J062636$-$425807}
The light curve for SUMSS J062636$-$425807 (see Appendix \ref{sec:light_curves_transients}) has a detection and a non-detection separated by almost a year. The optical counterpart is point like and the colours are red, with B-K = 5. The source is most likely an AGN scintillating above our noise threshold.

\subsubsection{J135304$-$363726}
The light curve for J135304$-$363726 (see Appendix \ref{sec:light_curves_transients}) is consistent with a flaring source, or a variable source occasionally appearing above our sensitivity limit. It is marginally detected in the SuperCOSMOS R plates and not detected in B or I, or any 2MASS images. This source is likely an optically faint AGN scintillating above our noise threshold.

\subsubsection{Unidentified variables}
Some 25 per cent of the variable sources have a clear point-like or marginally resolved optical counterpart most of which have red colours, with $0 < B-R < 2$. The optical point source and radio detection would typically imply AGN, but the colour distribution is unusual for AGN, for which the vast majority have $B-R<1$ \citep{Croom04}. The reddest sources are therefore possibly flaring M-dwarfs, but optical spectra are needed to confirm this.

\subsubsection{Unidentified transients}
The spectral luminosity of our transients is $L_{\nu} \simeq 10^{19} (D_{L}/1\unit{kpc})^2 \unit{erg ~s^{-1}Hz^{-1}}$, where $D_{L}$ is the luminosity distance. This spectral luminosity is an order of magnitude more luminous than those found by \citet{bower2007sta} at 5 and 8~GHz, but roughly consistent assuming a spectral index of $\alpha=-0.7$. The sampling interval of our light curves is typically much longer than those discussed by \citet{bower2007sta} (typically years as opposed to weeks) and we have not yet obtained dedicated optical follow-up of our sources, so our limits on optical counterparts are not very strong. The scarcity of data means we are unable to robustly classify a large fraction of our transient sources.

Of the unidentified transients, none have X-Ray counterparts in the \emph{ROSAT} X-ray All-Sky Survey Bright Source Catalogue \citep{ROSAT}. GRB afterglows without a gamma ray trigger (the so-called orphan GRB afterglows) are a possibility and might explain the higher average luminosity. Radio supernovae are also a possibility as long as the hosts are nearby and intrinsically faint, but distant RSN would appear to be ruled out by the high luminosity. Stellar events are less likely, but flares from nearby late type stars are not ruled out. Soft gamma repeaters and X-Ray binaries are also possibilities, although the rarity of such objects and our measured transient rate are incompatible.

The most likely source of variability from extragalactic sources at 843~MHz is refractive Interstellar Scintillation (ISS) of the compact components of AGN. By our definition, our transients could well be such  scintillating sources occasionally appearing above our flux threshold, although the light curves of the transient sources have $\mathcal{V} > 0.6$, which would make them the among most extremely variable sources in our sample (c.f.  Figure \ref{fig:v_histogram}). Star-forming galaxies, whose radio flux is from widespread star formation and diffuse synchrotron emission in the disk, are unlikely to be intrinsically variable on our time-scales (due to light travel time constraints), or extrinsically variable (due to larger angular size quenching the interstellar scintillation).

Another explanation for the transient sources is time-integrated emission from pulsars with on-times of greater than $\sim~12\unit{h}$. \citet{Ofek10} proposed pulsars as the source of the transients discovered by \citet{bower2007sta} and \citet{Niinuma07} based on population arguments, and there is some observational evidence for this. Intermittent pulsars such as PSR B1931$+$24 \citep{Kramer06}, have active periods of a few days and then turn completely off.  PSR J0941$-$39, which was discovered as a Rotating Radio Transient (RRAT) with 5 single pulses in one observation, exhibited standard pulsar emission during another observation \citep{BurkeSpolaor10}. Such objects might appear as transients in our survey if the time between mode-changes had a time-scale of 12~h or longer, and the time-integrated flux densities in both modes were compatible with our non-detection thresholds. Pulsars are also not likely to be optically detected at the SuperCOSMOS plate limits, which would explain the \change{lack of optical detections of some our transient sources.}

Finally, microlensing is an unlikely explanation for our unknown transients, as the implied number of sources to explain our sample is far too high \citep{bower2007sta}. Reflected solar flares are unlikely, as a 1~MJy solar flare would need to reflect of a 1000~km object at a distance of only 7000~km to be detected above our threshold.

\begin{landscape}
\begin{table}
\caption{Summary and classifications of the 15 radio sources classified as transient. }
\label{tab:transients_comments}
\centering
\begin{tabular}{lp{5cm}p{5cm}p{8cm}}
\hline
Source & Radio Lightcurve & Counterparts at other wavelengths & Classification \\
\hline
SN1987A &  Linear increase &  Bright point source in SuperCOSMOS R and I bands. & Supernova.\\

SUMSS J055712$-$381106 & Detection followed by a non-detection 6 days later. & Blend of three objects in SuperCOSMOS and 2MASS. & Unknown \\

J060938$-$333507 & Single detection. Non-zero PA possibly indicates variation during 12h synthesis. & 9~arcsec from the optical centre of Fairall 1138, a barred spiral galaxy at $z=0.037$. & Luminosity $L_{\nu} \simeq 6 \times 10^{29}\unit{erg~s^{-1}~Hz^{-1}}$. Possible ultra-bright RSN or GRB afterglow.\\

J061051$-$342404 &  Non-detection between two detections. & No known counterparts. & Unknown \\

SUMSS J062636$-$425807 & Single detection. & SuperCOSMOS and 2MASS colours typical of a quasar. & Probable scintillating AGN.\\

J062716$-$371736 & Non-detection between two detections. & No known counterparts. & Unknown \\

J064149$-$371706 & Single detection.  & No known counterparts. & Unknown\\

SUMSS J102641$-$333615 & Single detection. & No known counterparts. & Unknown \\

SUMSS J112610$-$330216 &  Detection and non-detection separated by 1 day. & No known counterparts. & Less likely to be an AGN or explosion. Possible flare star other short term variable. \\

Nova Muscae 1991 & Single detection. & Bright point source visible only in SuperCOSMOS R image. Crowded field. Point source in RASS. & Low mass X-ray binary.\\

J121032$-$381439 & Non-detection between two detections. & No known counterparts. & Unknown\\

J135304$-$363726 &  Non-detection between two detections & Faint ($M_{R1} = 20.4$ $M_{R2} = 20.7$) point source in SuperCOSMOS. & Probable scintillating AGN but faint, red detection could be a star.\\

J153613$-$332915 &  Non-detection between two detections. Non-zero PA possibly indicates variation during 12h synthesis. & No known counterparts. & Unknown \\

GRO 1655-40 & Power law decrease over 3 months, then non-detection after 11 years. & Point source in all SuperCOSMOS plates.  & GRO J1655-40. A low mass X-ray binary and microquasar.\\

SUMSS J224152$-$300823 &  Single detection & 6~arcsec from a blue point-like optical source. Possible association. & Probable scintillating AGN if optical association is correct, otherwise unknown.\\

\hline
\end{tabular}
\end{table}
\end{landscape}

\section{Conclusions}
We have conducted a 22 year survey for radio variability at 843~MHz and measured 29230 radio light curves. We have discovered 15 candidate transient sources and 53 candidate highly variable sources. Only 3 of the transients were previously known.

We conclude that many variable sources can be explained as scintillating AGN, and some that are associated with nearby galaxies may be over-luminous radio supernovae with atypical light curves. We conclude that at least 3 of the transients are unlike any known source and could belong to the class of radio transients without optical counterpart discovered by \citet{bower2007sta}, which have a number of possible explanations including giant M-dwarf flares, or flaring Galactic neutron stars.

We have also presented a number of statistical techniques to aid in future radio variability surveys. In particular, we have described techniques to remove systematic gain errors, to confirm errors in flux density  are correctly estimated and to choose a robust variability threshold.

Upcoming wide-field radio variability surveys will offer a new insight into this parameter space. We have shown that the radio sky is indeed variable so these surveys are likely to uncover many new sources. We have also confirmed that there is at least one class of transient radio source without optical counterparts. The time interval between radio activity and optical measurement in our cases was several years, highlighting the need for prompt multi-wavelength follow-up.

\section{Acknowledgements}
We would like to thank the past and present staff at the MOST for making these observations possible, in particular Duncan Campbell-Wilson and Barbara Piestrzynska. KB thanks Elaine Sadler for her useful comments on optical counterparts \change{and Greg Madsen for his help with IDL. We thank the anonymous referee for their extremely helpful comments.}

This research has made use of several facilities and tools including:

\begin{enumerate}

\item  The MOST, which is operated by the School of Physics, University of Sydney with support from the Australian Research Council.

\item The NASA/IPAC Extragalactic Database (NED) which is operated by the Jet Propulsion Laboratory, California Institute of Technology, under contract with the National Aeronautics and Space Administration.

\item The SIMBAD database, operated at CDS, Strasbourg, France.

\item SAOImage DS9, developed by Smithsonian Astrophysical Observatory.

\item The Aladin sky atlas \citep{Bonnarel00}.

\item Data obtained through the High Energy Astrophysics Science Archive Research Center Online Service, provided by the NASA/Goddard Space Flight Center.

\item  Data obtained from the Leicester Database and Archive Service at the Department of Physics and Astronomy, Leicester University, UK

\item The Q3C spatial indexing extension for Postgresql by \citet{q3c}

\item The Cosmology Calculator \citep{Wright06}.

\end{enumerate}

KB acknowledges the support of an Australian Postgraduate Award. This research has been supported by the Australian Research Council through grant DP0987072.

\bibliographystyle{scemnras}
\bibliography{Master.bib}

\appendix
\section{Light curves of variable sources}
\label{sec:light_curves_variables}
\begin{figure*}
\centering
\includegraphics[width=\linewidth]{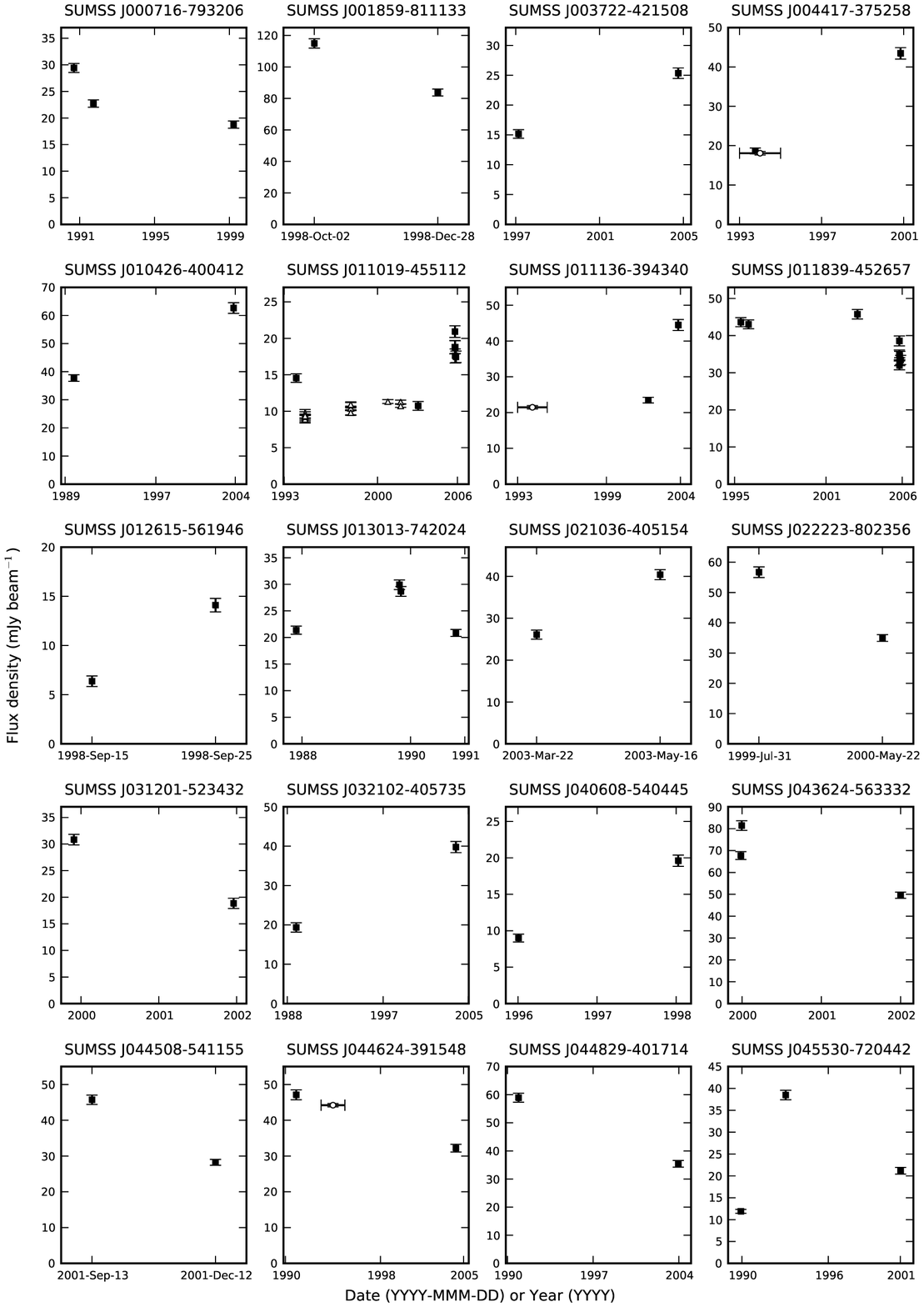}
\caption{Light curves of variable sources}
\label{fig:variables0}
\end{figure*}

This appendix contains radio light curves of all variable sources shown in Table~\ref{tab:variables}, in RA order (Fig. \ref{fig:variables0}). 843~MHz flux density measurements from MOST are shown as black \change{squares}. Flux density measurements or limits from NVSS, where available, are shown as \change{open circles with $x$} errors indicating the observation duration of the NVSS survey. For SUMSS J011019$-$455112, flux density measurements at 1.4~GHz are shown as open triangles. For SUMSS J223225$-$615308 the Parkes-MIT-NRAO (PMN) flux density measured at 4.85~GHz \citep{Wright94PMN} is shown as an open triangle.

\label{app:var_light_curves}

\begin{figure*}
\centering
\includegraphics[width=\linewidth]{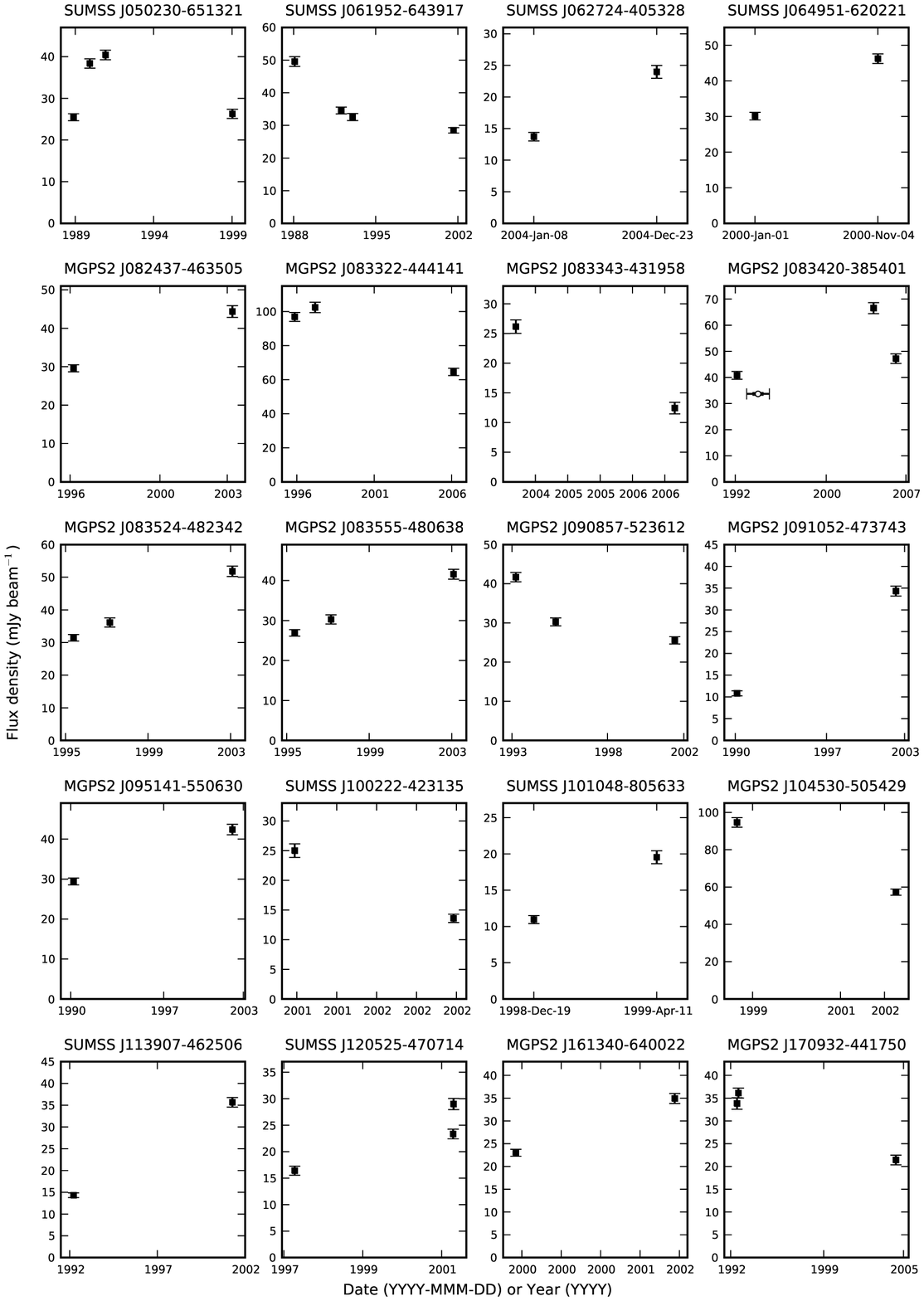}
{Figure \ref{fig:variables0} (continued)}
\end{figure*}

\begin{figure*}
\centering
\includegraphics[width=\linewidth]{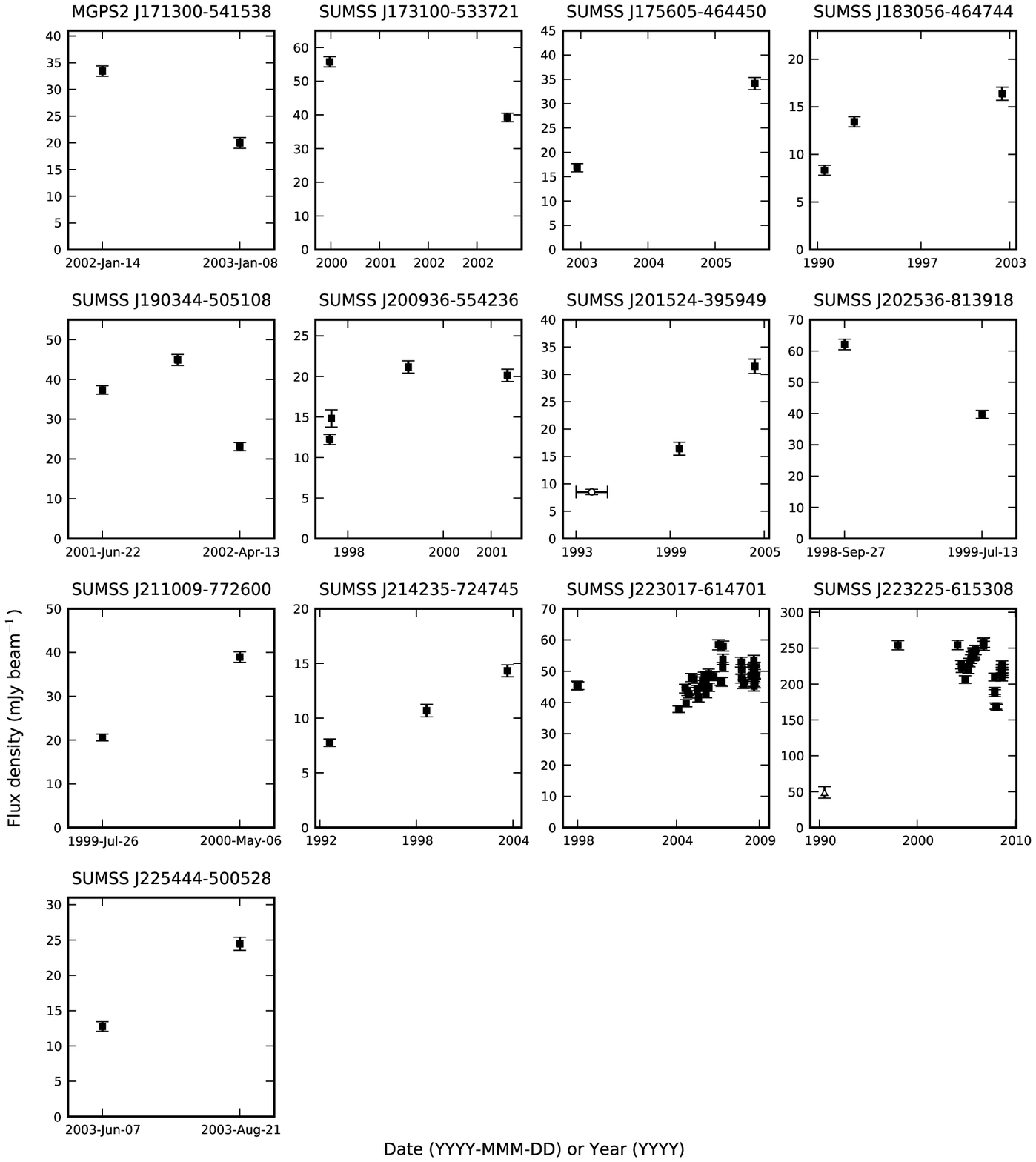}
{Figure \ref{fig:variables0} (continued)}
\label{fig:variables2}
\end{figure*}

\section{Transient sources}
\label{sec:light_curves_transients}
\begin{figure*}
\begin{minipage}{0.3\textwidth}
\centering
\includegraphics[width=1\textwidth]{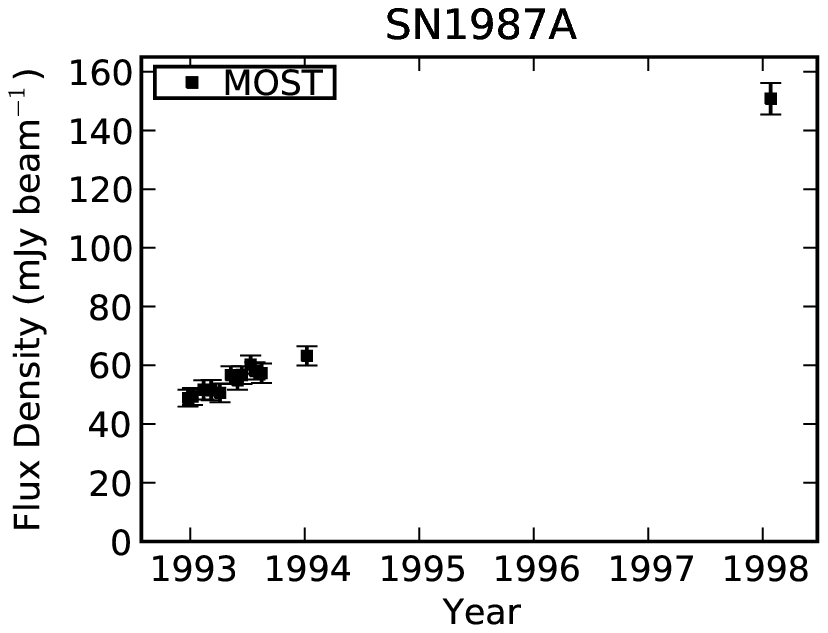}
\end{minipage}
\qquad
\begin{minipage}{0.3\textwidth}
\centering
\includegraphics[width=1\textwidth]{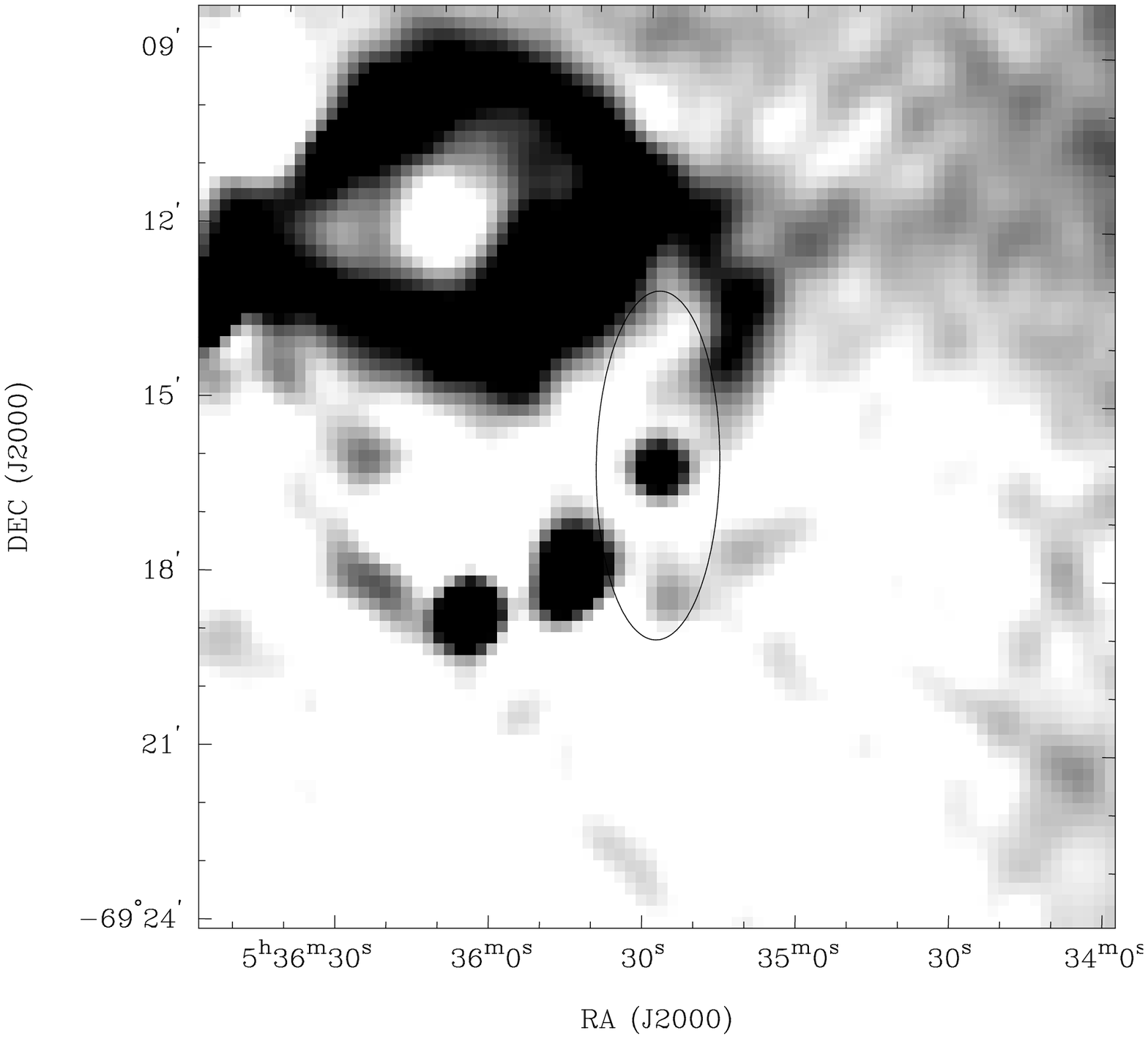}
\centering
\end{minipage}
\qquad
\begin{minipage}{0.3\textwidth}
\includegraphics[width=1\textwidth]{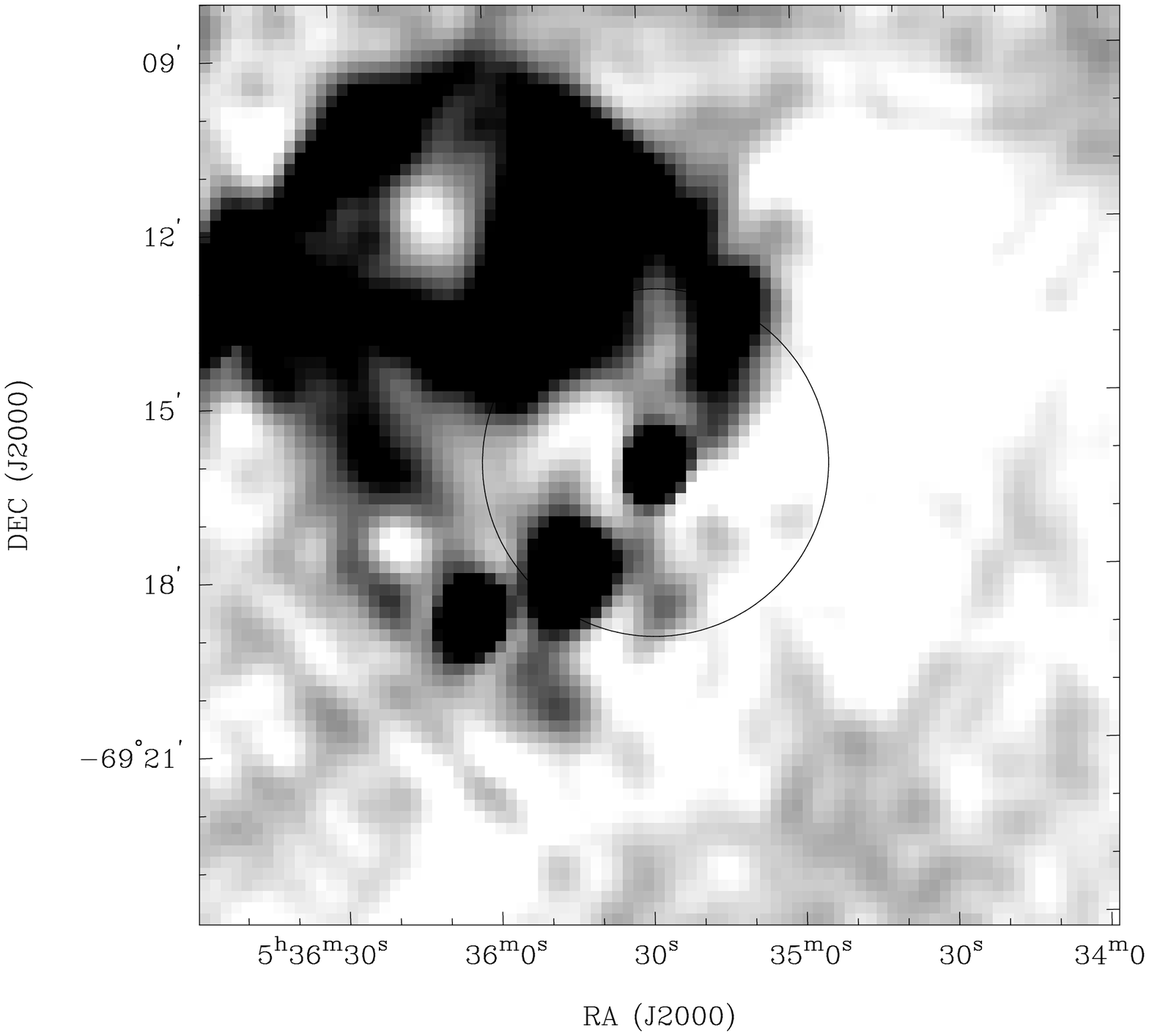}
\centering
\end{minipage}
\qquad
\begin{minipage}{0.3\textwidth}
\centering
\includegraphics[width=1\textwidth]{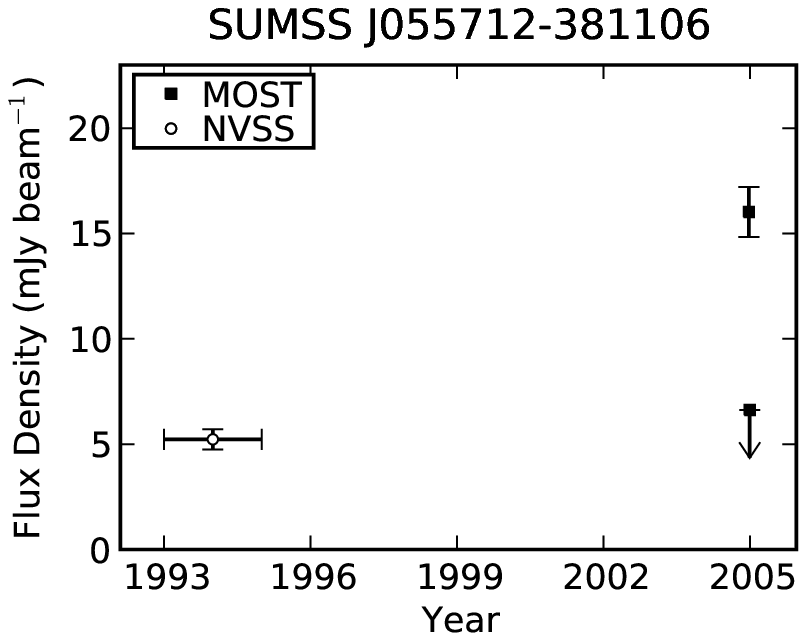}
\end{minipage}
\qquad
\begin{minipage}{0.3\textwidth}
\centering
\includegraphics[width=1\textwidth]{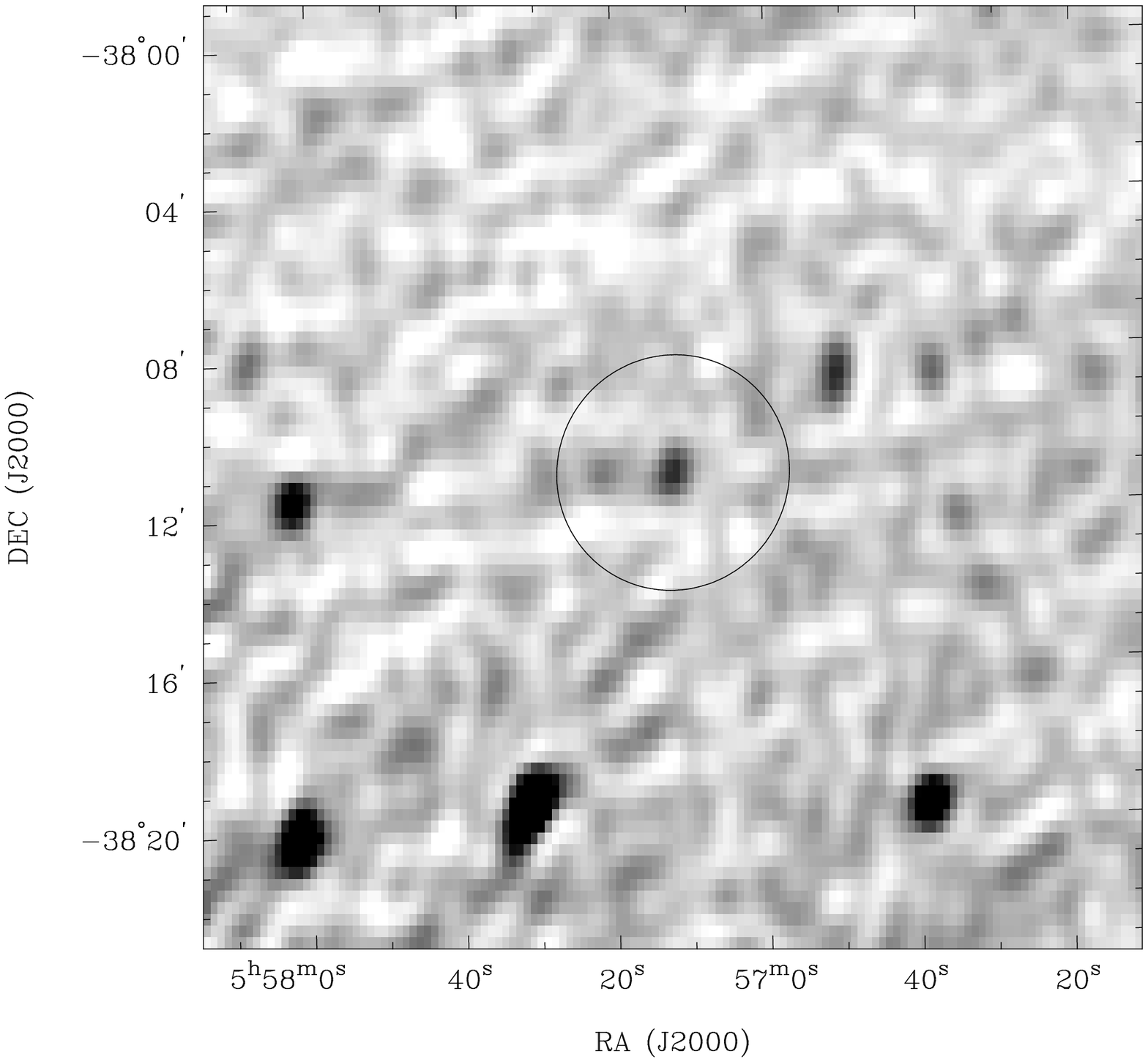}
\centering
\end{minipage}
\qquad
\begin{minipage}{0.3\textwidth}
\includegraphics[width=1\textwidth]{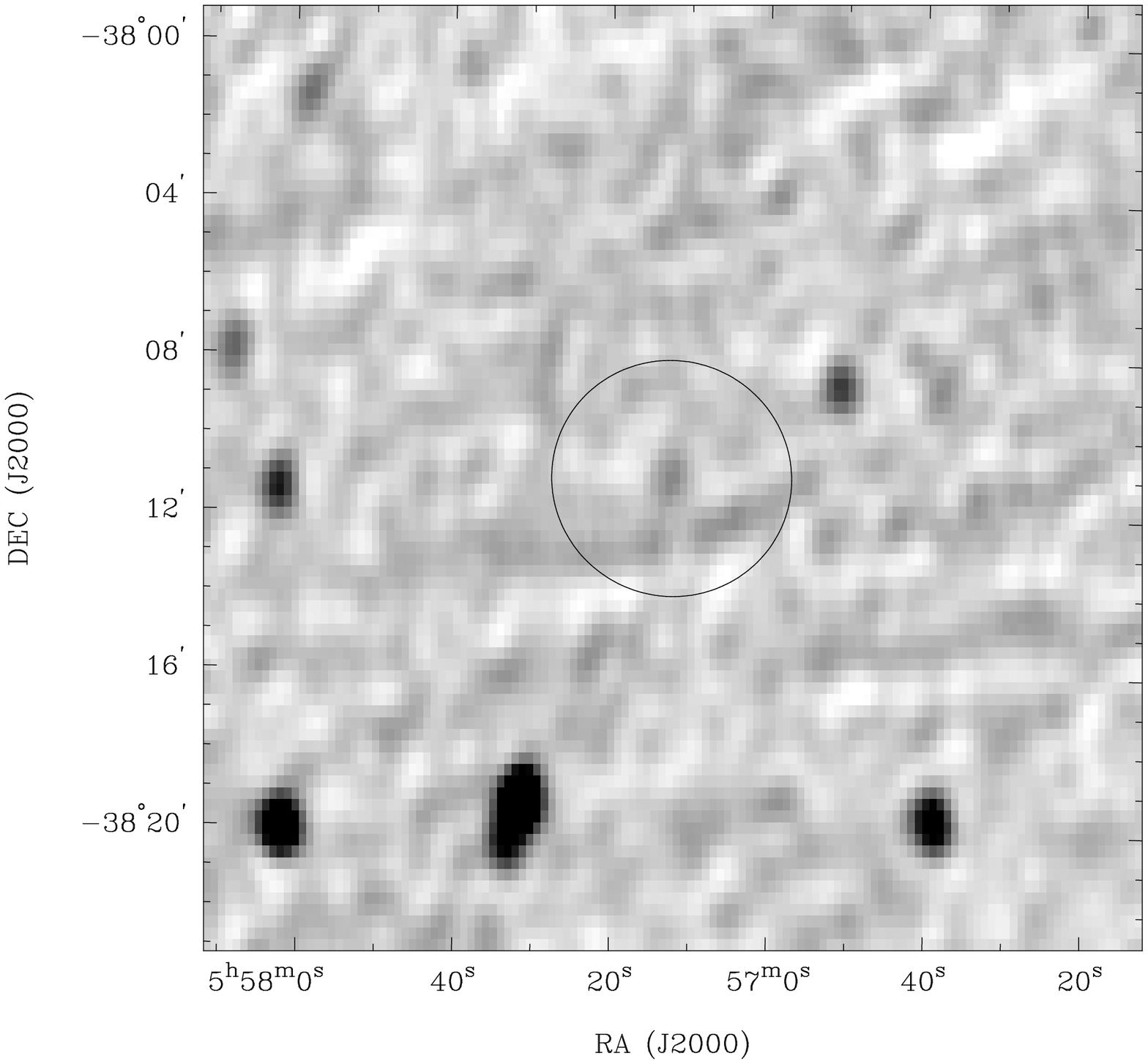}
\centering
\end{minipage}
\qquad
\begin{minipage}{0.3\textwidth}
\centering
\includegraphics[width=1\textwidth]{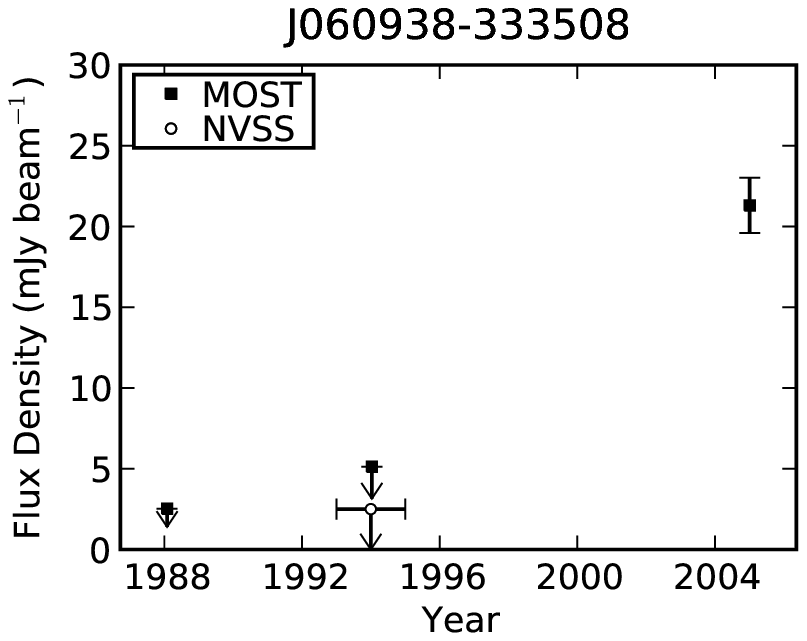}
\end{minipage}
\qquad
\begin{minipage}{0.3\textwidth}
\centering
\includegraphics[width=1\textwidth]{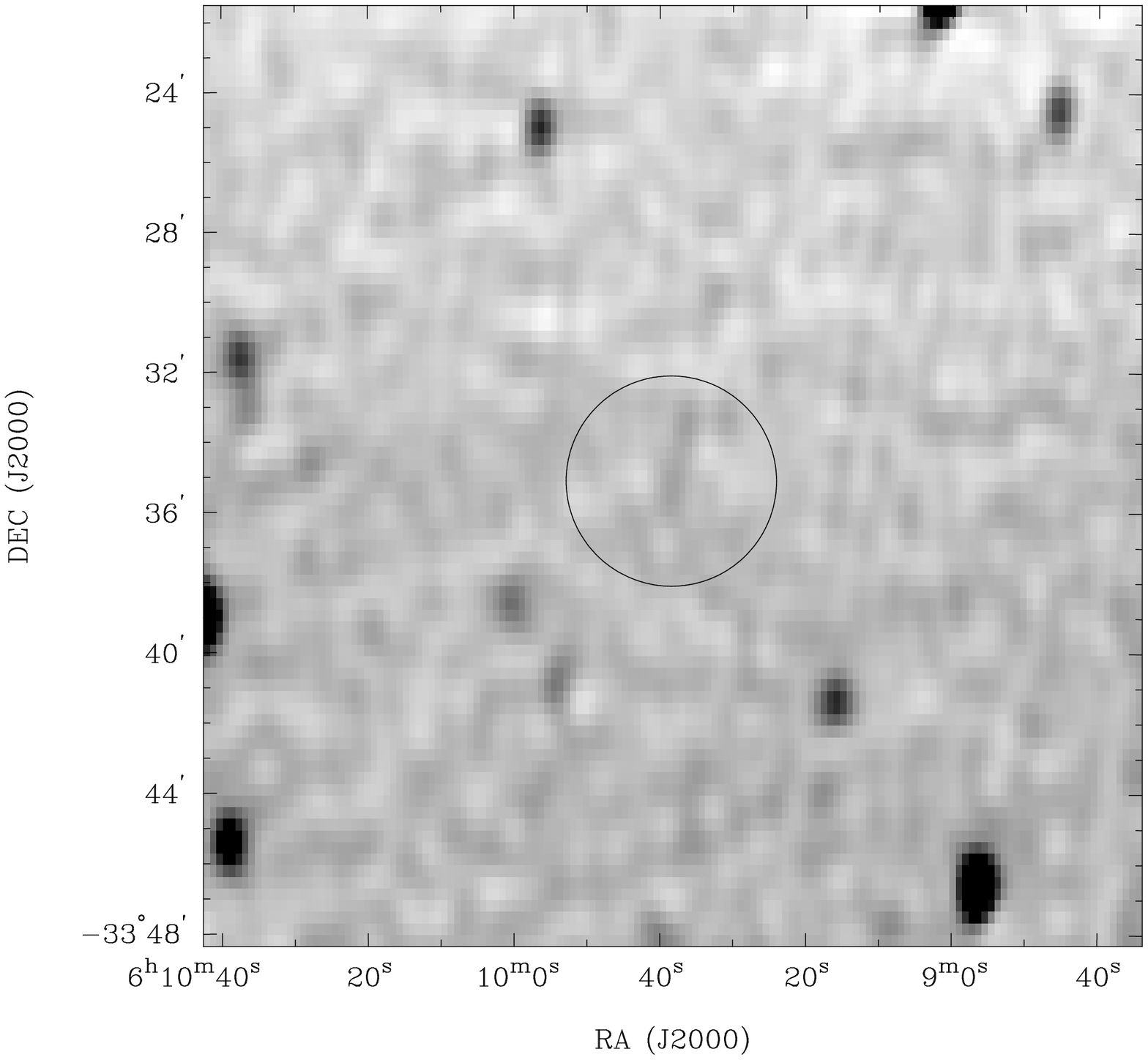}
\centering
\end{minipage}
\qquad
\begin{minipage}{0.3\textwidth}
\includegraphics[width=1\textwidth]{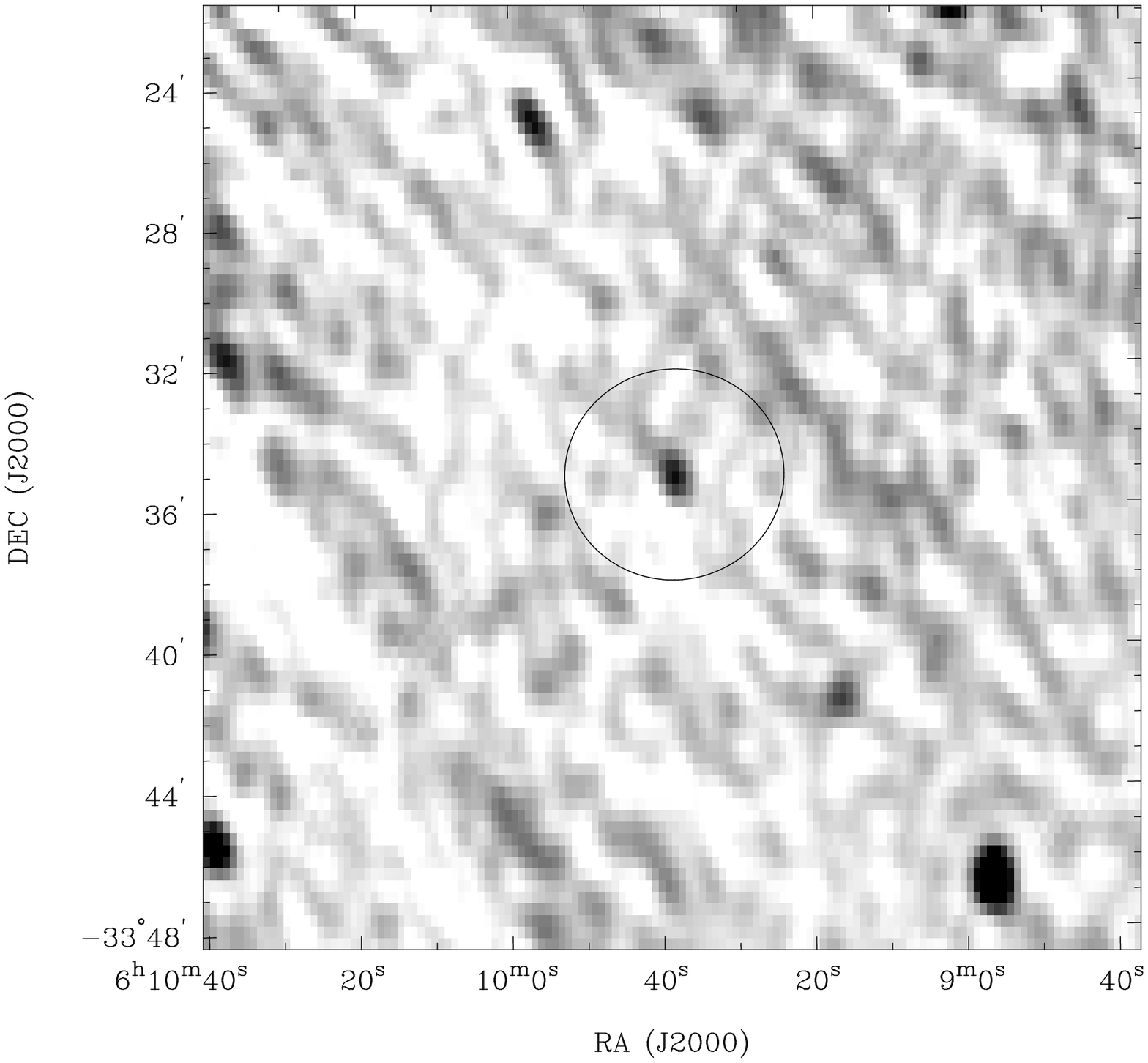}
\centering
\end{minipage}
\qquad
\begin{minipage}{0.3\textwidth}
\centering
\includegraphics[width=1\textwidth]{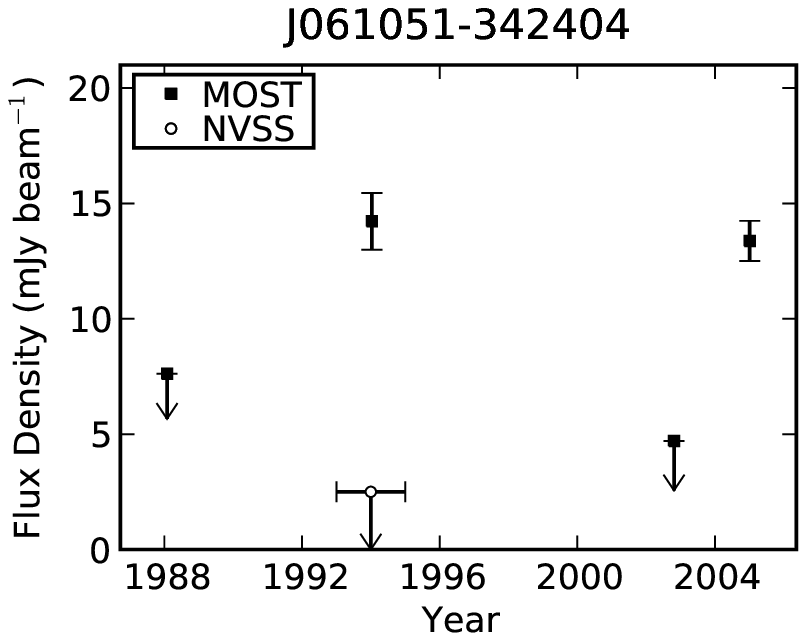}
\end{minipage}
\qquad
\begin{minipage}{0.3\textwidth}
\centering
\includegraphics[width=1\textwidth]{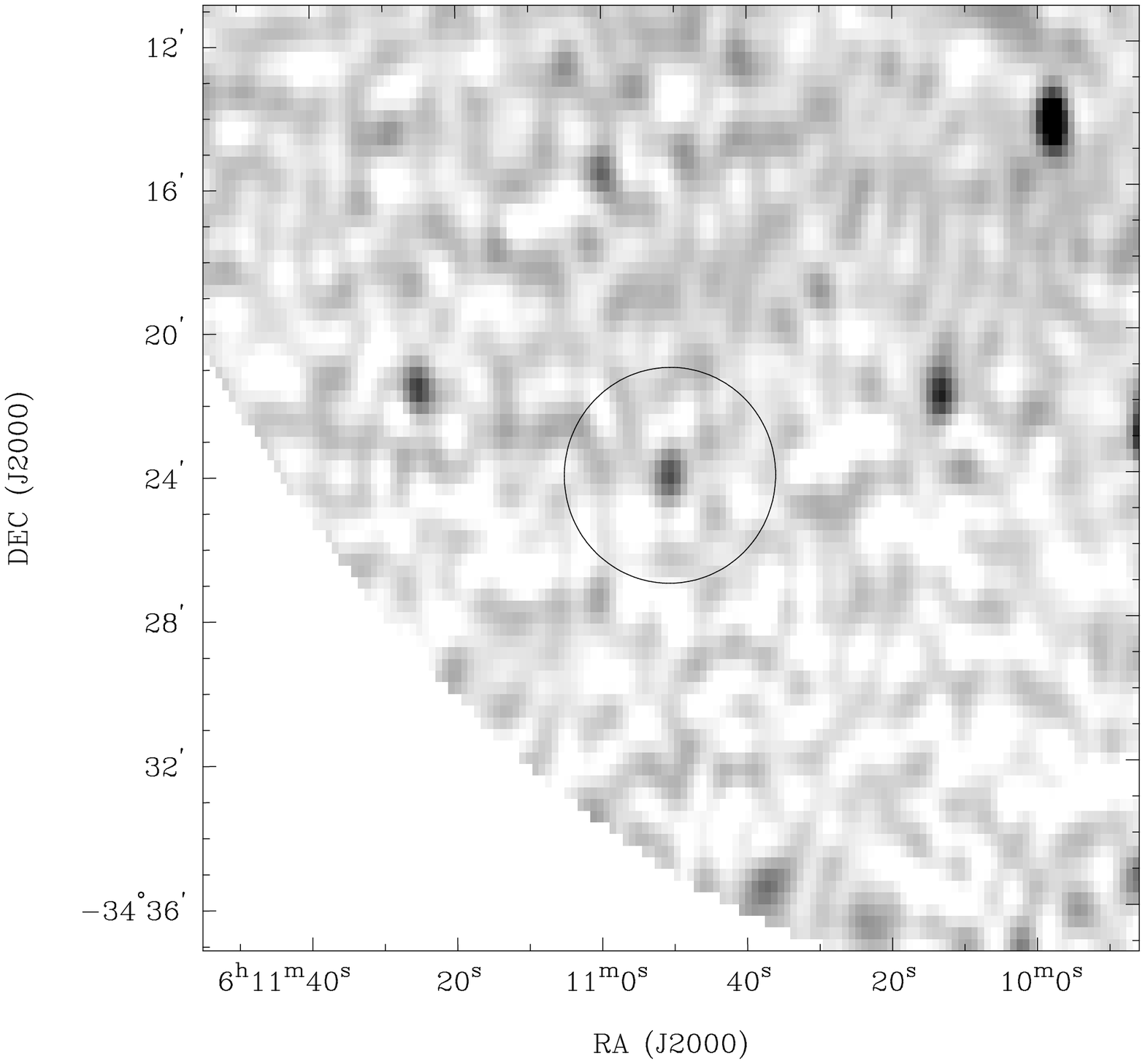}
\centering
\end{minipage}
\qquad
\begin{minipage}{0.3\textwidth}
\includegraphics[width=1\textwidth]{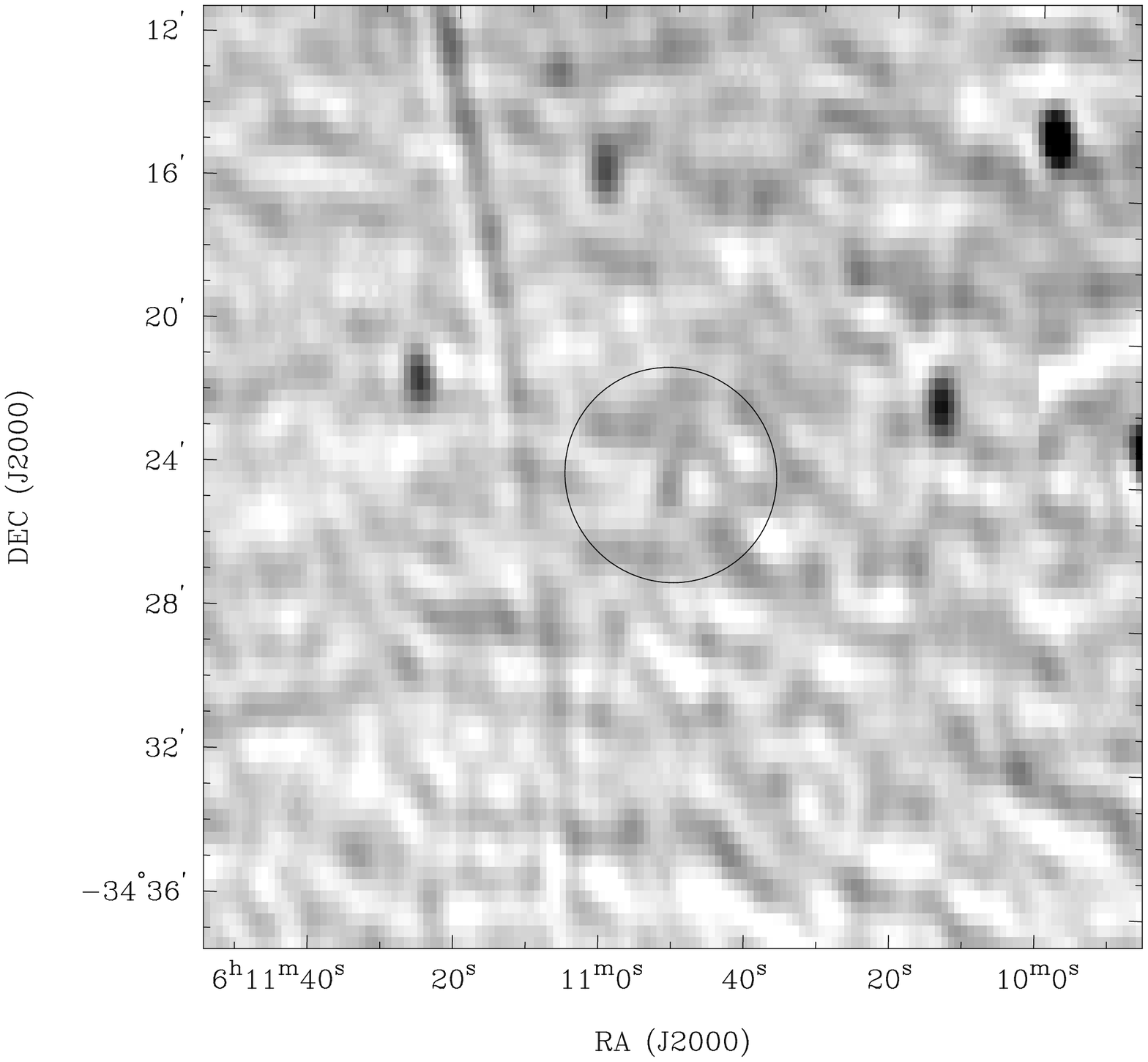}
\centering
\end{minipage}
\qquad
\caption{Transient sources.  Left panel: radio light curve. The centre and right panels show the images of the source at the minimum and maximum detection significance, in time order from left to right, with the position of the source encircled}.
\label{fig:transients_stamps}

\end{figure*}

This appendix contains radio light curves of all variable sources shown in Table~\ref{tab:transients}, in RA order (Fig. \ref{fig:transients_stamps}). Each source is shown on a single row. The leftmost panel is the radio light curve.  843~MHz flux density measurements from MOST are shown as black \change{squares}. Flux density measurements or limits from NVSS, where available, are shown as \change{open circles with $x$} errors indicating the observation duration of the NVSS survey. The centre and right panels show the images of the source at the minimum and maximum detection significance, in time order from left to right, with the position of the source encircled. The grey scale of all images is linear from $-5$ to $20\unit{mJy~beam^{-1}}$.

\begin{figure*}
\begin{minipage}{0.3\textwidth}
\centering
\includegraphics[width=1\textwidth]{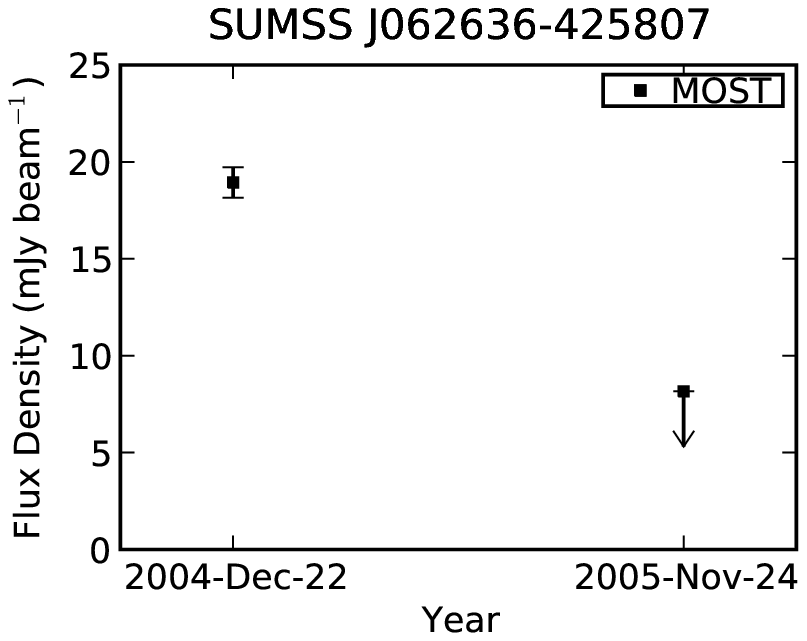}
\end{minipage}
\qquad
\begin{minipage}{0.3\textwidth}
\centering
\includegraphics[width=1\textwidth]{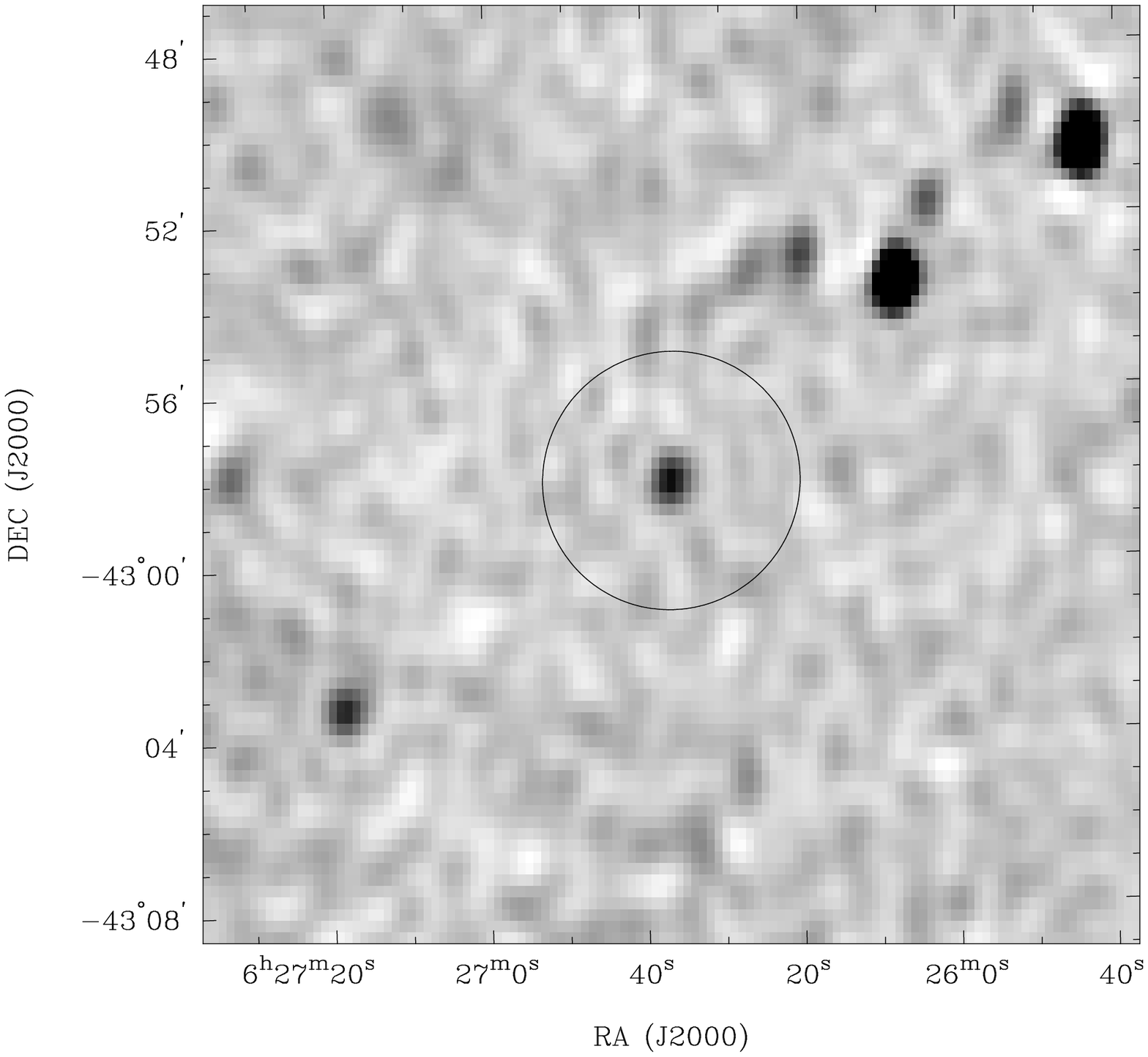}
\centering
\end{minipage}
\qquad
\begin{minipage}{0.3\textwidth}
\includegraphics[width=1\textwidth]{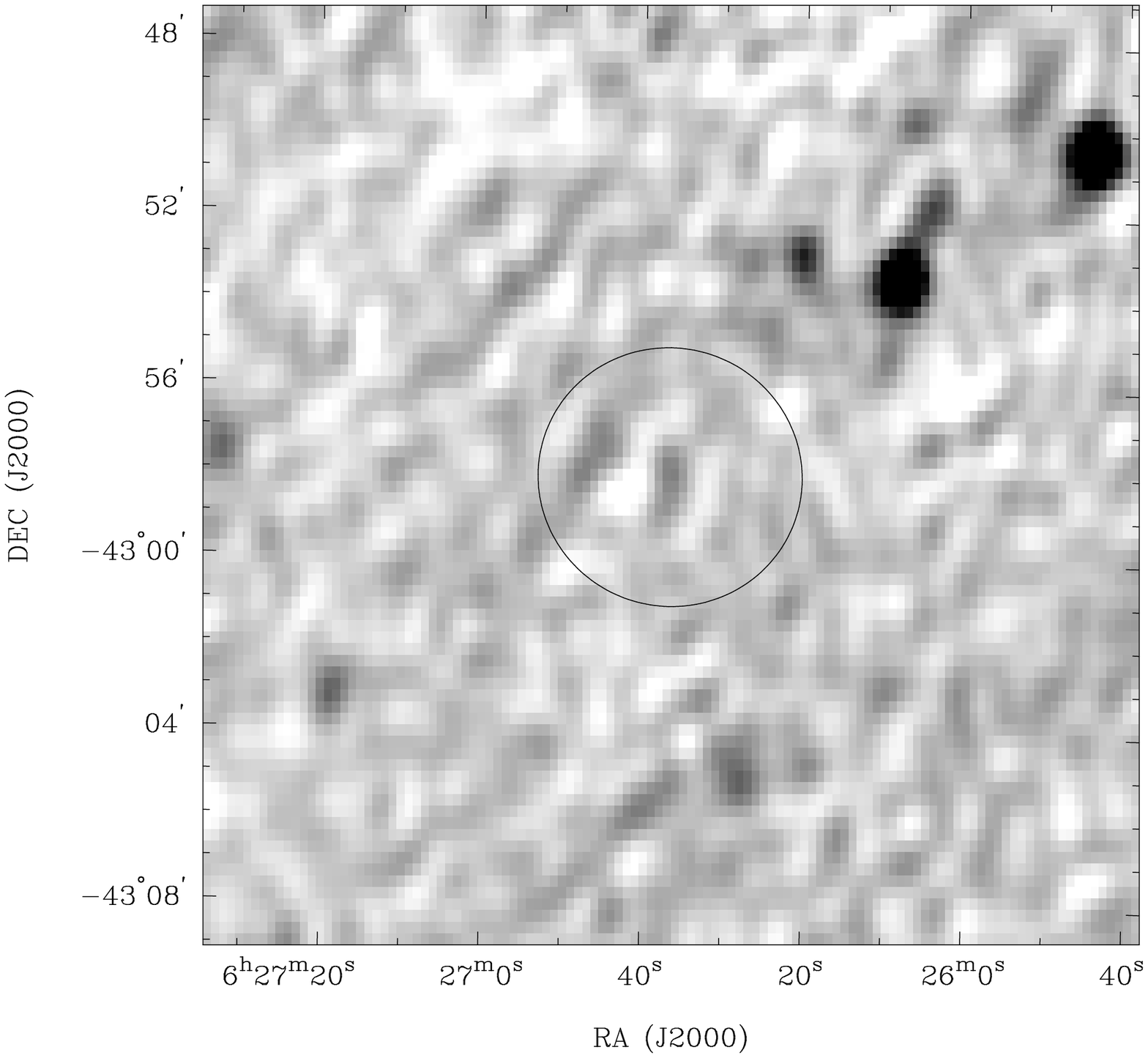}
\centering
\end{minipage}
\qquad
\begin{minipage}{0.3\textwidth}
\centering
\includegraphics[width=1\textwidth]{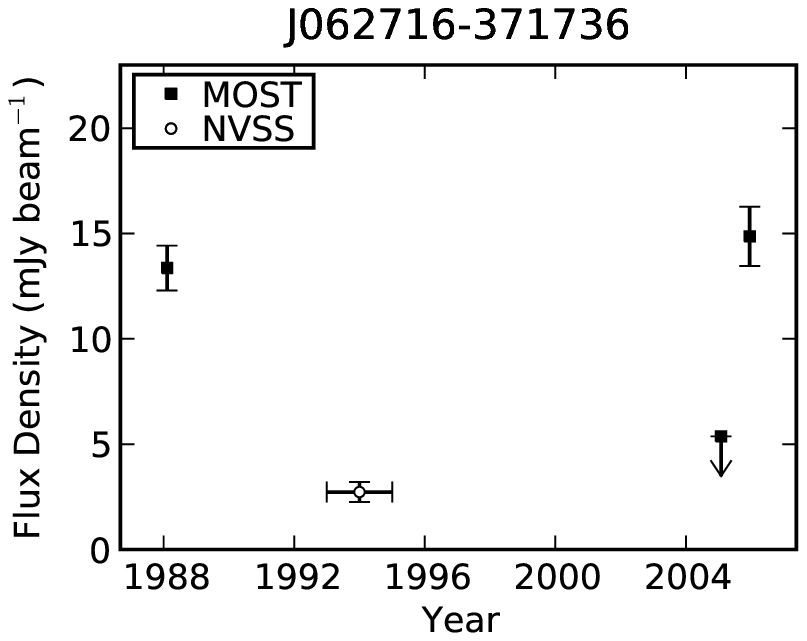}
\end{minipage}
\qquad
\begin{minipage}{0.3\textwidth}
\centering
\includegraphics[width=1\textwidth]{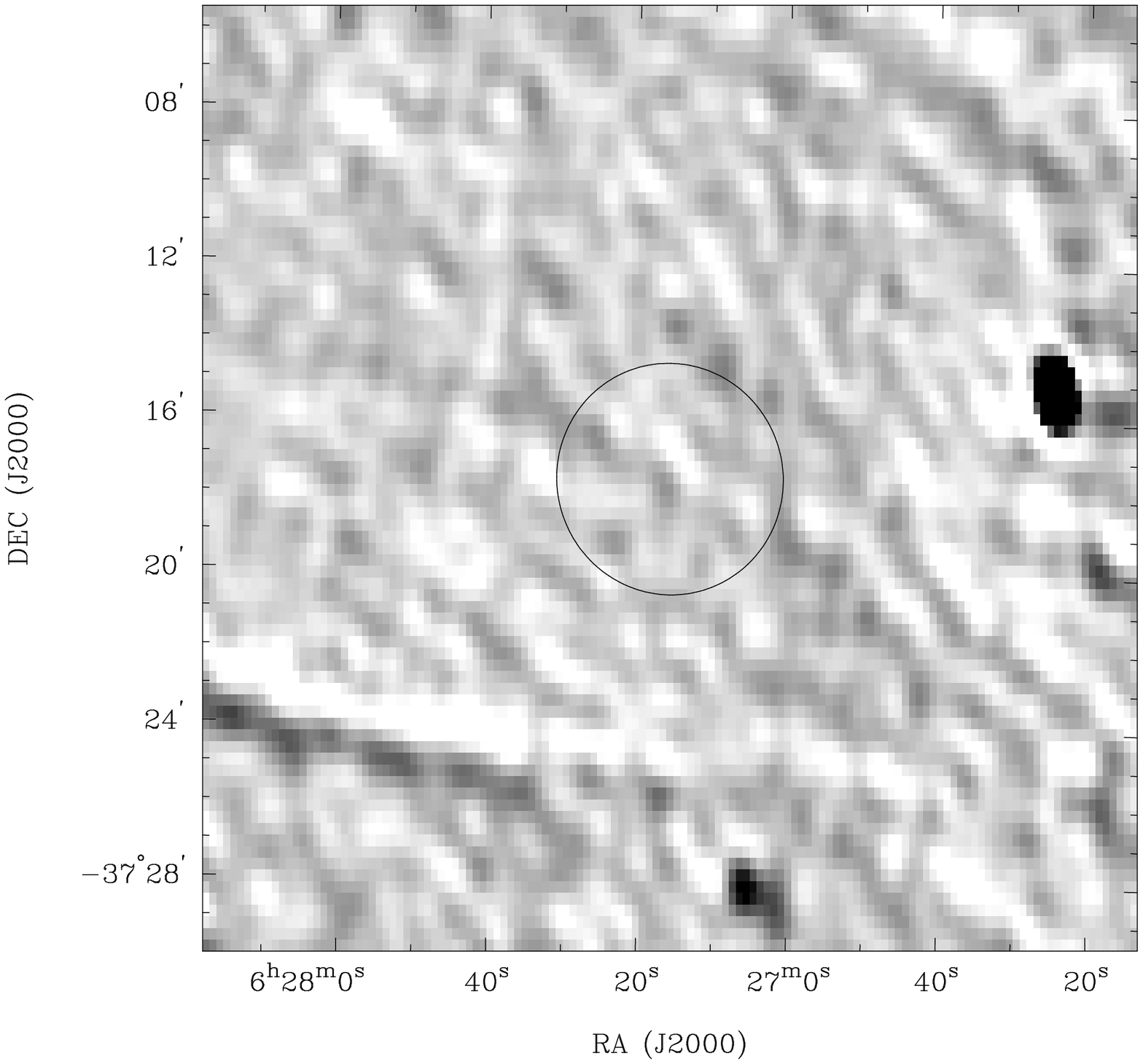}
\centering
\end{minipage}
\qquad
\begin{minipage}{0.3\textwidth}
\includegraphics[width=1\textwidth]{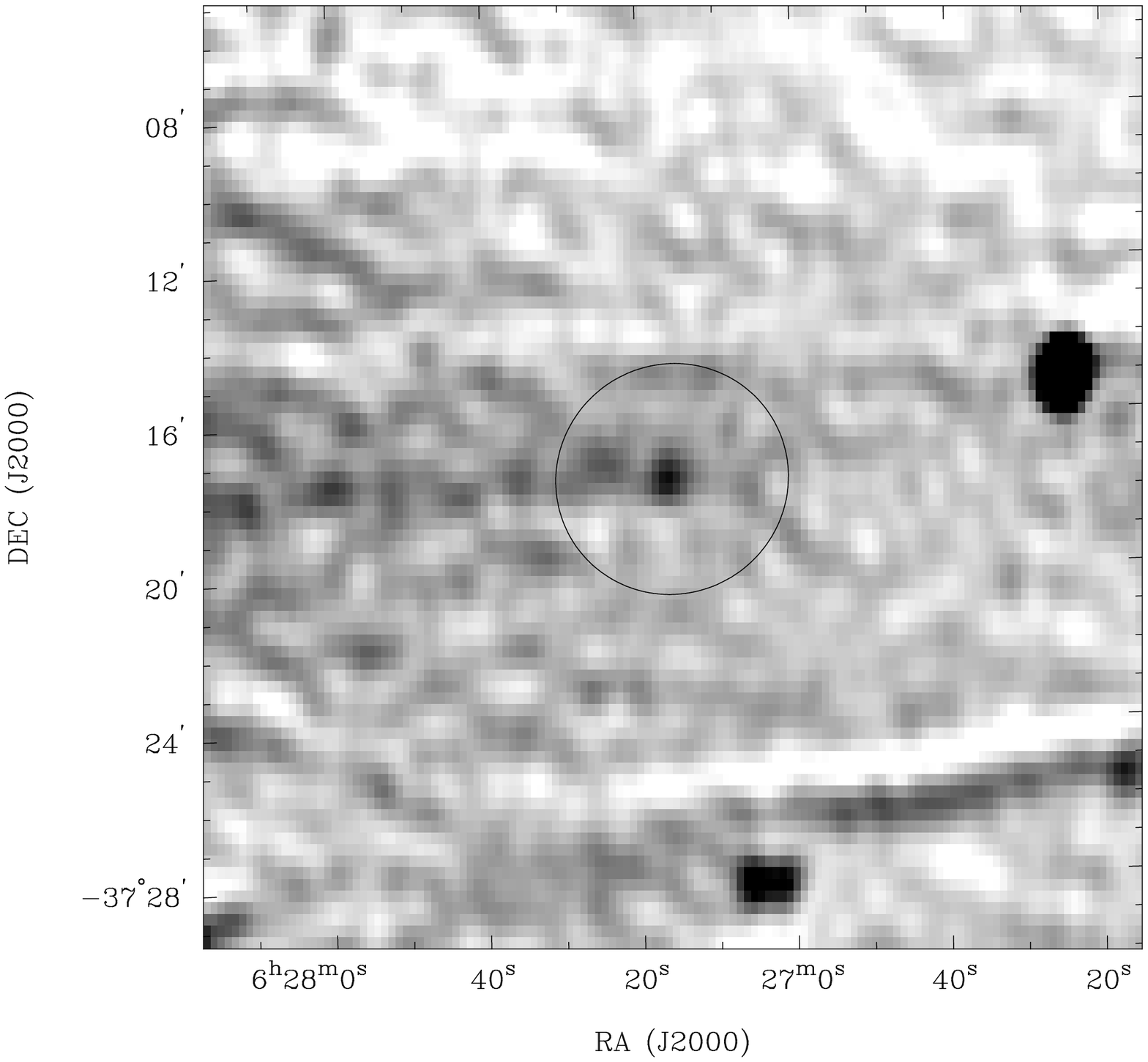}
\centering
\end{minipage}
\qquad
\begin{minipage}{0.3\textwidth}
\centering
\includegraphics[width=1\textwidth]{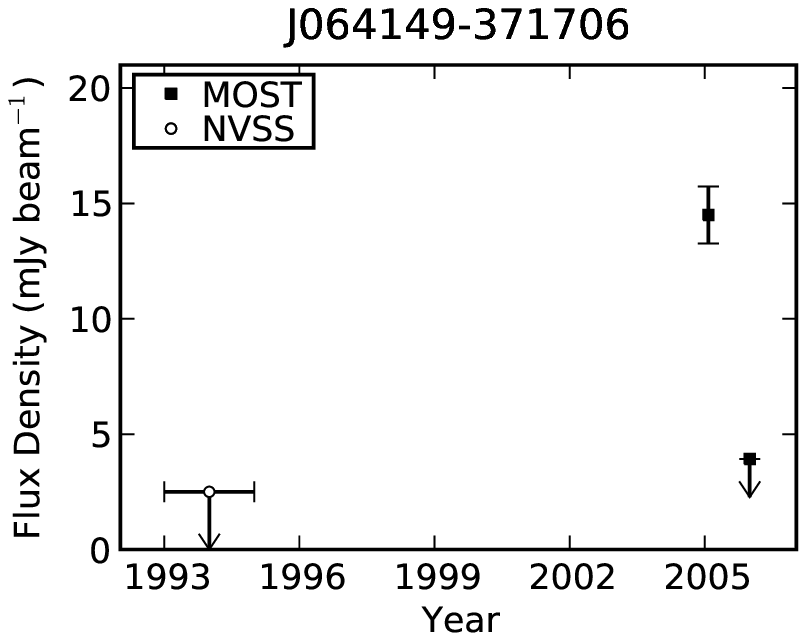}
\end{minipage}
\qquad
\begin{minipage}{0.3\textwidth}
\centering
\includegraphics[width=1\textwidth]{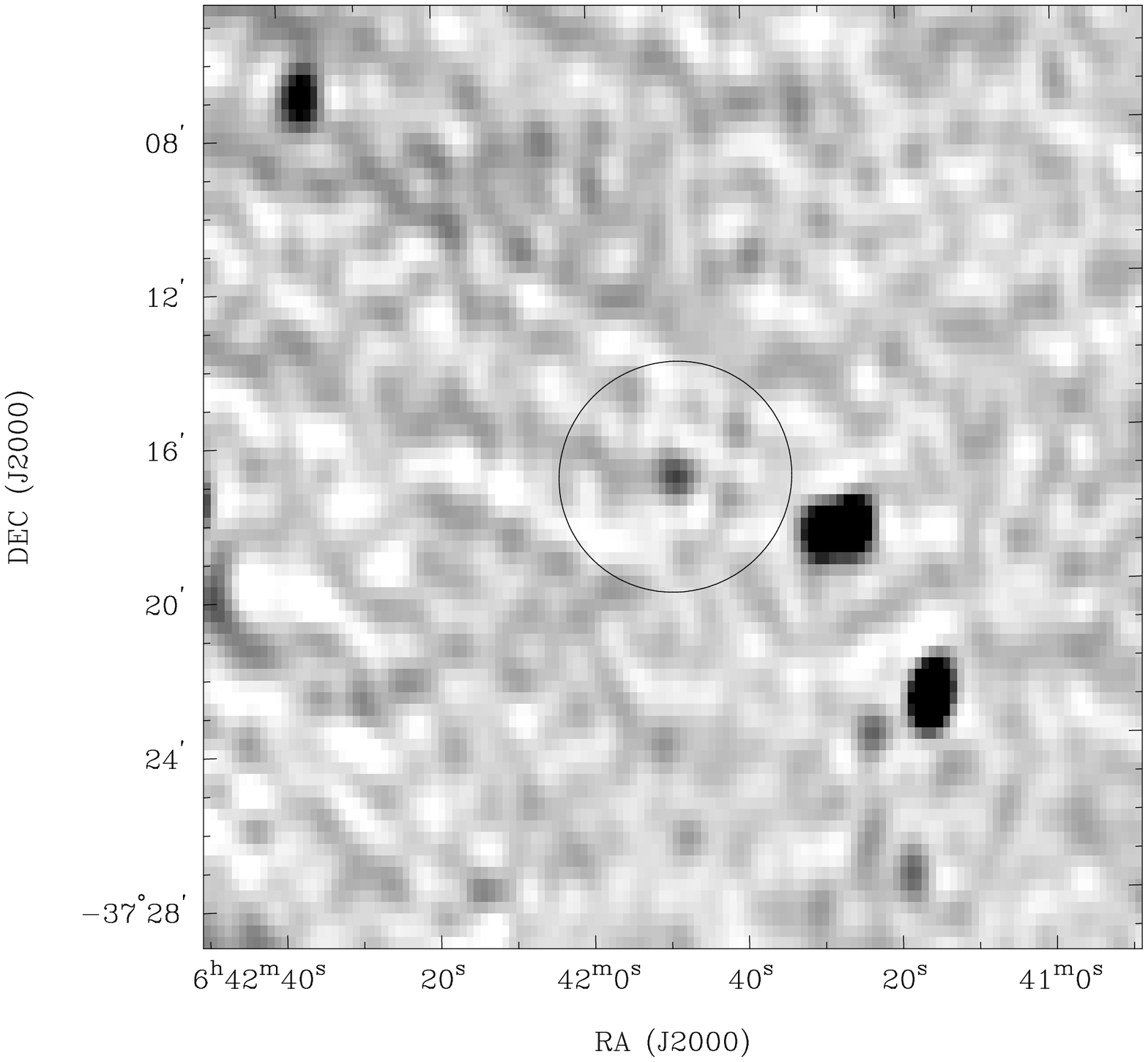}
\centering
\end{minipage}
\qquad
\begin{minipage}{0.3\textwidth}
\includegraphics[width=1\textwidth]{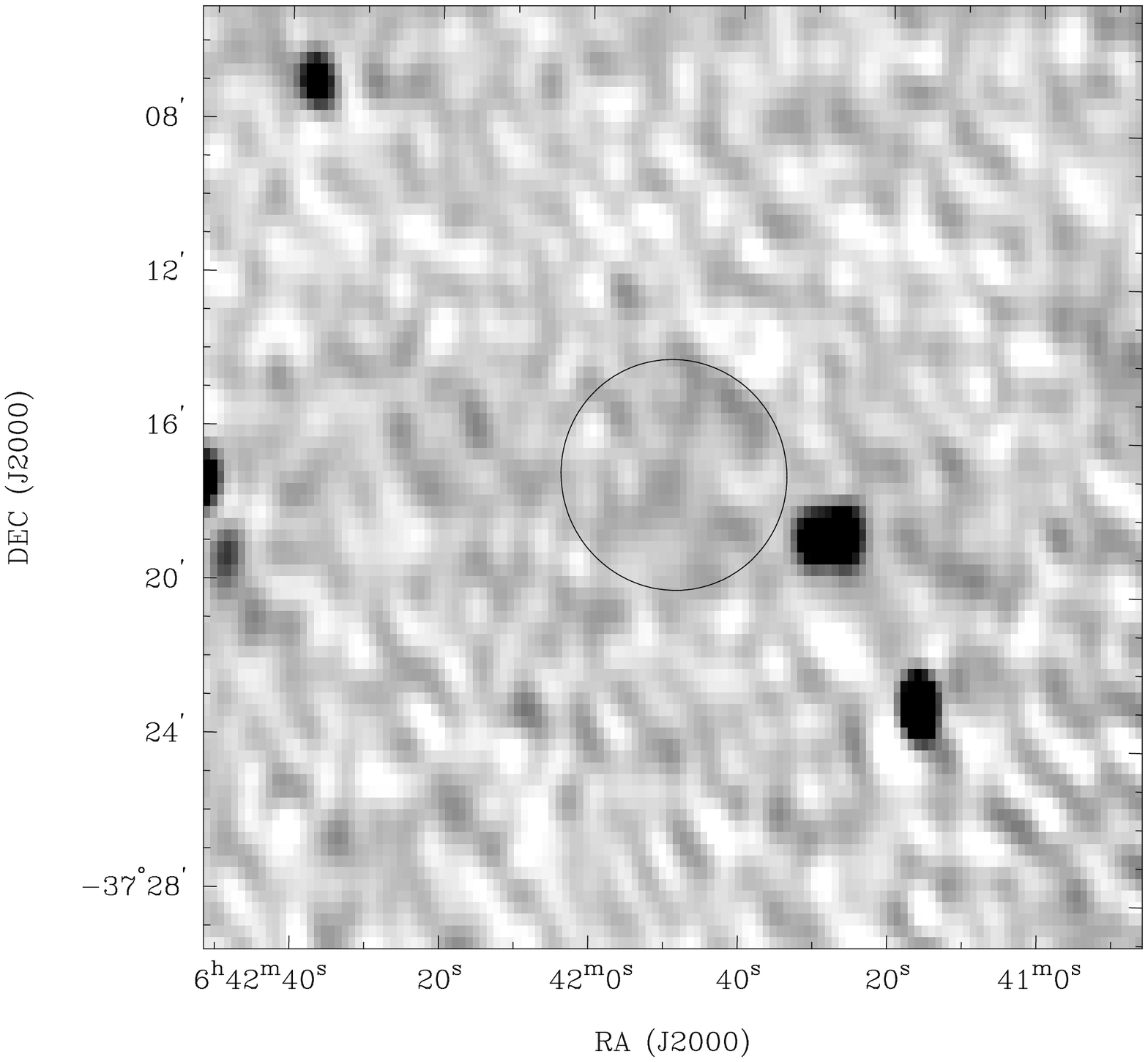}
\centering
\end{minipage}
\qquad
\begin{minipage}{0.3\textwidth}
\centering
\includegraphics[width=1\textwidth]{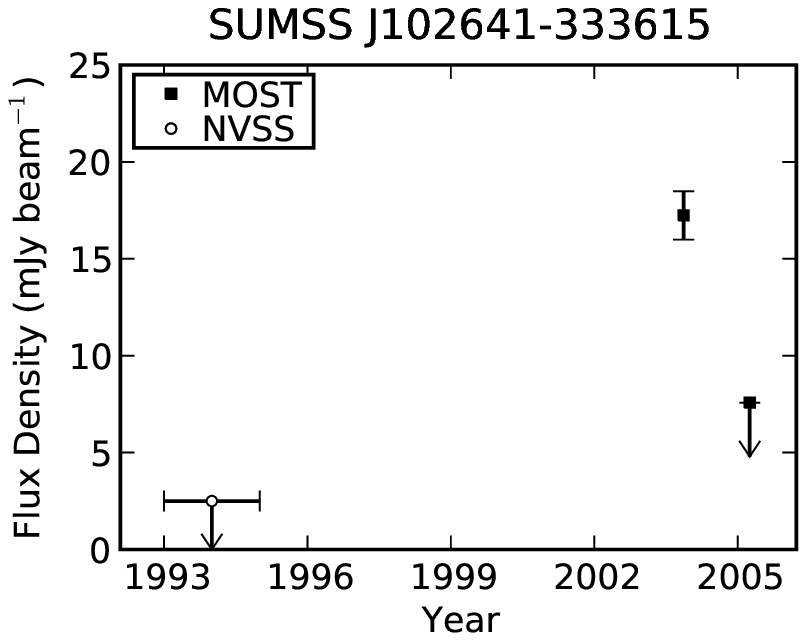}
\end{minipage}
\qquad
\begin{minipage}{0.3\textwidth}
\centering
\includegraphics[width=1\textwidth]{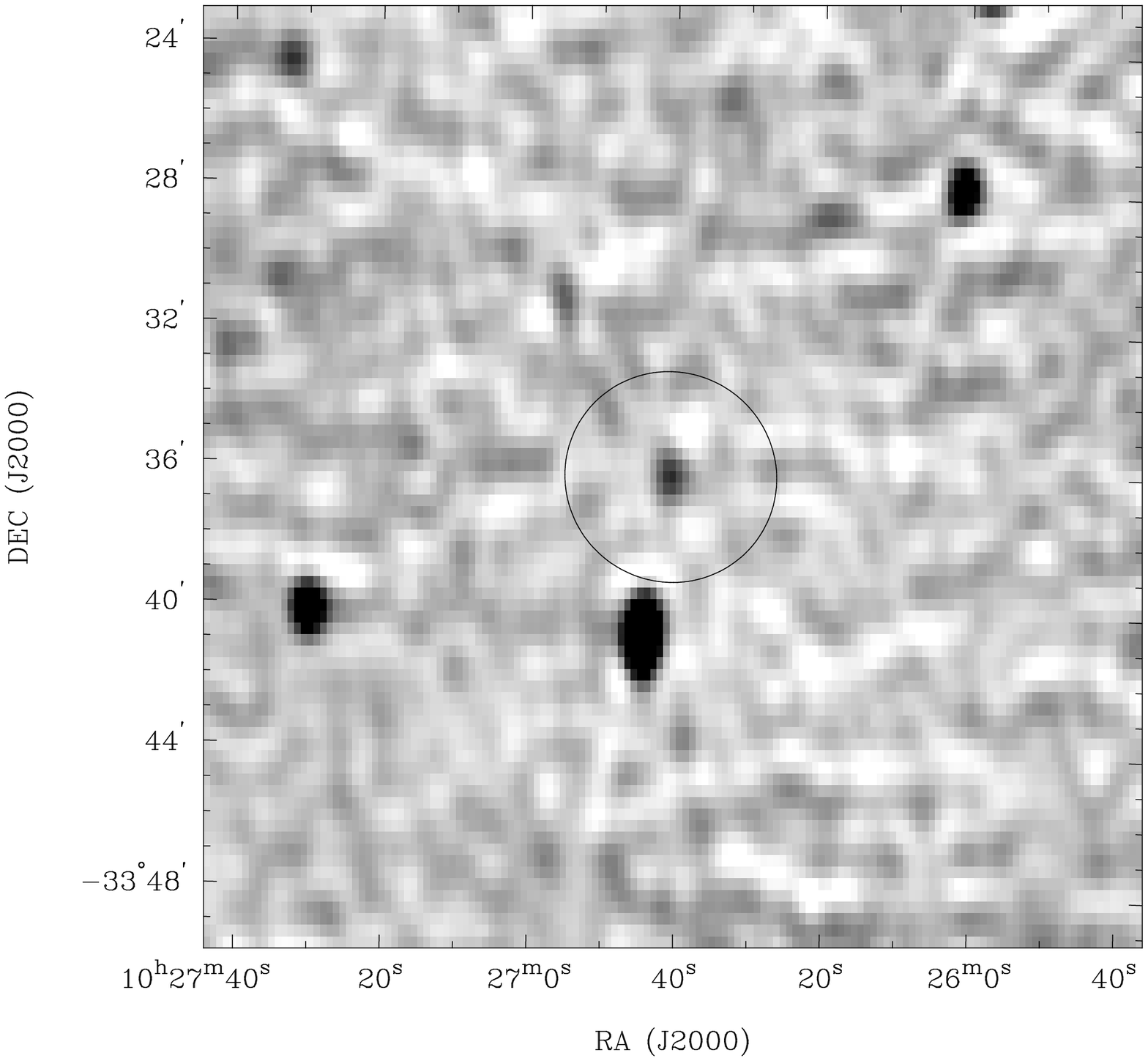}
\centering
\end{minipage}
\qquad
\begin{minipage}{0.3\textwidth}
\includegraphics[width=1\textwidth]{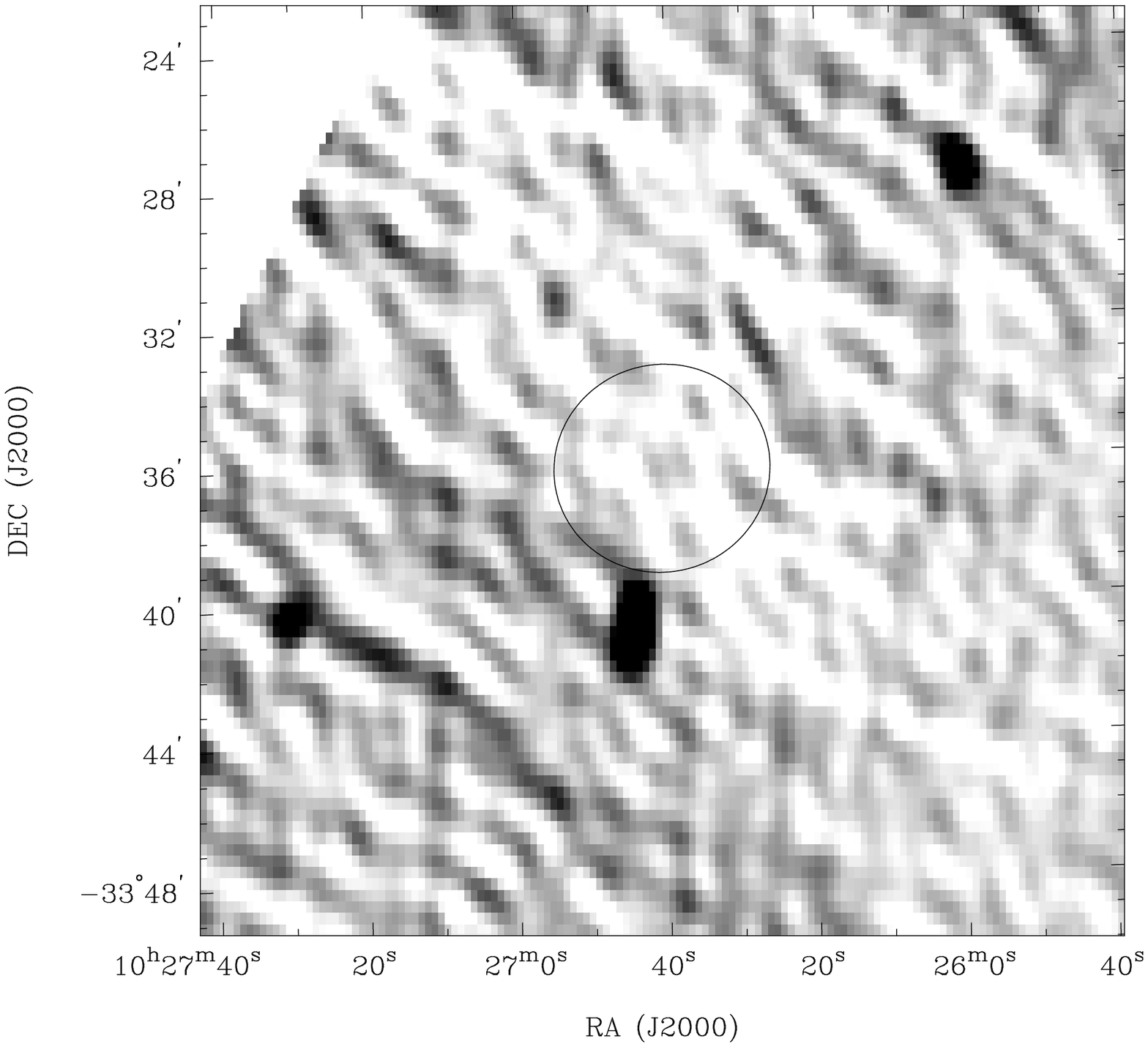}
\centering
\end{minipage}
\qquad
{Fig.~\ref{fig:transients_stamps} (continued)}.
\end{figure*}

\begin{figure*}
\begin{minipage}{0.3\textwidth}
\centering
\includegraphics[width=1\textwidth]{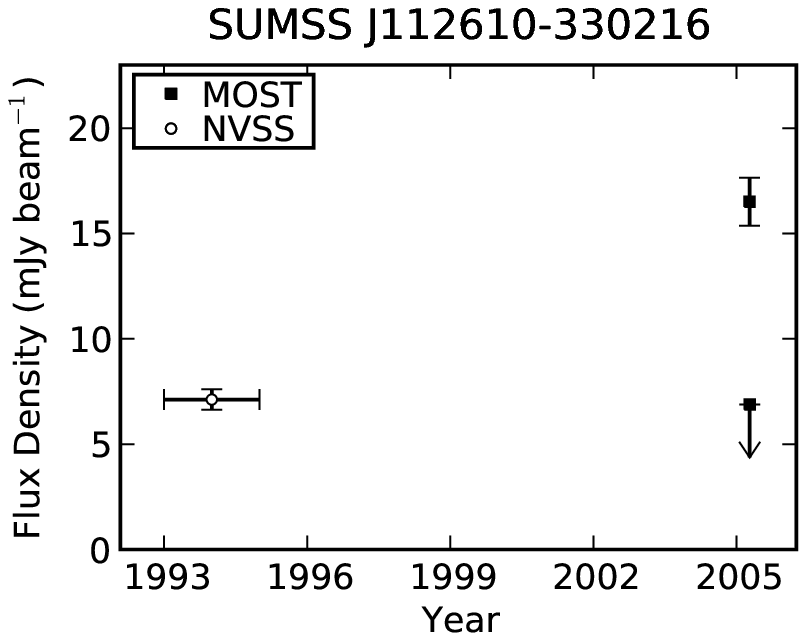}
\end{minipage}
\qquad
\begin{minipage}{0.3\textwidth}
\centering
\includegraphics[width=1\textwidth]{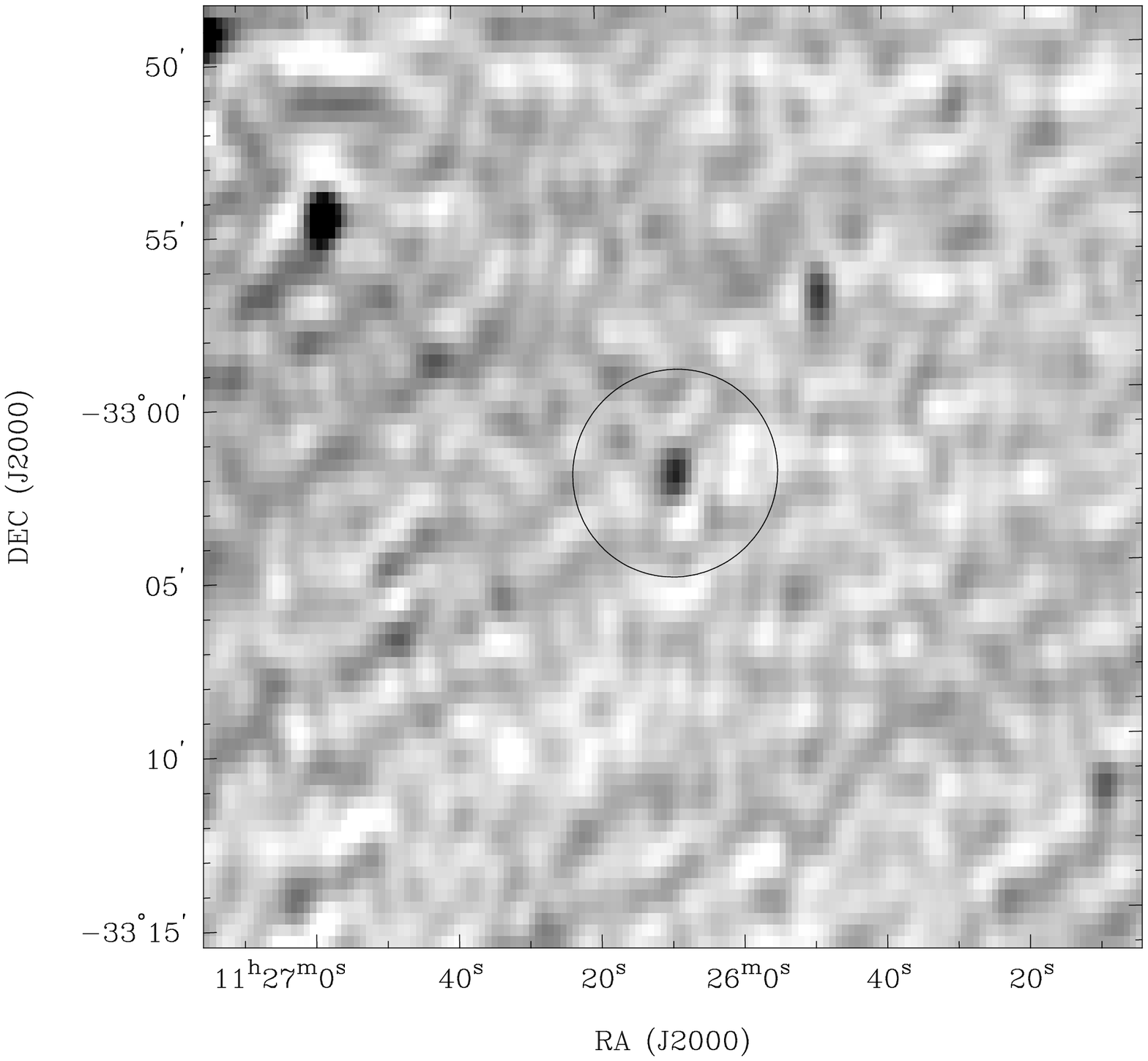}
\centering
\end{minipage}
\qquad
\begin{minipage}{0.3\textwidth}
\includegraphics[width=1\textwidth]{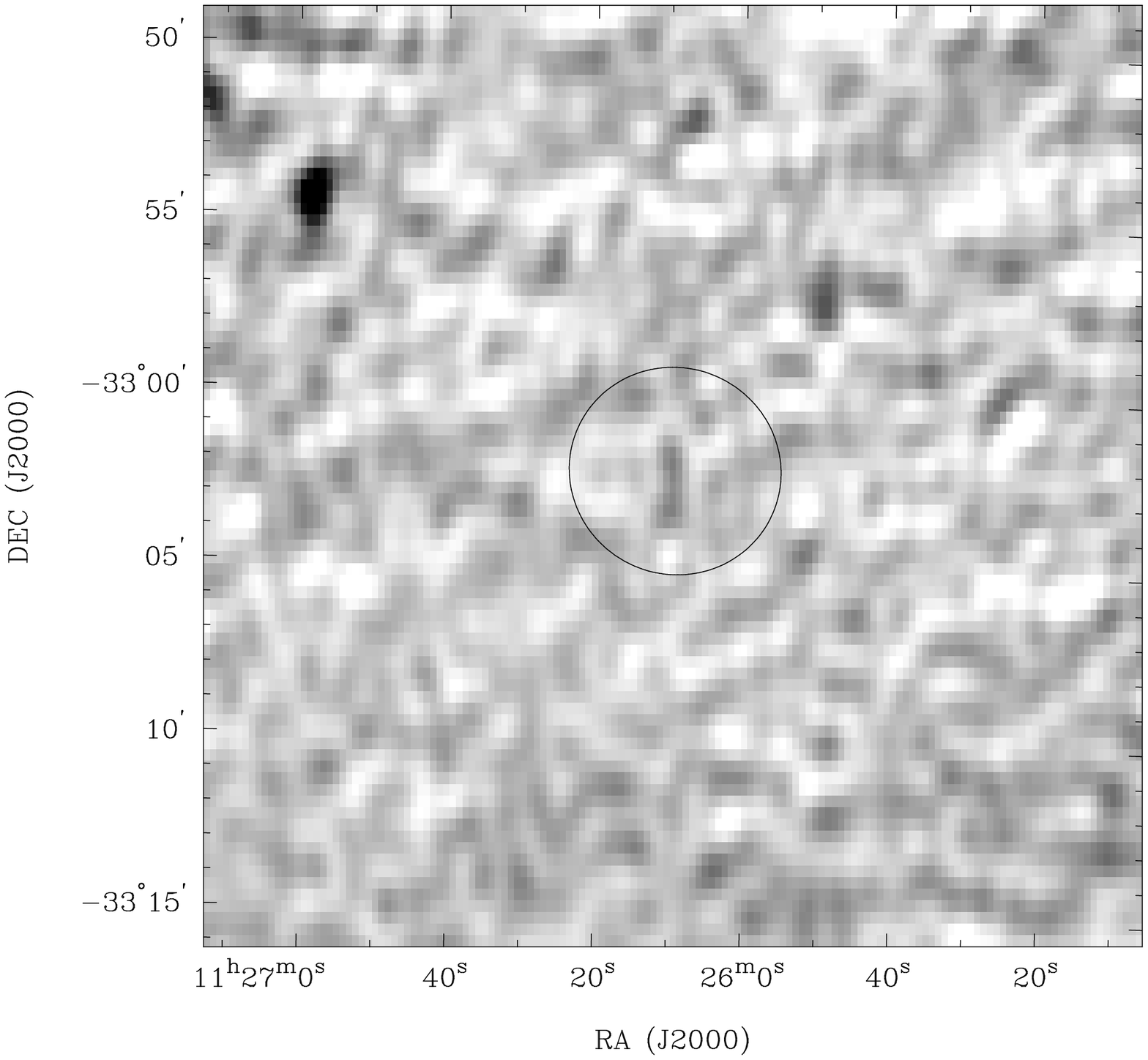}
\centering
\end{minipage}
\qquad
\begin{minipage}{0.3\textwidth}
\centering
\includegraphics[width=1\textwidth]{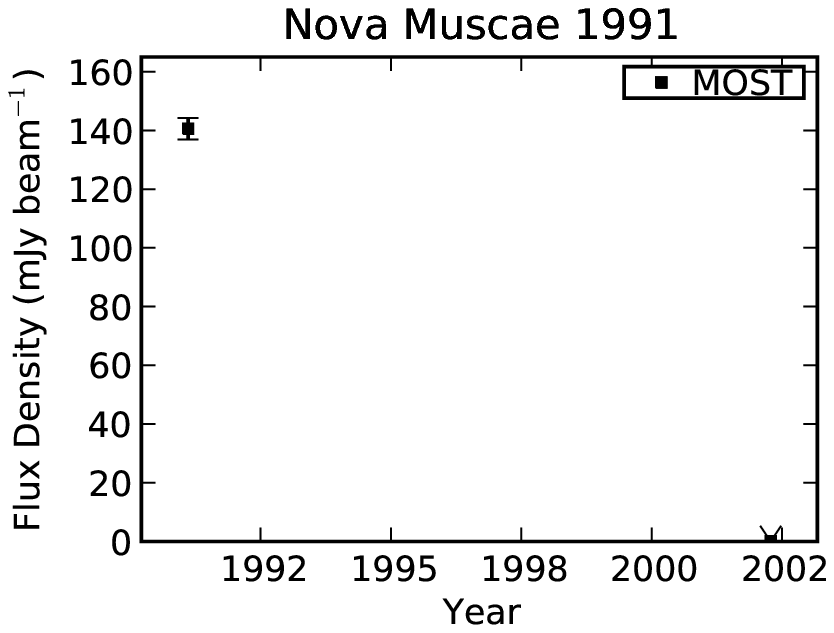}
\end{minipage}
\qquad
\begin{minipage}{0.3\textwidth}
\centering
\includegraphics[width=1\textwidth]{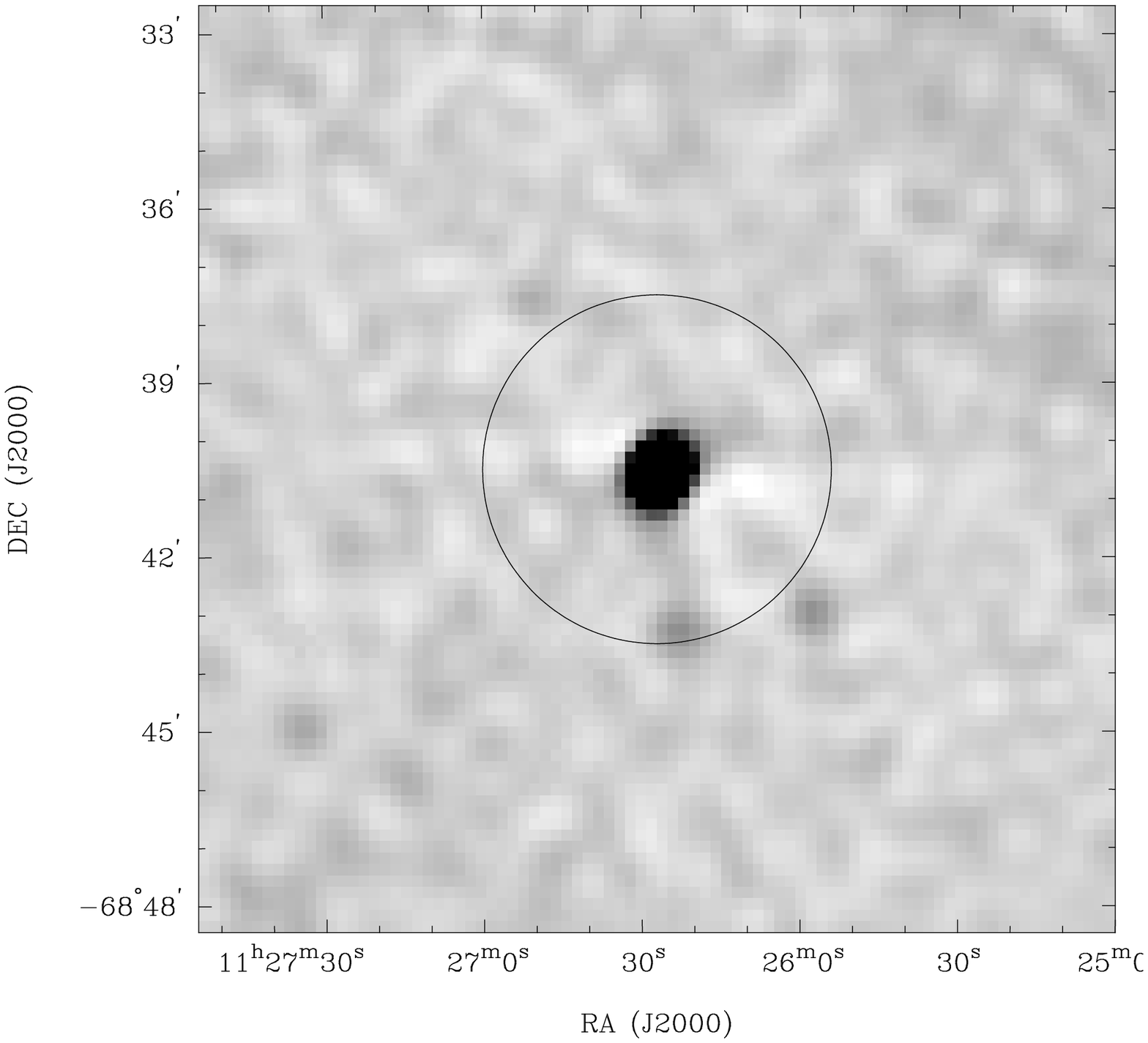}
\centering
\end{minipage}
\qquad
\begin{minipage}{0.3\textwidth}
\includegraphics[width=1\textwidth]{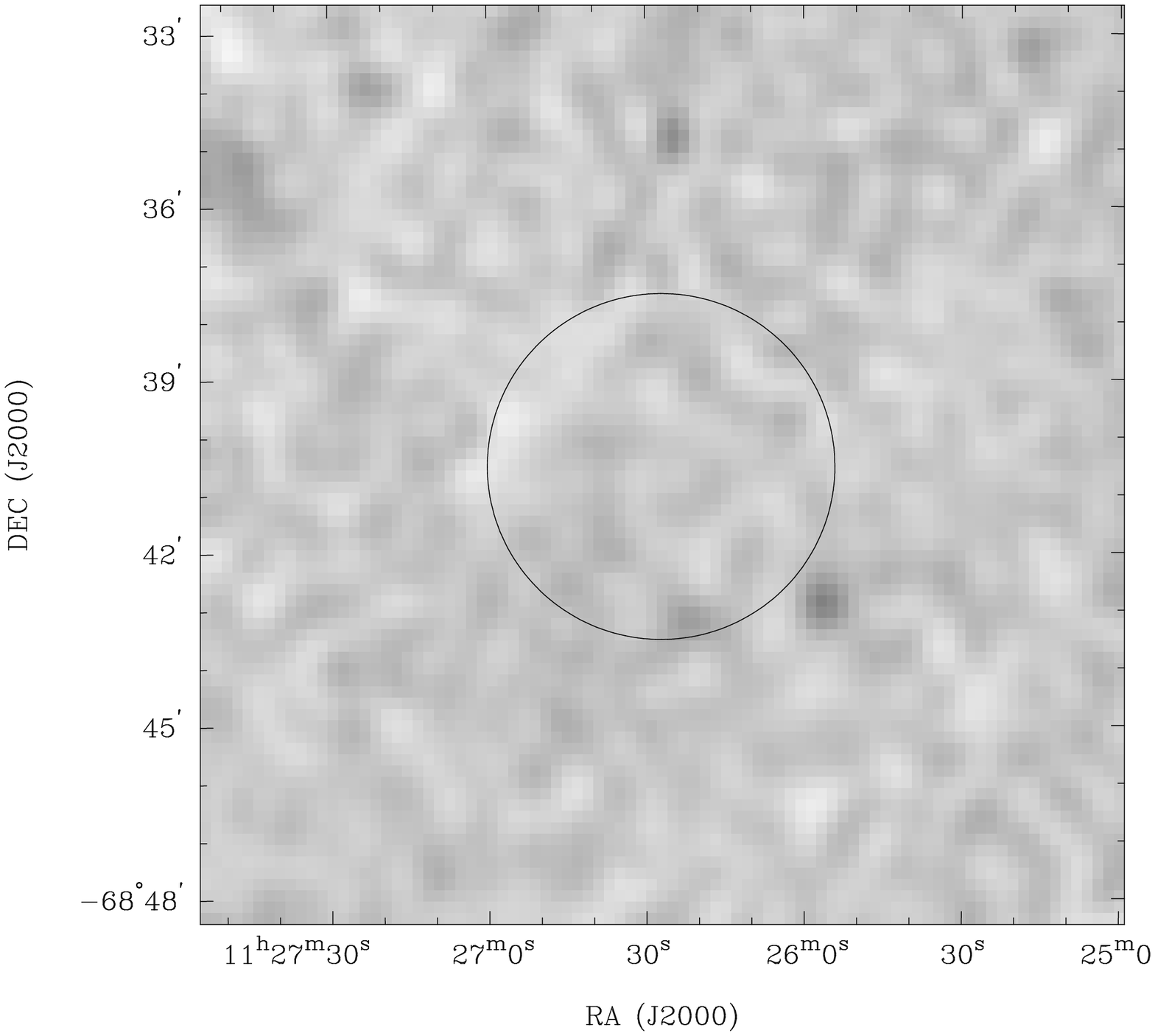}
\centering
\end{minipage}
\qquad
\begin{minipage}{0.3\textwidth}
\centering
\includegraphics[width=1\textwidth]{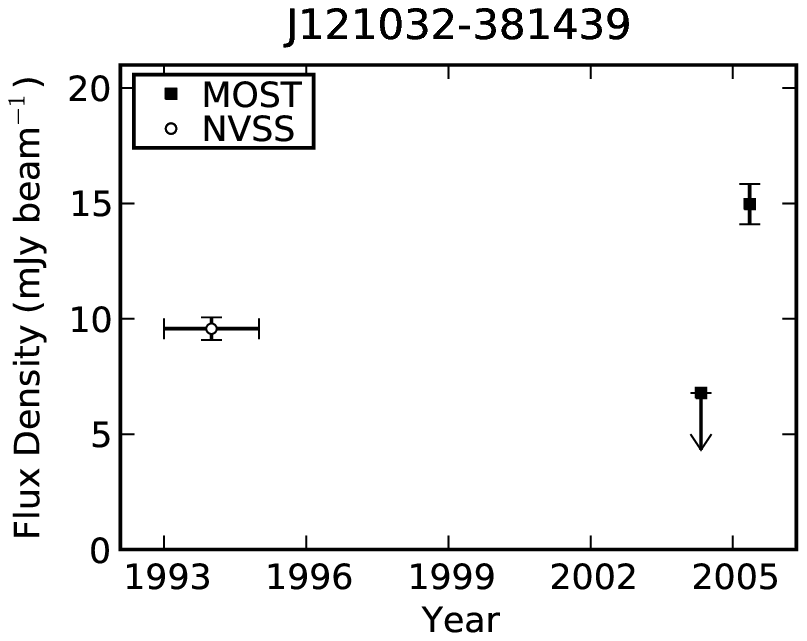}
\end{minipage}
\qquad
\begin{minipage}{0.3\textwidth}
\centering
\includegraphics[width=1\textwidth]{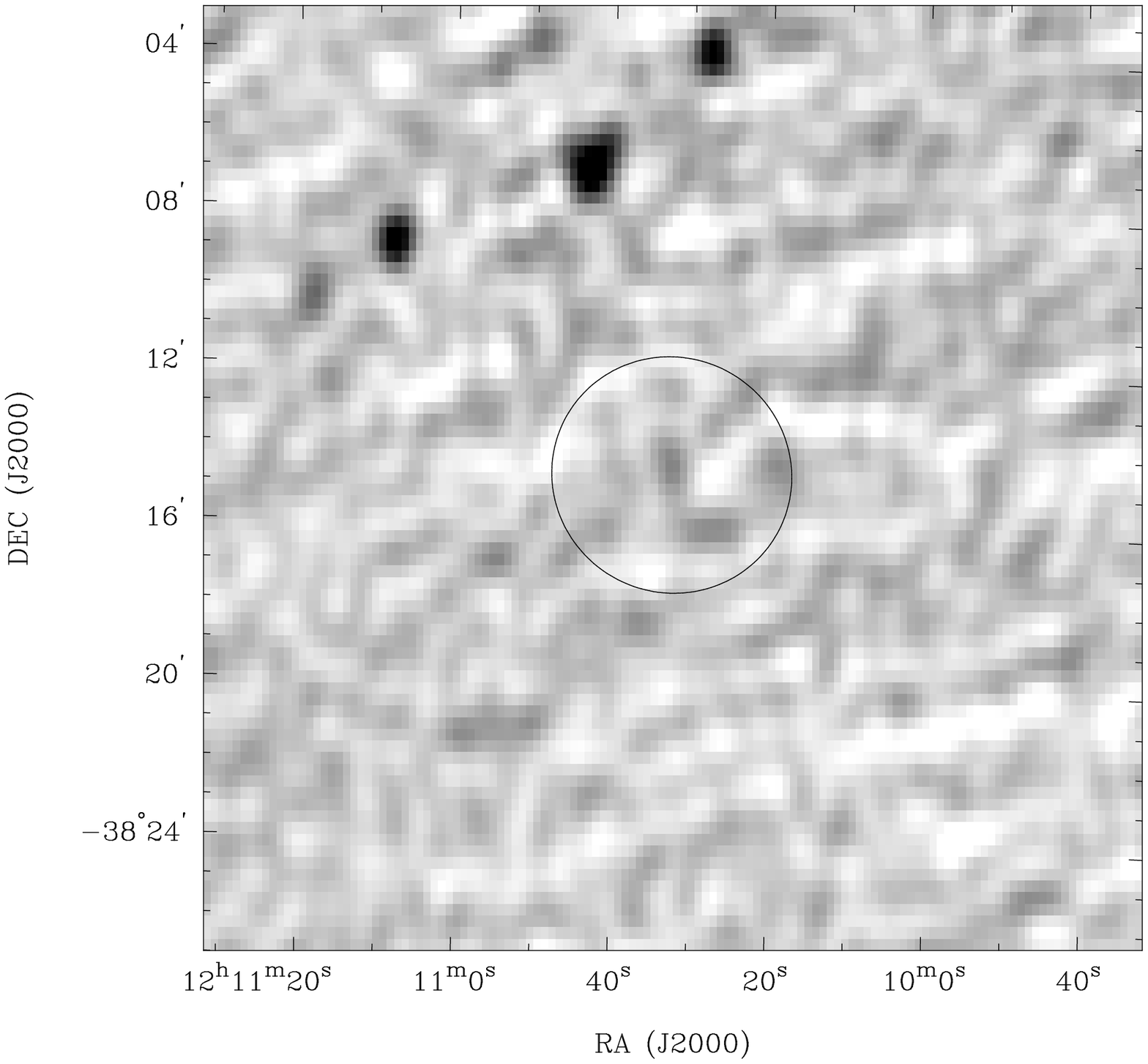}
\centering
\end{minipage}
\qquad
\begin{minipage}{0.3\textwidth}
\includegraphics[width=1\textwidth]{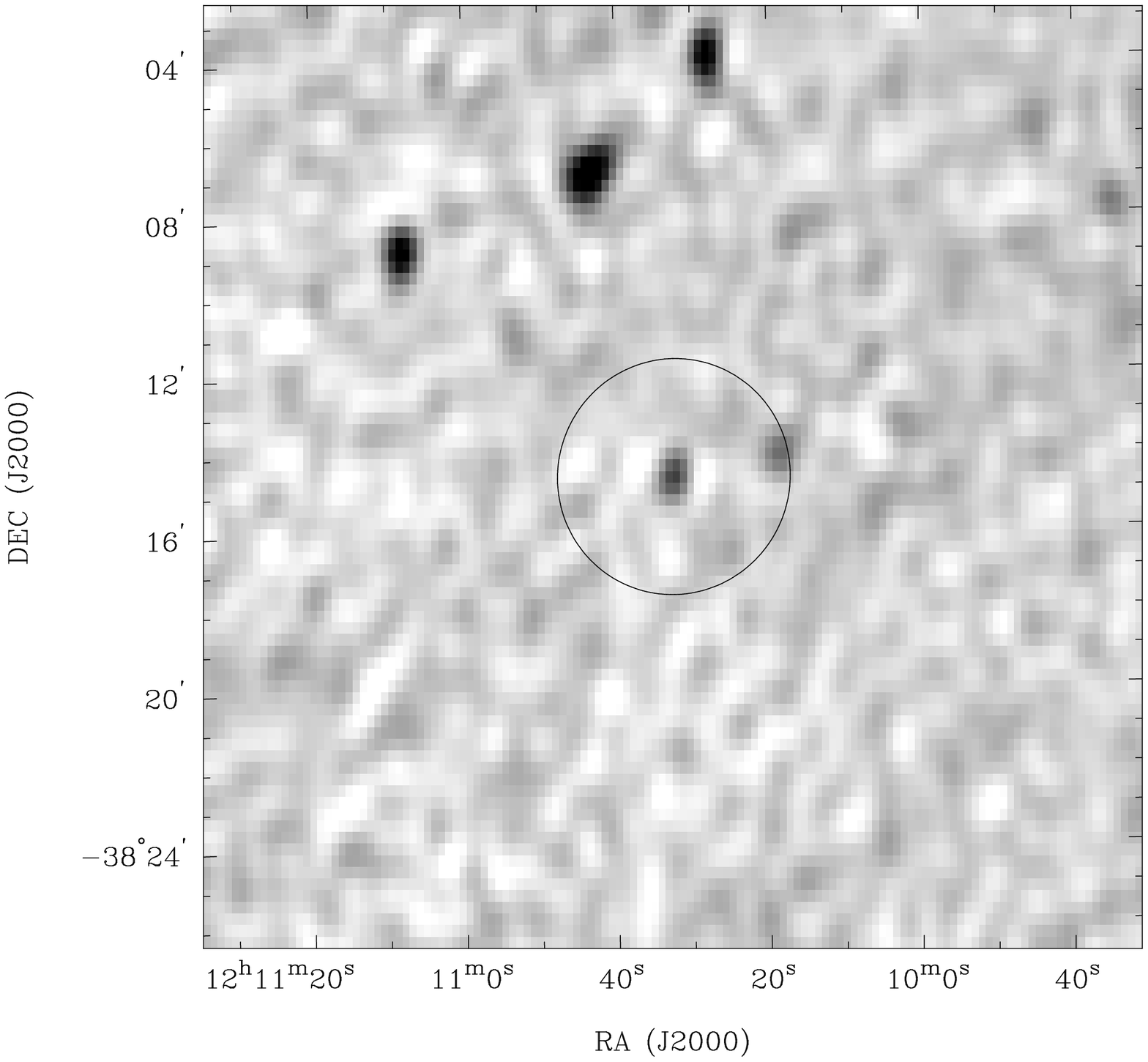}
\centering
\end{minipage}
\qquad
\begin{minipage}{0.3\textwidth}
\centering
\includegraphics[width=1\textwidth]{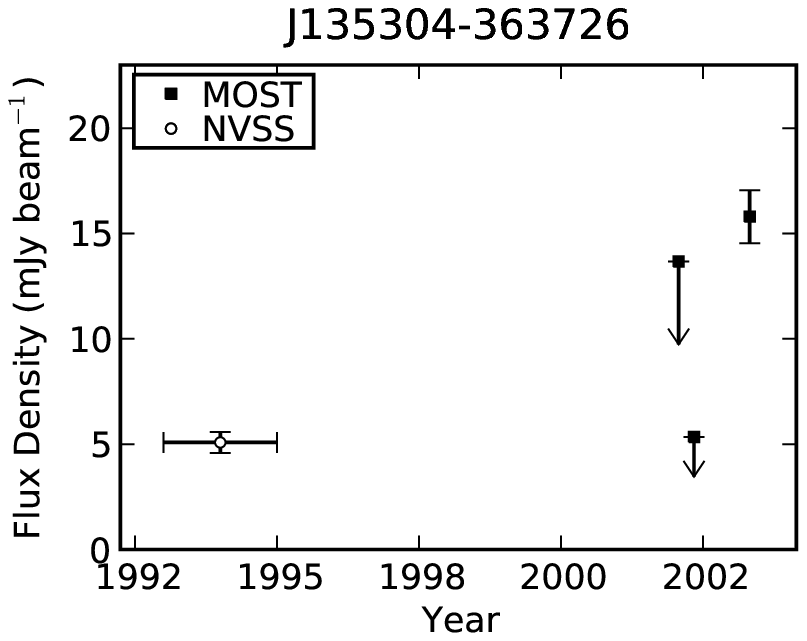}
\end{minipage}
\qquad
\begin{minipage}{0.3\textwidth}
\centering
\includegraphics[width=1\textwidth]{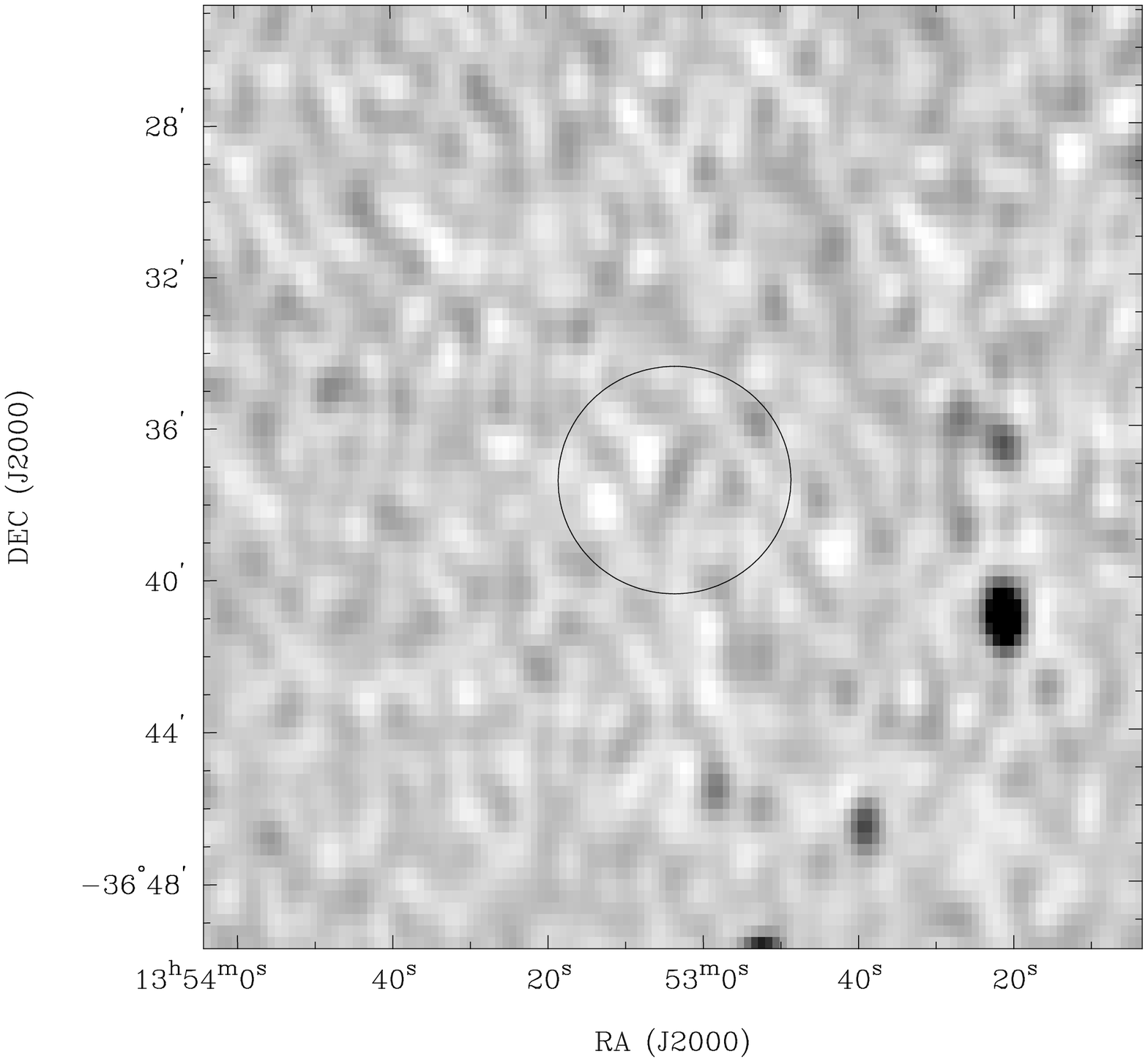}
\centering
\end{minipage}
\qquad
\begin{minipage}{0.3\textwidth}
\includegraphics[width=1\textwidth]{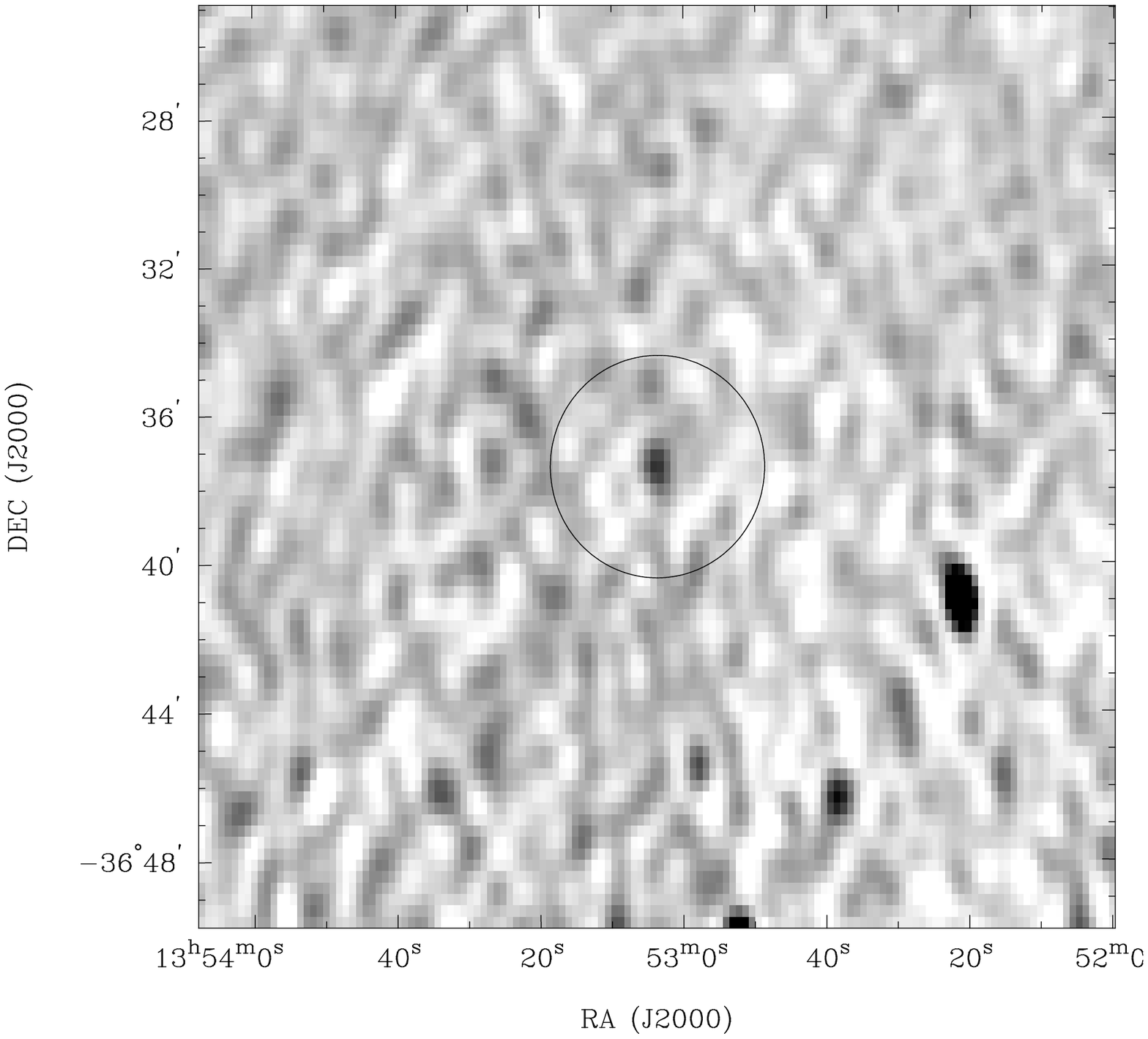}
\centering
\end{minipage}
\qquad

{Fig.~\ref{fig:transients_stamps} (continued)}.

\end{figure*}

\begin{figure*}
\begin{minipage}{0.3\textwidth}
\centering
\includegraphics[width=1\textwidth]{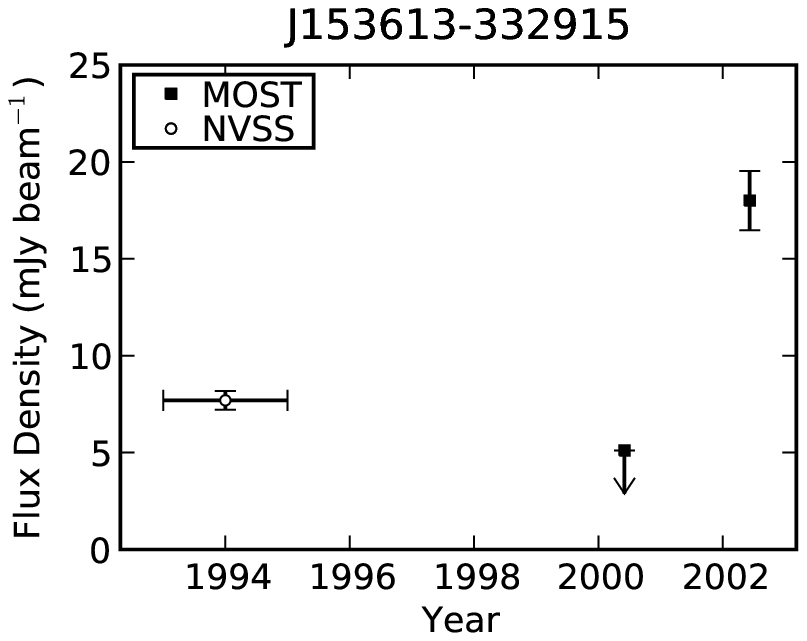}
\end{minipage}
\qquad
\begin{minipage}{0.3\textwidth}
\centering
\includegraphics[width=1\textwidth]{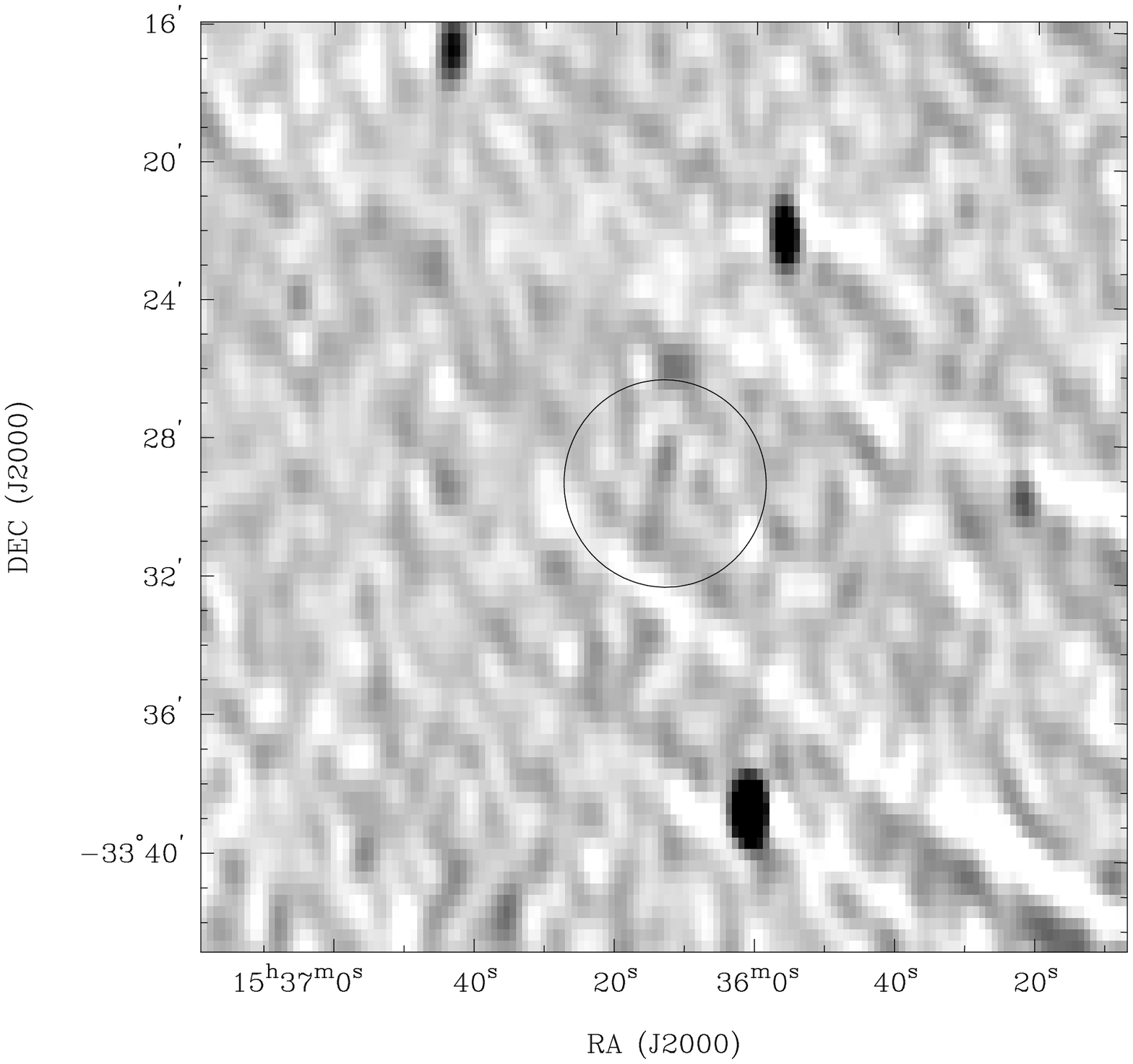}
\centering
\end{minipage}
\qquad
\begin{minipage}{0.3\textwidth}
\includegraphics[width=1\textwidth]{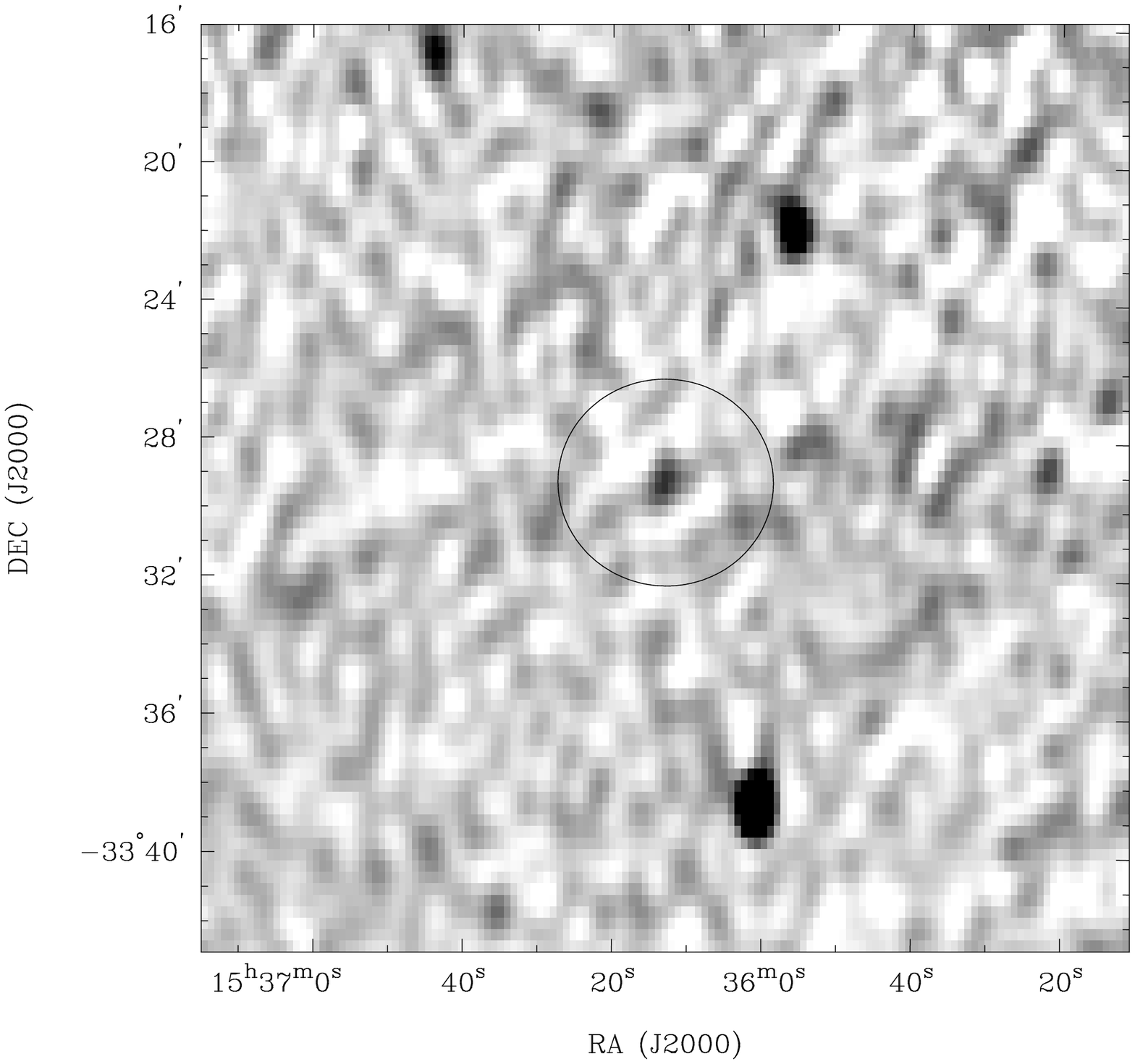}
\centering
\end{minipage}
\qquad
\begin{minipage}{0.3\textwidth}
\centering
\includegraphics[width=1\textwidth]{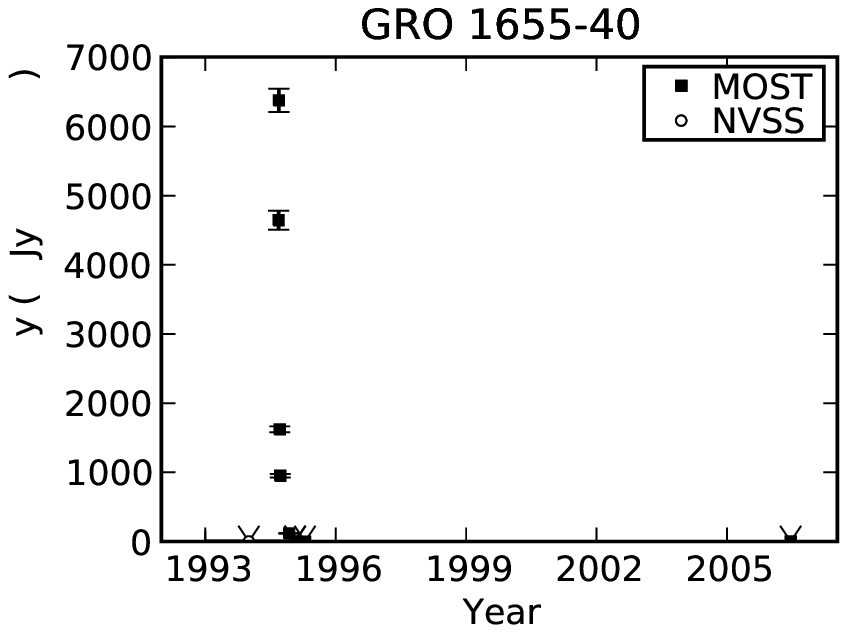}
\end{minipage}
\qquad
\begin{minipage}{0.3\textwidth}
\centering
\includegraphics[width=1\textwidth]{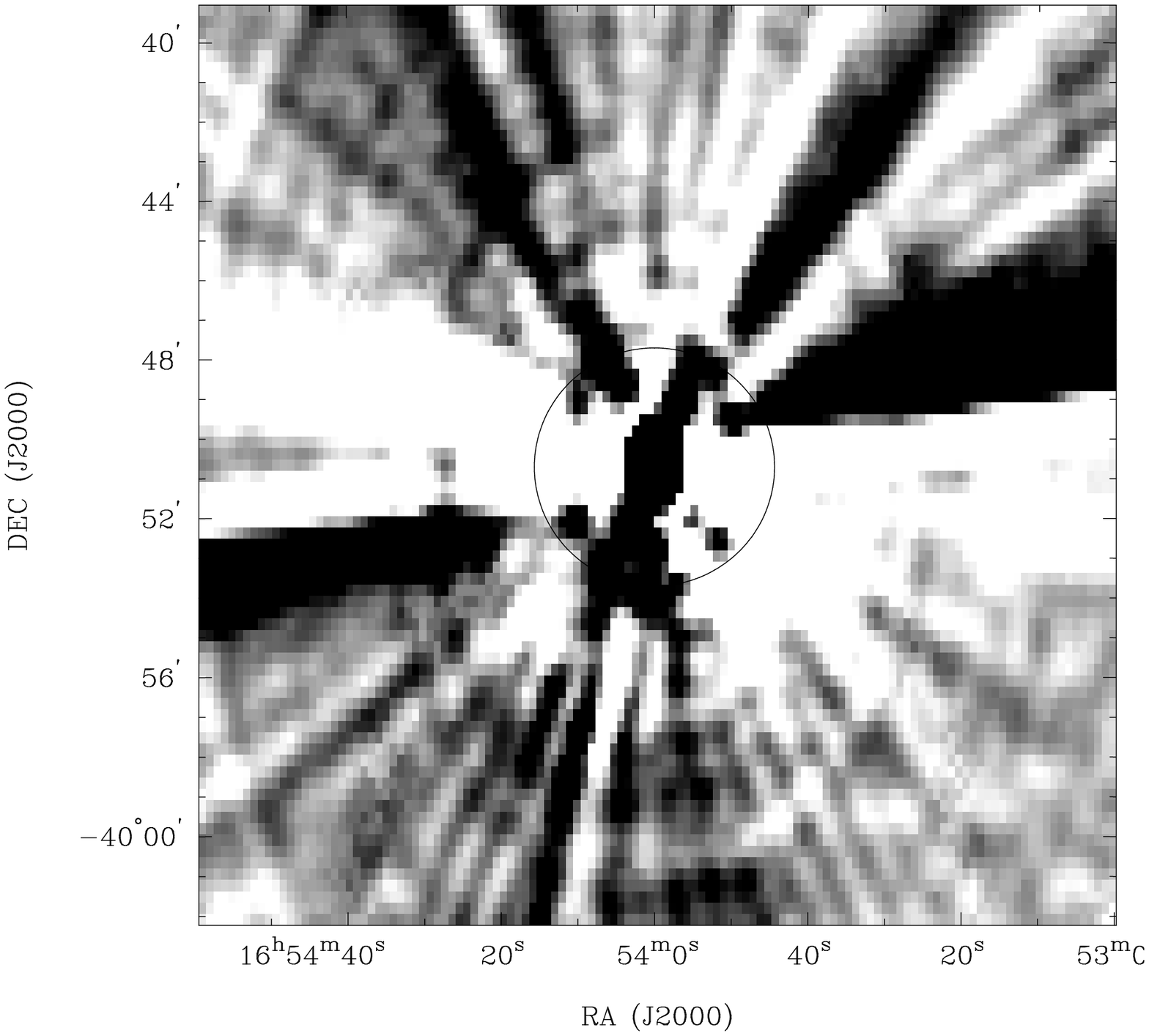}
\centering
\end{minipage}
\qquad
\begin{minipage}{0.3\textwidth}
\includegraphics[width=1\textwidth]{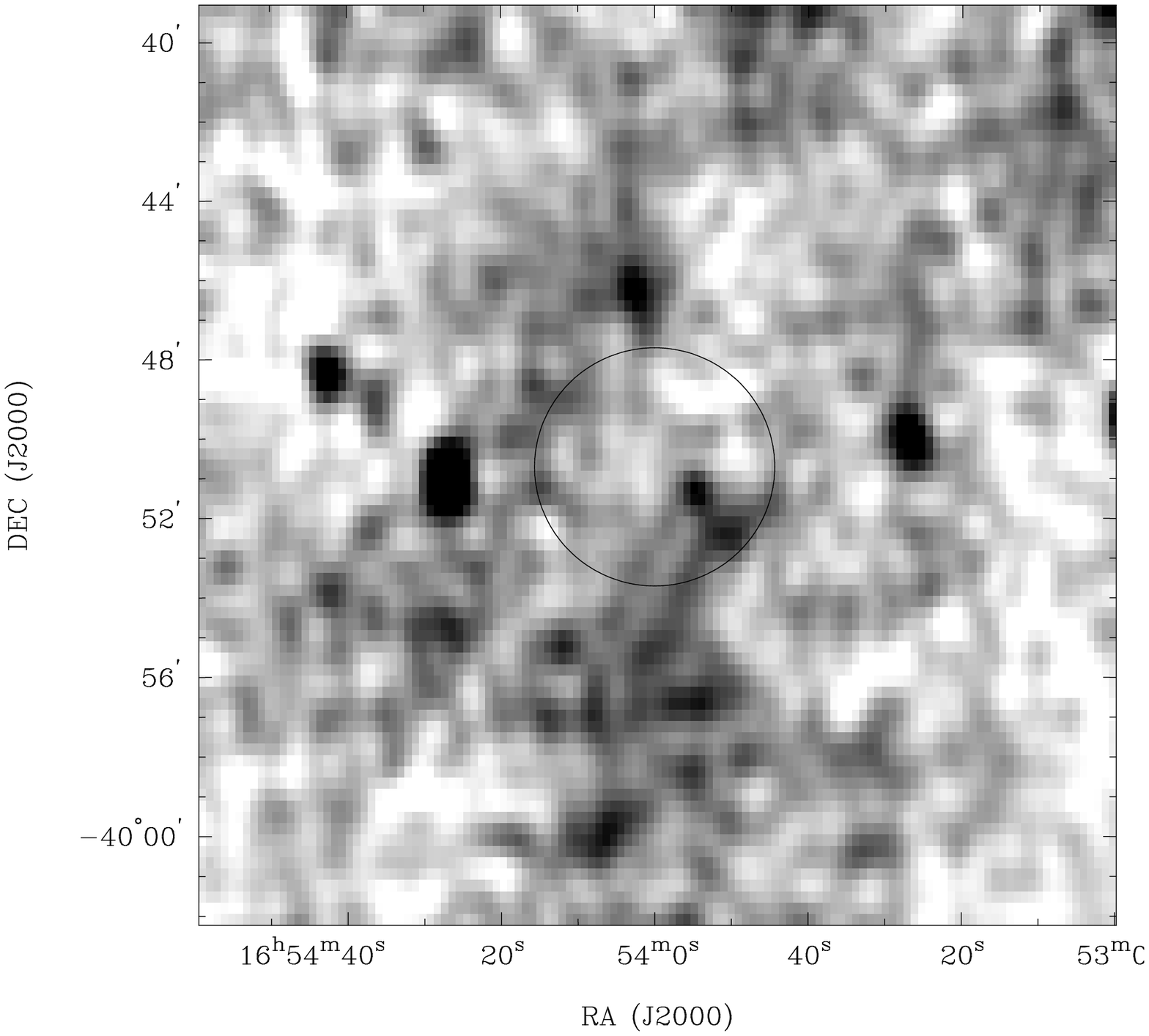}
\centering
\end{minipage}
\qquad
\begin{minipage}{0.3\textwidth}
\centering
\includegraphics[width=1\textwidth]{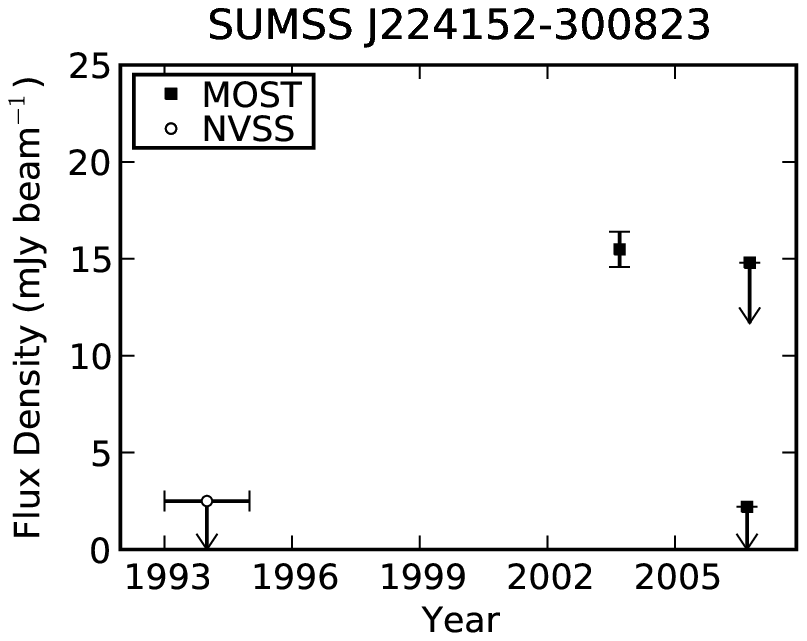}
\end{minipage}
\qquad
\begin{minipage}{0.3\textwidth}
\centering
\includegraphics[width=1\textwidth]{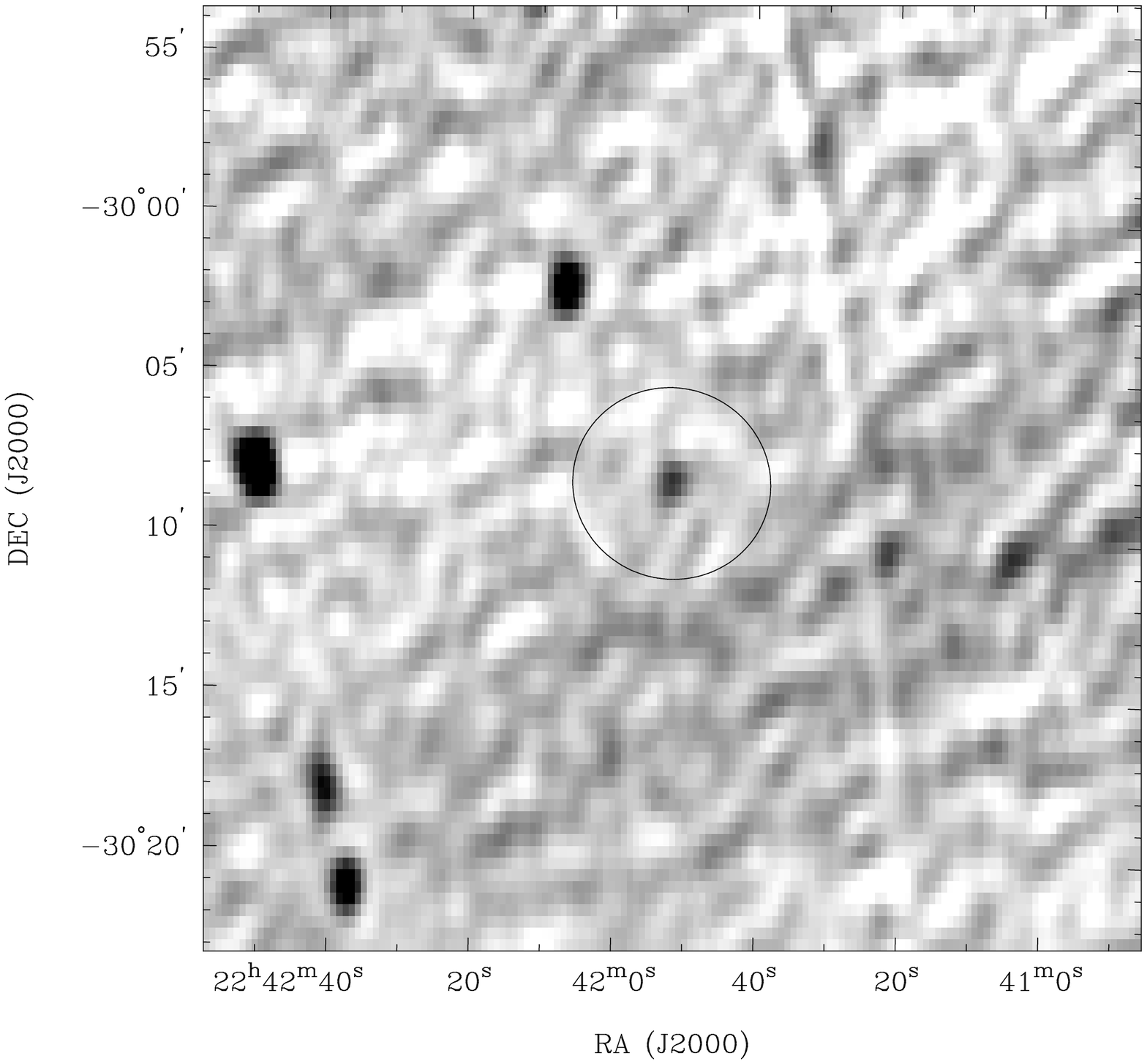}
\centering
\end{minipage}
\qquad
\begin{minipage}{0.3\textwidth}
\includegraphics[width=1\textwidth]{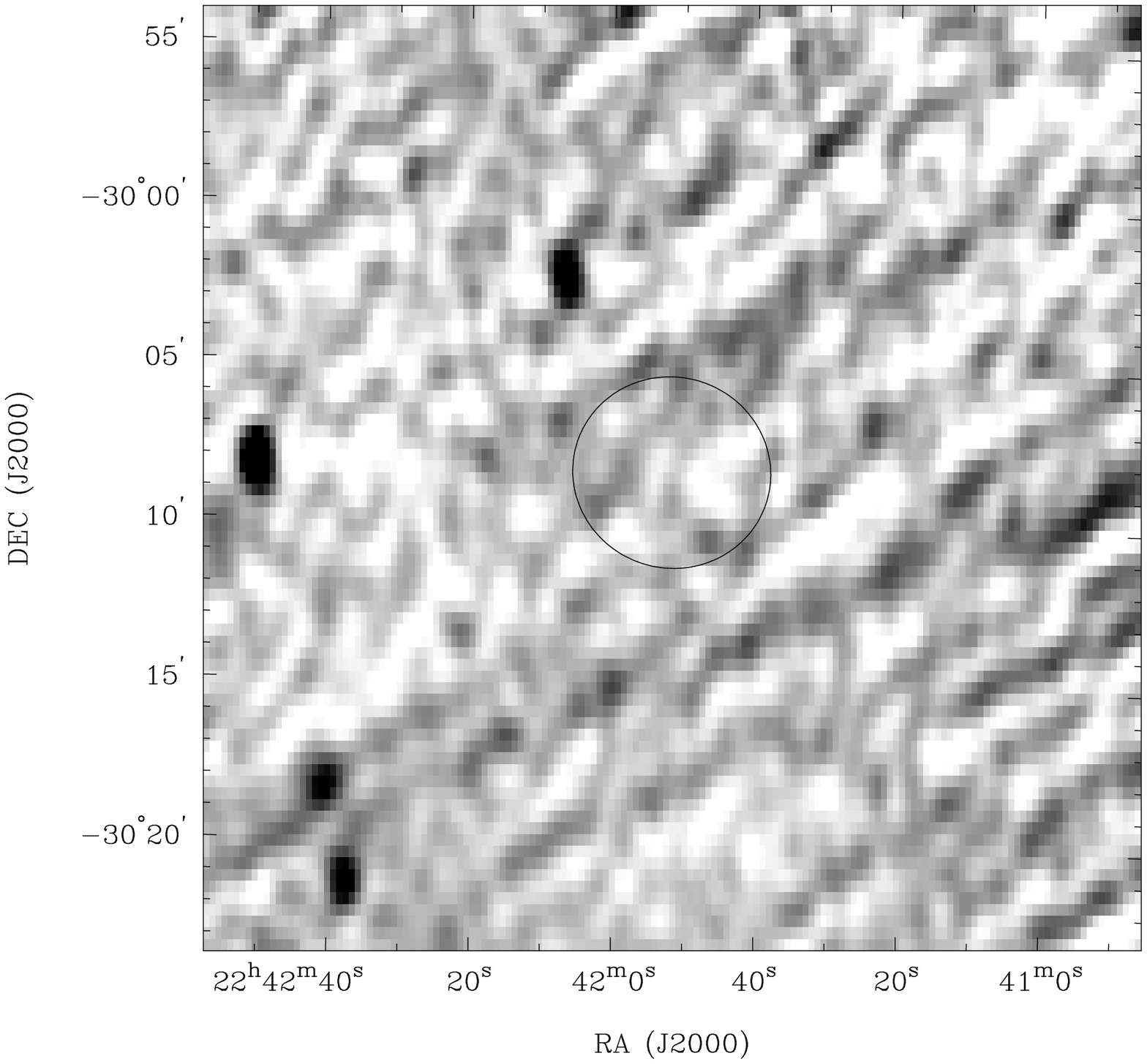}
\centering
\end{minipage}
\qquad

{Fig. \ref{fig:transients_stamps} (continued)}.

\end{figure*}

\bsp

\label{lastpage}
\end{document}